\PassOptionsToPackage{unicode}{hyperref}
\PassOptionsToPackage{hyphens}{url}
\PassOptionsToPackage{dvipsnames,svgnames,x11names}{xcolor}
\documentclass[12pt]{article}

\usepackage{appendix}
\usepackage{minitoc}
\usepackage{bibunits}
\usepackage{enumitem}
\usepackage{amsmath,amssymb,mathtools}
\usepackage{amsthm}
\usepackage{iftex}
\usepackage{algorithm, algpseudocode}
\usepackage{bm}
\usepackage{float}
\usepackage{mathrsfs}

\ifPDFTeX
  \usepackage[T1]{fontenc}
  \usepackage[utf8]{inputenc}
  \usepackage{textcomp} 
\else 
  \usepackage{unicode-math}
  \defaultfontfeatures{Scale=MatchLowercase}
  \defaultfontfeatures[\rmfamily]{Ligatures=TeX,Scale=1}
\fi
\usepackage{lmodern}
\ifPDFTeX\else  
\fi
\IfFileExists{upquote.sty}{\usepackage{upquote}}{}
\IfFileExists{microtype.sty}{
  \usepackage[]{microtype}
  \UseMicrotypeSet[protrusion]{basicmath} 
}{}
\makeatletter
\@ifundefined{KOMAClassName}{
  \IfFileExists{parskip.sty}{%
    \usepackage{parskip}
  }{
    \setlength{\parindent}{0pt}
    \setlength{\parskip}{6pt plus 2pt minus 1pt}}
}{
  \KOMAoptions{parskip=half}}
\makeatother
\usepackage{xcolor}
\makeatletter
\ifx\paragraph\undefined\else
  \let\oldparagraph\paragraph
  \renewcommand{\paragraph}{
    \@ifstar
      \xxxParagraphStar
      \xxxParagraphNoStar
  }
  \newcommand{\xxxParagraphStar}[1]{\oldparagraph*{#1}\mbox{}}
  \newcommand{\xxxParagraphNoStar}[1]{\oldparagraph{#1}\mbox{}}
\fi
\ifx\subparagraph\undefined\else
  \let\oldsubparagraph\subparagraph
  \renewcommand{\subparagraph}{
    \@ifstar
      \xxxSubParagraphStar
      \xxxSubParagraphNoStar
  }
  \newcommand{\xxxSubParagraphStar}[1]{\oldsubparagraph*{#1}\mbox{}}
  \newcommand{\xxxSubParagraphNoStar}[1]{\oldsubparagraph{#1}\mbox{}}
\fi
\makeatother

\usepackage{longtable,booktabs,array}
\usepackage{calc} 
\usepackage{etoolbox}
\makeatletter
\patchcmd\longtable{\par}{\if@noskipsec\mbox{}\fi\par}{}{}
\makeatother
\IfFileExists{footnotehyper.sty}{\usepackage{footnotehyper}}{\usepackage{footnote}}
\makesavenoteenv{longtable}
\usepackage{graphicx}
\makeatletter
\def\maxwidth{\ifdim\Gin@nat@width>\linewidth\linewidth\else\Gin@nat@width\fi}
\def\maxheight{\ifdim\Gin@nat@height>\textheight\textheight\else\Gin@nat@height\fi}
\makeatother
\setkeys{Gin}{width=\maxwidth,height=\maxheight,keepaspectratio}
\makeatletter
\def\fps@figure{htbp}
\makeatother

\makeatletter
\@ifpackageloaded{caption}{}{\usepackage{caption}}
\AtBeginDocument{%
\ifdefined\contentsname
  \renewcommand*\contentsname{Table of contents}
\else
  \newcommand\contentsname{Table of contents}
\fi
\ifdefined\listfigurename
  \renewcommand*\listfigurename{List of Figures}
\else
  \newcommand\listfigurename{List of Figures}
\fi
\ifdefined\listtablename
  \renewcommand*\listtablename{List of Tables}
\else
  \newcommand\listtablename{List of Tables}
\fi
\ifdefined\figurename
  \renewcommand*\figurename{Figure}
\else
  \newcommand\figurename{Figure}
\fi
\ifdefined\tablename
  \renewcommand*\tablename{Table}
\else
  \newcommand\tablename{Table}
\fi
}
\@ifpackageloaded{float}{}{\usepackage{float}}
\floatstyle{ruled}
\@ifundefined{c@chapter}{\newfloat{codelisting}{h}{lop}}{\newfloat{codelisting}{h}{lop}[chapter]}
\floatname{codelisting}{Listing}

\makeatother
\makeatletter
\@ifpackageloaded{caption}{}{\usepackage{caption}}
\@ifpackageloaded{subcaption}{}{\usepackage{subcaption}}
\makeatother

\ifLuaTeX
  \usepackage{selnolig}  
\fi
\usepackage[]{natbib}
\usepackage{bookmark}

\IfFileExists{xurl.sty}{\usepackage{xurl}}{} 
\hypersetup{
  pdftitle={Title},
  pdfauthor={Author 1; Author 2},
  pdfkeywords={3 to 6 keywords, that do not appear in the title},
  colorlinks=true,
  linkcolor={blue},
  filecolor={Maroon},
  citecolor={Blue},
  urlcolor={Blue},
  pdfcreator={LaTeX via pandoc}}

\usepackage{cleveref}
\newcommand{\anon}{1}

\DeclareMathOperator{\dom}{dom}
\DeclareMathOperator*{\argmin}{arg\,min}
\DeclareMathOperator*{\argmax}{arg\,max}
\DeclareMathOperator{\col}{col}
\DeclareMathOperator{\var}{Var}
\DeclareMathOperator{\cov}{Cov}
\DeclareMathOperator{\vol}{vol}
\DeclareMathOperator{\diag}{diag}

\DeclareMathOperator*{\esssup}{ess\,sup}
\newcommand{\erm}{\mathrm{ERM}}
\newcommand{\dro}{\mathrm{DRO}}
\newcommand{\op}{\mathrm{op}}
\newcommand{\hvg}{\mathrm{hvg}}
\newcommand{\readd}{\mathrm{READ}}
\newcommand{\init}{\mathrm{init}}

\newtheorem{theorem}{Theorem}[section]
\newtheorem{lemma}{Lemma}[section]
\newtheorem{proposition}{Proposition}[section]
\newtheorem{corollary}{Corollary}[section]

\theoremstyle{definition}
\newtheorem{definition}{Definition}[section]
\newtheorem{assumption}{Assumption}[section]

\theoremstyle{remark}
\newtheorem{remark}{Remark}[section]

\crefname{theorem}{Theorem}{Theorems}
\crefname{lemma}{Lemma}{Lemmas}
\crefname{proposition}{Proposition}{Propositions}
\crefname{corollary}{Corollary}{Corollaries}
\crefname{definition}{Definition}{Definitions}
\crefname{assumption}{Assumption}{Assumptions}
\crefname{example}{Example}{Examples}
\crefname{remark}{Remark}{Remarks}
\crefname{equation}{Equation}{Equations}
\crefalias{enumi}{assumption}

\crefname{figure}{Figure}{Figures}
\Crefname{figure}{Figure}{Figures}

\crefname{table}{Table}{Tables}
\Crefname{table}{Table}{Tables}

\crefname{algorithm}{Algorithm}{Algorithms}
\Crefname{algorithm}{Algorithm}{Algorithms}

\doparttoc
\mtcsetrules{parttoc}{off}

\defaultbibliographystyle{agsm}
\defaultbibliography{bibliography}

\begin{document}
\faketableofcontents

\def\spacingset#1{\renewcommand{\baselinestretch}%
{#1}\small\normalsize} \spacingset{1}


\if1\anon
{
  \title{\bf Representation-Aware Distributionally Robust Optimization: A Knowledge Transfer Framework}
  \author{Zitao Wang \\
    Department of Statistics, Columbia University. \\
    Nian Si \\
    Department of Industrial Engineering and Decision Analytics, \\ 
    Hong Kong University of Science and Technology. \\
    Molei Liu\footnote{Correspondence to: Nian Si (niansi@ust.hk) and Molei Liu (moleiliu@bjmu.edu.cn).} \\
    Department of Biostatistics, Peking University Health Science Center. \\
    Beijing International Center for Mathematical Research, Peking University.}
  \maketitle
} \fi

\if0\anon
{
  \bigskip
  \bigskip
  \bigskip
  \begin{center}
    {\LARGE\bf Representation-Aware Distributionally Robust Optimization: A Knowledge Transfer Framework}
\end{center}
  \medskip
} \fi

\bigskip
\begin{abstract}
Distributionally robust optimization (DRO) protects statistical learning against distributional shifts by optimizing the worst-case performance over a set of perturbed distributions. However, standard DRO formulations often treat all feature perturbations equally. This can be unnecessarily conservative when external knowledge suggests that the predictive signal is embedded in a low-dimensional representation of the covariates. Using such structures, we propose \textbf{RE}presentation-\textbf{A}ware \textbf{D}istributionally Robust Estimation (READ), a Wasserstein DRO framework that uses external representations to guide the geometry of robustness. Rather than uniformly perturbing all covariate directions, READ increases the transport cost of perturbations that change representation coordinates, thereby reshaping the dual regularization toward the representation subspace. Meanwhile, it preserves protection against variations orthogonal to the representation. We study READ in two regimes. First, for inference on the current target, we characterize our estimator asymptotically and develop a Wasserstein profile inference approach to construct representation-aligned confidence regions while enabling automatic hyperparameter tuning. Second, for deployment to future populations that differ from the current target but are generated from the same representation-invariant random-coefficient model, we show that the resulting regions achieve higher coverage of future model parameters than standard methods.
Simulations and a single-cell multi-omics application demonstrate the advantages of READ in multi-source and multitask transfer learning settings.
\end{abstract}

\noindent%

\spacingset{1.25} 
\setlength{\bibsep}{0pt}

\section{Introduction}
\label{sec:intro}

Conventional statistical learning methods based on empirical risk minimization (ERM) often suffer from overfitting and poor generalization, particularly in high-dimensional, small-sample settings. This has motivated methods that account for uncertainty beyond the empirical distribution, including distributionally robust optimization (DRO) \citep{kuhn2025dro}. In particular, DRO perturbs the observed empirical distribution to form an uncertainty set of plausible distributions, and optimizes the worst-case empirical risk over this set to improve robustness and out-of-distribution (OOD) generalization. Among its variants, the Wasserstein DRO (WDRO) \citep{kuhn2019wasserstein} is particularly prominent, as it provides a tractable formulation that models the local distributional shift through a Wasserstein uncertainty set around the observed empirical distribution.  

However, the standard WDRO is often overly conservative as it perturbs all features in the same way, regardless of whether they are informative or not. 
In many applications, the predictive signal is often concentrated in a lower-dimensional \emph{representation} of the covariates, which can be viewed as a projection onto a subspace capturing the directions most relevant for prediction. 
Such representations are often derived from external sources of information, such as scientific knowledge or auxiliary datasets, and may correspond to latent factors or sufficient statistics. 
By incorporating information beyond the target sample, they capture the feature directions most relevant for prediction and thereby encode a structure that is more intrinsic and stable across populations than the raw covariates.

This perspective suggests that the uncertainty set in WDRO can be guided by the representation structure, enabling the incorporation of external knowledge. In particular, by imposing stronger restrictions along representation-aligned directions, one can improve generalization while avoiding the excessive conservatism of unguided robust optimization, which treats all perturbations uniformly. 
Motivated by this insight, we aim to develop a framework that leverages externally derived representations to reshape the geometry of distributional uncertainty in WDRO.

\subsection{Related Works}

\textbf{Wasserstein DRO.} Wasserstein DRO has attracted substantial attention due to its computational tractability \citep{mohajerin2018data, blanchet2019quantifying, blanchet2019rwpi, gao2022, gao2023distributionally}. For generalized linear models, WDRO recovers standard regularization-based estimators: under squared loss, the square-root lasso \citep{belloni2011sqrtlasso} and related norm regularizations \citep{blanchet2019rwpi, chu2022squarerootreg, gao2022}; under logistic and hinge losses, regularized classification methods \citep{abadeh2015drologistic, abadeh2019regularizationviamass, blanchet2019rwpi, gao2022}. Beyond these optimization connections, WDRO has also motivated a growing statistical literature on generalization guaranties \citep{gao2023finite, olea2026distributionally} and asymptotic inference 
\citep{blanchet2019rwpi,blanchet2021statistical}. In particular, \citet{blanchet2019rwpi} introduced \emph{robust Wasserstein profile inference} (RWPI), which calibrates the size of the uncertainty set in a principled way and connects it to broader results in hyperparameter tuning \citep{bickel2009simultaneous, belloni2011sqrtlasso}. Building on this framework, \citet{blanchet2022confidence} established asymptotic results for WDRO estimators, including valid confidence regions and asymptotic optimality; see also \citet{blanchet2021statistical} for a review of the statistical properties of WDRO. However, this line of work largely focuses on uniform perturbation geometry, without incorporating low-dimensional predictive representations. 

\textbf{Transfer Learning.}
Transfer learning improves prediction in a target population with limited size of labeled samples by borrowing strength from related source datasets while adjusting for the distributional shifts between the source and the target. Existing approaches include proxy-based transfer \citep{bastani2020predicting}, shrinkage-based transfer \citep{li2021translasso, tian2023transglm, lin2024profiled, gu2025angle-based}, covariate-shift adaptation and model-assisted transfer \citep{he2024transfusion, zhou2024model}, semi-supervised transfer \citep{cai2025semi}, Bayesian framework \citep{lai2026bayesian}, and model-averaging approaches \citep{xyz2023semiparametric_ma}. 
Among these, \citet{li2021translasso} and \citet{tian2023transglm} achieved improved efficiency over the target-only estimator when the source and target coefficients exhibit a sparser contrast than themselves. In a similar context, \citet{gu2025angle-based} developed an angle-based transfer learning approach that uses the directions of source coefficients rather than their scales.

\textbf{Distributionally Robust Learning Guided by External Knowledge.}
Recent work has studied how to incorporate external information into distributionally robust learning. In the DRO group literature, \citet{xiong2023distributionally} studied domain adaptation using a DRO formulation on convex mixtures of source distributions. \citet{zhan2024domain} extended this idea to the adaptation of the mixture-population domain by combining a source-mixture model with a group-adversarial robustness step. \citet{rychener2024wasserstein} studied Wasserstein DRO with smaller uncertainty sets obtained by intersecting Wasserstein balls around multiple heterogeneous data sources, and \citet{gui2024distributionally} analyzed distributionally robust risk evaluation under isotonic constraints. However, these works primarily addressed point estimation and prediction, but not statistical inference.


\textbf{Multitask Representation Learning.}
A related stream of literature leveraged shared representation structures in multiple regression tasks to improve the efficiency of statistical learning. For example, \citet{lounici2011oracle} studied shared group sparsity patterns in model coefficients across tasks. \citet{duan2023adaptive} developed an adaptive multitask learning method that remains robust to outlier tasks. In the context of representation learning, \citet{maurer2016benefit} characterized the conditions under which multitask representation learning improved over learning tasks separately, and \citet{tian2025similar-rep} established an adaptive learning framework under representation heterogeneity when tasks had similar but non-identical representations. Meanwhile, \citet{hanneke2022nofreelunch} showed that the gains from pooling tasks were not automatic without sufficient structural information.

\textbf{Causal Invariance Learning.}
Another related literature aims at improving generalizability across environments through causal invariance, with the central idea that generalizable predictors should rely on structures remaining stable under interventions or environmental changes. Invariant causal prediction \citep{peters2016invariant} formalized this idea for causal variable selection and inference, while anchor regression \citep{rothenhaeusler2021anchor} translated it into a prediction-oriented robustness criterion for learning with heterogeneous data sources. More recently, \citet{shen2025causality} developed a causality-oriented robustness framework for linear models based on general noise interventions that included anchor regression as a special case, and \citet{sola2025causality} extended this perspective to nonlinear models through representation learning with neural networks. 

\subsection{Our Contributions}
In this paper, we introduce \emph{\textbf{RE}presentation-\textbf{A}ware \textbf{D}istributionally Robust Learning} (READ), a Wasserstein DRO framework that incorporates an external representation structure into the DRO. Instead of perturbing all covariate dimensions through transport cost indiscriminately, READ reduces the degree of perturbations in the directions aligned with the external representation, while still pursuing robustness to variations orthogonal to the representation. We study READ in two complementary regimes: estimation and inference at the {\bf current target distribution}, and generalization to {\bf future distribution} shifting around the target. For inference on the current target, we study the  asymptotic properties of the proposed estimator and develop a Wasserstein profile inference procedure that enables hyperparameter tuning and yields representation-adaptive confidence regions. For deployment to future populations that may have different model coefficients from the current target but preserve the representation-invariant structure, we show that the resulting robust confidence regions achieve higher limiting coverage than the standard methods, under a representation-invariant random-coefficient model. In general, READ provides a principled framework for statistical inference that integrates knowledge transfer with distributional robustness. We summarize the main advantages of our approach and compare it with existing work as follows.

\textbf{Adaptive Knowledge Transfer.}
READ provides an adaptive way to transfer external representation information into Wasserstein DRO. 
Instead of imposing a full transfer from the representation, READ introduces multi-dimensional alignment parameters that control the amount of transfer along each representation direction.
This allows the method to automatically borrow strength from informative representation components while down-weighting uninformative or misleading ones. Our RWPI-based tuning strategy for the Wasserstein radius parameter is also adaptive to the quality of external representation.

\textbf{Statistical Inference.}
READ is not only a point estimation method, but also provides a valid inferential framework. 
For the current target distribution, we characterize the asymptotic behavior of the READ estimator and develop a robust Wasserstein profile inference (RWPI) procedure that yields representation-aligned confidence regions whose geometry reflects the external representation structure. Moreover, the selection of the Wasserstein radius parameter (size of the uncertainty set) is designed to enable adaptive knowledge transfer while preserving asymptotic validity of our confidence region, thereby avoiding ad hoc regularization choices and enabling principled statistical inference. 

\textbf{Out-of-Distribution Generalization.}
Beyond inference for the current target distribution, READ constructs confidence regions that are robust to shifts toward future populations preserving a similar low-rank representation structure.
In \Cref{sec:robustness}, we study a model shift scenario in which the stable component lies in the representation subspace, while the population-specific shift occurs mainly in its orthogonal directions. 
Under this model, the confidence region of READ shows a higher chance of covering future population parameters than the standard WDRO based on RWPI. 

\textbf{Tractable Formulation.}
READ remains computationally tractable despite modifying the Wasserstein geometry. For both linear regression and binary classification problems, READ admits an equivalent formulation of finite-dimensional regularized regression. 


Finally, note that an earlier conference version of this work appeared as \citet{wang2025kgwdro}, which introduced a single-source knowledge-guided Wasserstein DRO formulation for point estimation. The present paper substantially extends that preliminary version in several directions. First, we study the transfer of knowledge from multidimensional external representations, while the earlier version focused on a special case of a single source. Second, we move beyond point estimation by developing a robust Wasserstein profile inference procedure. Third, we study deployment to future populations under a representation-invariant model-shift regime.

The remainder of the paper is organized as follows. \Cref{sec:prelim} reviews the standard WDRO and its tractable reformulation. \Cref{sec:read} introduces the READ framework with our adaptive tuning and inference approaches. \Cref{sec:inference} develops the asymptotic theory underlying the approaches proposed in \Cref{sec:read}. \Cref{sec:robustness} introduces the confidence region and studies generalization guaranties under future distributional shift. \Cref{sec:simulation} and \Cref{sec:data} present simulation and real-data results. The discussion of future directions is presented in \Cref{sec:diss}. 

\section{Seminorm Regularizations as WDRO}
\label{sec:prelim}

For a positive integer $k$, define $[k] = \{1,\ldots,k\}$. Positive integers $n$, $m$, and $d$ denote the target sample size, the number of sources, and the dimension of the covariates. We index the samples by $i$ and index the sources by $j$. The exponents \(p\) and \(q \in [1, \infty]\) are reserved for pairs of H\"{o}lder conjugates, satisfying \(p^{-1} + q^{-1} = 1\) for \(p, q \in (1, \infty)\), as well as the pair $1$ and $\infty$. For the target data distribution \(\mathbb{P}_*\), let \(\mathbb{P}_n\) denote its empirical measure based on a sample of size \(n\). For a random variable $X\sim \mathbb{P}$ supported in $\mathbb{R}^d$, let $\mathbb{E}_\mathbb{P}[X]$ denote its expected value. For a vector \(v \in \mathbb{R}^d\), \(\|v\|\) denotes a norm, while \(\|v\|_p\) denotes the \(p\)-norm, where \(p \in [1, \infty]\), and $v^\top$ denotes the transpose of $v$. All vectors are assumed to be column vectors. The span of a vector $v$ or the column space of a matrix $M$ is denoted as $\col{(v)}$ or $\col{(M)}$. The extended real line is denoted by $\bar{\mathbb{R}}$. For any linear subspace $U \subseteq \mathbb{R}^d$, let $P_U:\mathbb{R}^d \to U$ be the orthogonal projection onto $U$.

For a function \(f: \mathbb{R}^d \to \bar{\mathbb{R}}\), we denote its domain by $\dom{f} \coloneqq \{x \in \mathbb{R}^d : f(x) < \infty\}$. The \emph{convex conjugate} of \(f\) is defined as $f^*(x) = \sup_{y \in \dom{f}} \left\{y^\top x - f(y)\right\}$. The \emph{support function} of a convex set \(C \subset \mathbb{R}^d\) is defined as $\sigma_C(x) = \sup_{y \in C} y^\top x$. The \emph{dual norm} associated with \(\|\cdot\|\) is $\|x\|_* \coloneqq \sup_{y : \|y\| \leq 1} y^\top x$. A function \(g: \mathbb{R}^d \to [0,\infty]\) is called an \emph{extended norm} if it is a norm that can take the value \(\infty\), with the convention that \(0 \cdot \infty = 0\). A function \(r: \mathbb{R}^d \to [0,\infty)\) is called a \emph{seminorm} if it is a convex, positively homogeneous, nonnegative function. The definition of a dual norm extends to extended norms and seminorms.

Next, we introduce the Wasserstein DRO (WDRO) problem and its dual formulation. WDRO replaces empirical risk minimization (ERM) by a criterion that protects against perturbations of the empirical distribution, providing a principled notion of robustness under distributional uncertainty. We show that, in settings including linear regression with root mean squared error and binary classification with cross-entropy or hinge loss, the estimator admits an equivalent form of seminorm-regularized regression. 

\subsection{Wasserstein Distributionally Robust Optimization}
\label{sec:wdro_form}

Suppose that the target population has an unknown data-generating distribution $\mathbb{P}_*$ of the pair $(X,Y) \in \mathbb{R}^{d+1}$, with samples $\{(x_i,y_i)\}_{i=1}^n$ observed. The standard empirical risk minimization (ERM) approach solves:
\[
\hat\beta_{\erm}\in \argmin_\beta \mathbb{E}_{\mathbb{P}_n}[\ell(X,Y;\beta)] \eqqcolon \argmin_{\beta}\frac{1}{n} \sum_{i=1}^n \ell(x_i,y_i;\beta),
\]
where the risk function $\ell(x,y;\beta)$ measures the prediction or classification loss at the parameter $\beta$ and is convex in \(\beta\). Because ERM relies solely on the finite empirical distribution $\mathbb{P}_n$, it can be unstable when the sample size is limited and may generalize poorly when the future deployment distribution deviates from the current target distribution $\mathbb{P}_*$. DRO addresses these issues by solving:
\[
\hat\beta_{\dro}\in\argmin_{\beta} \sup_{\mathbb{P} \in \mathcal{P}} \mathbb{E}_\mathbb{P}[\ell(X,Y;\beta)],
\]
where $\mathcal{P}$ is an \emph{uncertainty set} of plausible distributions constructed by ``perturbing'' the observed empirical distribution $\mathbb{P}_n$. 

For distributions $\mathbb{P},\mathbb{Q}$ on $\mathbb{R}^{d+1}$, we define their Wasserstein distance as
\[
    d_c(\mathbb{P},\mathbb{Q})
    \coloneqq
    \inf_{\pi \in \Pi(\mathbb{P},\mathbb{Q})} \left\{ \mathbb{E}_\pi[c(U,V)] : U\sim \mathbb{P}, V\sim \mathbb{Q}\right\},
\]
where \(c(\cdot,\cdot):\mathbb{R}^{d+1}\times \mathbb{R}^{d+1}\to [0,\infty]\) is a cost function to be specified later and $\Pi(\mathbb{P},\mathbb{Q})$ denotes the set of couplings of $\mathbb{P}$ and $\mathbb{Q}$, i.e., distributions on $\mathbb{R}^{d+1}\times \mathbb{R}^{d+1}$ with their marginal distributions remaining as $\mathbb{P}$ and $\mathbb{Q}$. In Wasserstein DRO, the set of perturbed distributions \(\mathcal{P}\) is taken to be a Wasserstein ball of radius \(\delta > 0\) centered around the observed empirical distribution \(\mathbb{P}_n\): 
\[
\mathcal{B}_\delta (\mathbb{P}_n;c) \coloneqq \left\{ \mathbb{P}: d_c(\mathbb{P},\mathbb{P}_n) \leq \delta \right\}.
\]


\subsection{Tractable Reformulation for Costs with Extended Norms}

The Wasserstein DRO problem involves an inner maximization over probability measures and may therefore appear computationally intractable. However, the linear programming structure of optimal transport yields a strong dual reformulation that reduces the worst-case risk to a finite-dimensional optimization problem.

\begin{proposition}[Strong Duality \citep{mohajerin2018data}]
\label{prop:duality}
    Let $c:\mathbb{R}^{d+1}\times \mathbb{R}^{d+1}\to [0,\infty]$ be a lower semi-continuous cost function satisfying $c\big((x,y),(u,v)\big) = 0$ whenever $(x,y) = (u,v)$. Then
    \[
    \label{obj:wdro}
    \tag{WDRO}
    \inf_{\beta\in\mathbb{R}^d} \sup_{\mathbb{P}\in \mathcal{B}_\delta(\mathbb{P}_n;c)} \mathbb{E}_{\mathbb{P}}\left[\ell(X,Y;\beta) \right]
    =
    \min_{\beta\in\mathbb{R}^d,\lambda\geq 0}\left\{\lambda \delta + \dfrac{1}{n}\sum_{i=1}^n \phi_\lambda(x_i,y_i;\beta)\right\},
    \]
    where $\phi_\lambda(x_i,y_i;\beta) =\sup_{(u,v)\in \mathbb{R}^{d+1}} \big\{ \ell(u,v;\beta) - \lambda c\big((u,v),(x_i,y_i) \big)\big\}$.
\end{proposition}

Let \(g : \mathbb{R}^d \to [0,\infty]\) be an extended norm that is lower semi-continuous. \cref{lem:seminorm_duality} shows that the dual norm of \(g\) is a seminorm on \(\mathbb{R}^d\) with null space $V_g^\perp$, the orthogonal complement of \(V_g \coloneqq \dom{g}\). Combined with Toland's duality \citep{toland1978duality_nonconvex} and \cref{prop:duality}, this yields tractable reformulations of \eqref{obj:wdro} for linear regression and binary classification.

\begin{proposition}[Linear Regression]
\label{prop:linear_regression}
Suppose the cost function \(c : \mathbb{R}^{d+1}\times \mathbb{R}^{d+1} \to [0,\infty]\) is $c\big((x,y),(u,v)\big) = \{g(x - u)\}^2 + \infty \cdot |y - v|$, where \(g\) is an extended norm on \(\mathbb{R}^d\). Then,
\[
\inf_{\beta \in \mathbb{R}^d} \sup_{\mathbb{P} \in \mathcal{B}_\delta(\mathbb{P}_n;c)} \mathbb{E}_{\mathbb{P}} \left[(Y - X^\top \beta)^2\right]
=
\min_{\beta \in \mathbb{R}^d} \left\{ \sqrt{\mathrm {MSE}_n(\beta)} + \sqrt{\delta} \, g_*(\beta) \right\}^2,
\]
where \( g_*\) is the dual seminorm of \(g\) and $\mathrm{MSE}_n(\beta) = n^{-1}\sum_{i=1}^n (y_i-x_i^\top \beta)^2$.
\end{proposition}

\begin{proposition}[Binary Classification]
\label{prop:binary_classification}
Suppose \(Y\in\{-1,1\}\) and the loss function \(\ell(x, y; \beta)\) is either the logistic loss \(\log{\big(1 + e^{-yx^\top \beta}\big)}\) or the hinge loss \((1 - yx^\top \beta)^+\). Let the cost function \(c : \mathbb{R}^{d+1}\times\mathbb{R}^{d+1} \to [0,\infty]\) be $c\big((x,y),(u,v)\big) = g(x - u) + \infty \cdot |y - v|$, where \(g\) is an extended norm on \(\mathbb{R}^d\). Then,
\[
\inf_{\beta \in \mathbb{R}^d} \sup_{\mathbb{P} \in \mathcal{B}_\delta(\mathbb{P}_n;c)} \mathbb{E}_{\mathbb{P}} \left[\ell(X, Y; \beta)\right]
=
\min_{\beta \in \mathbb{R}^d} \left\{ \frac{1}{n} \sum_{i=1}^n \ell(x_i, y_i; \beta) + \delta g_*(\beta) \right\},
\]
where \( g_*\) is the dual seminorm of \(g\).
\end{proposition}

In \cref{prop:linear_regression,prop:binary_classification}, the transport cost perturbs only the covariates while keeping the responses fixed. This is conventional in WDRO \citep{blanchet2019rwpi} and typically leads to cleaner reformulations than costs that also perturb \(Y\). In addition, the penalty for dual regularized problems is entirely determined by the Wasserstein radius parameter \(\delta\). 

\begin{remark}
The dual representation yields two complementary interpretations of the same estimator. The WDRO form (primal) in \Cref{sec:wdro_form} corresponds to the objective of generalization to future data, as it protects against distributional shifts around the observed empirical distribution. The equivalent regularized regression forms (dual) in \cref{prop:linear_regression,prop:binary_classification} correspond to stabilizing the regression with the induced penalty $g_{*}(\cdot)$, which can improve the estimation, especially when the sample size is small and the dimensionality of $X$ is relatively high.
\end{remark}

\section{Representation-Aware Wasserstein DRO}
\label{sec:read}

In the WDRO literature, the transport cost in \eqref{obj:wdro} is typically $c(x,u) = \|x-u\|_q^k$, for some \(q \geq 1\) and \(k \geq 1\). This imposes a simple geometry that is agnostic to the dependence between \(X\) and \(Y\). In many applications, the predictive signal in \(X\) is assumed to concentrate in a lower-dimensional transformation \(r(X)\). We call \(r(\cdot):\mathbb{R}^d\to\mathbb{R}^m\), a \emph{representation} of \(X\) for $Y$ if it compresses the covariates (satisfying \(m<d\)) while preserving most of the predictive signal, in the sense that $\var(\mathbb{E}[Y| r(X)]) \approx \var(\mathbb{E}[Y| X])$. In this work, we focus on linear representations of the form \(r(x)=\Theta^\top x\), where \(\Theta\in\mathbb{R}^{d\times m}\) is external knowledge constructed independently of the target sample. We do not strictly impose an assumption on the signal strength measured by \(\var(\mathbb{E}[Y | \Theta^\top X])\), as our method is expected to adapt to the non-informativeness of \(\Theta\). Nevertheless, it is expected that the efficiency gain of our method will be pronounced when \(\Theta\) is informative.Such representations arise in several common settings summarized as follows.

\textbf{Principal component regression \citep{bair2006prediction}.} Assume that the data is generated from a latent-factor model in which a low-dimensional, unobserved variable $U$ linearly generates both $X$ and $Y$ with $Y\perp X\mid U$. In this case, one can extract \(\Theta\) as the leading eigenvectors of the covariance structure of $X$ obtained using a large unlabeled sample and then use the representation $\Theta^\top X$ to approximate $U$ which is ``sufficient'' to predict $Y$. 

\textbf{Transfer of knowledge from external sources \citep{li2021translasso,tian2023transglm,gu2025angle-based}.} In transfer learning, \(\Theta\) can be taken as pretrained coefficients \(\{\theta_1,\ldots,\theta_m\}\) learned from $m$ external source datasets assumed to have regression model coefficients for $Y\sim X$ that are similar to the current target. 


\textbf{Multitask learning \citep{duan2023adaptive,tian2025similar-rep}.} Suppose that there are $k$ related prediction tasks. Each task \(j\in[k]\) has an outcome \(Y_j\) and the goal is to improve performance on the target task $Y\sim X$ using $Y_j$'s. Each task, including the target, admits a decomposition of the model coefficients: $\beta_j = \Theta\kappa_j + \varepsilon_j$, where \(\varepsilon_j\) satisfies $\Theta^\top \varepsilon_j = 0$ and captures task-specific deviation from the central representation \(\Theta\) with dimension $m\leq k$. 

In all three scenarios introduced above, external knowledge \(\Theta\) could be informative to the current target parameter
\[
\beta_* := \argmin_{\beta\in\mathbb{R}^d}\mathbb{E}_{\mathbb{P}_*}[\ell(X,Y;\beta)],
\]
in the sense that $\beta_*$ is close to the linear column space \(\col(\Theta)\). Moreover, \(\Theta\) may also encode some intrinsic and stable structure that remains invariant in both the current target and future populations that may shift from the target distribution.

    
\subsection{Representation-aware Formulation and Point Estimation}

Our key idea is to leverage \(\Theta\) to reshape the Wasserstein geometry characterized by the cost function $c(\cdot)$, so that the resulting uncertainty set favors distributions aligned with the observed empirical distribution $\mathbb{P}_n$ in the directions $\Theta^\top X$. In specific, given \(\Theta=[\theta_1,\ldots,\theta_m]\), we introduce the representation-aware transport cost
\[
c_{q,\Lambda}(x,u)
=\|x-u\|_q^2 + \sum_{j=1}^m \lambda_j \big(\theta_j^\top(x-u)\big)^2
=\|x-u\|_q^2 + \|\Theta^\top(x-u)\|_{\Lambda}^2,
\]
where \(\lambda_j \in [0,\infty]\) is an alignment hyperparameter for direction \(\theta_j\) and $\Lambda=\diag(\lambda_1,\ldots,\lambda_m)$. The matrix \(\Lambda\) determines how strongly invariance is enforced along each direction of representation. A higher value of $\lambda_j$ will place a greater cost on perturbations of \(\theta_j^\top X\). Unlike the standard WDRO perturbing $X$ uniformly, \(c_{q,\Lambda}\) allows the ``degree of robustness'' in each direction of $X$ to be adjusted according to \(\Theta\). This cost function reflects the belief that the representation is relatively stable in the current target and future populations, while spurious components orthogonal to $\Theta$ are more likely to shift between different populations.

Using the cost $c_{q,\Lambda}$, the READ estimator is obtained through the WDRO problem:
\[
\label{obj:read}
\tag{READ}
\hat{\beta}_{\readd}
\in
\argmin_{\beta\in\mathbb{R}^d}
\sup_{\mathbb{P}\in\mathcal{B}_\delta(\mathbb{P}_n;c_{q,\Lambda})}
\mathbb{E}_{\mathbb{P}}[\ell(X,Y;\beta)].
\]

This estimator still admits an equivalent tractable regularized formulation. The cost function \(c_{q,\Lambda}\) is the square of a norm when all \(\lambda_j<\infty\), and the square of an extended norm otherwise. Hence, by the dual reformulations in \cref{prop:linear_regression,prop:binary_classification}, our defined problem \eqref{obj:read} is equivalent to a finite-dimensional optimization problem with a regularizer determined by the convex conjugate of \(c_{q,\Lambda}\). Proposition \ref{prop:read_estimator} provides the specific form of this problem.

\begin{proposition}[Dual Form of READ]
\label{prop:read_estimator}
(I) For the linear regression version of \eqref{obj:read},
    \[
    \inf_{\beta \in \mathbb{R}^d}
    \sup_{\mathbb{P} \in \mathcal{B}_\delta(\mathbb{P}_n;c_{q,\Lambda})}
    \mathbb{E}_{\mathbb{P}} \bigl[(Y- X^\top \beta)^2\bigr]
    = \min_{\beta\in\mathbb{R}^d} \left\{
    \sqrt{\mathrm{MSE}_n(\beta)} + \sqrt{\delta} \varphi_{p,\Lambda}(\beta)
    \right\}^2,
    \]
    where $\varphi_{p,\Lambda}(\beta) = \min_{\kappa\in \mathbb{R}^m}
    \left\{\sqrt{ \|\beta-\Theta\kappa\|_p^2 + \|\kappa\|_{\Lambda^{-1}}^2}
    \right\}$. 

(II) For binary classification, suppose that the loss \(\ell(x,y;\beta)\) is either the logistic loss \(\log(1+e^{-yx^\top\beta})\) or the hinge loss \((1-yx^\top\beta)^+\) and the cost function is taken as \(\sqrt{c_{q,\Lambda}}\). Then,
    \[
    \inf_{\beta \in \mathbb{R}^d}
    \sup_{\mathbb{P} \in \mathcal{B}_\delta(\mathbb{P}_n;\sqrt{c_{q,\Lambda}})}
    \mathbb{E}_{\mathbb{P}} \bigl[\ell(X,Y;\beta)\bigr]
    = \min_{\beta \in \mathbb{R}^d}
    \left\{ \frac{1}{n}\sum_{i=1}^n \ell(x_i,y_i;\beta) + \delta \varphi_{p,\Lambda}(\beta)\right\}.
    \]
\end{proposition}

The regularizer \(\varphi_{p,\Lambda}\) introduced in Proposition \ref{prop:read_estimator} admits a natural interpretation. It characterizes how well \(\beta\) can be approximated by the linear subspace \(\col(\Theta)\), together with a ridge penalty on the coefficient vector \(\kappa\). Thus, READ encourages the estimator to align with \(\col(\Theta)\), while penalizing the directions orthogonal to that subspace more heavily. The role of the alignment hyperparameter \(\Lambda\) is transparent from the primal formulation: perturbations that substantially change \(\Theta^\top X\) are assigned a higher cost by \(\Lambda\), especially along directions with large \(\lambda_j\). In the dual problem, this becomes a weaker regularization penalty for coefficients aligned with those directions. 

We conclude this subsection with remarks on \(\varphi_{p,\Lambda}\). When \(\lambda_j=0\), we have \(\lambda_j^{-1}\kappa_j^2<\infty\) if and only if \(\kappa_j=0\), under the convention \(0\cdot\infty=0\). Thus, \(\lambda_j=0\) means that there is no transfer of knowledge in the direction \(\theta_j\). In contrast, when \(\lambda_j=\infty\), we have \(\lambda_j^{-1}\kappa_j^2=0\) for all \(\kappa_j\), so the corresponding \(\theta_j\) is utilized without penalty. In the special case \(p=q=2\), $\varphi_{2,\Lambda}^2(\beta)$ admits the closed-form expression
\[
    \inf_{\kappa} \left\{ \|\beta-\Theta\kappa\|_2^2 + \|\kappa\|_{\Lambda^{-1}}^2 \right\}
    =\beta^\top \left(\mathbb{I}_d -\Theta(\Theta^\top\Theta+\Lambda^{-1})^{-1}\Theta^\top \right)\beta.
\]
Finally, to solve the dual formulation of \eqref{obj:read} in \cref{prop:read_estimator}, in \Cref{sec:read_reformulation} we introduce a conic reformulation of the READ objective and use standard convex optimizers to efficiently compute the solution.

\subsection{Tuning and Inference Based on RWPI}
\label{sec:adaptive_tuning}

This subsection introduces the pipelines for tuning and inference. For a fixed alignment parameter \(\Lambda\), we select the Wasserstein radius \(\delta\) and construct the confidence region for \(\beta_*\) using robust Wasserstein profile inference (RWPI) \citep{blanchet2019rwpi}. We then introduce a data-adaptive tuning strategy for \(\Lambda\) that preserves the validity of our confidence region.

Conventionally, the radius parameter $\delta$ used in DRO is chosen to ensure that the target population distribution $\mathbb{P}_*$ is within the uncertainty set. However, due to the curse of dimensionality, this criterion is often overly conservative. RWPI addresses this issue from a regression-oriented perspective: instead of requiring the target distribution $\mathbb{P}_*$ to be covered by the uncertainty set, it ensures with a certain confidence level that the target population parameter $\beta_*$ can be induced by some distribution in the uncertainty set. 

We now introduce RWPI in detail. For fixed \(\delta, \Lambda\), we define the set of optimal parameters induced by the uncertainty set as
\[
\Omega_n(\delta)\coloneqq
\bigcup_{\mathbb{P}\in\mathcal{B}_{\delta}(\mathbb{P}_n;c_{q,\Lambda})}
\argmin_{\beta\in\mathbb{R}^d}\mathbb{E}_{\mathbb{P}}[\ell(X,Y;\beta)].
\]
The set $\Omega_n(\delta)$ is random due to the empirical distribution $\mathbb{P}_n$ and can therefore be viewed as a confidence set that is intended to cover $\beta^*$ with a certain probability. 
Thus, for a given confidence level $1-\alpha$ and the alignment parameter $\Lambda$, the Wasserstein radius $\delta$ can be obtained accordingly: 
$$
\delta_{\alpha,n}(\Lambda)
\coloneqq \inf\left\{\delta>0:{\rm Pr}(\beta_*\in\Omega_n(\delta))\ge 1-\alpha\right\}.
$$
With this choice, \(\Omega_n(\delta_{\alpha,n}(\Lambda))\) becomes a confidence region for \(\beta_*\) with confidence level \((1-\alpha)\). Note that $\Omega_n(\delta)$ can be equivalently represented as 
\begin{align*}
   \Omega_n(\delta)  =\bigg\{\tilde\beta\in\mathbb{R}^d:~& \text{there exists a distribution }\mathbb{Q} \text{ such that }  d_{c_{q,\Lambda}}(\mathbb{P}_n,\mathbb{Q})\leq \delta   \\
   & \text{ and } \tilde\beta \in \argmin_{\beta\in\mathbb{R}^d} \mathbb{E}_{\mathbb{Q}}[\ell(X,Y;\beta)]  \bigg \}.
\end{align*}
Therefore, we define the \emph{Robust Wasserstein Profile} (RWP) function
\[
\tag{RWP}
\label{eq:rwp}
R_n(\beta)\coloneqq \inf_{\mathbb{Q}} \left\{ d_{c_{q,\Lambda}}(\mathbb{P}_n,\mathbb{Q}) :\mathbb{E}_{\mathbb{Q}}[h(X,Y;\beta)]=0 \right\},
\]
where $h(x,y;\beta)\coloneqq -\nabla_\beta \ell(x,y;\beta)$. Note that the moment condition \(\mathbb{E}_{\mathbb{Q}}[h(X,Y;\tilde\beta)]=0\) is equivalent to \(\tilde\beta\in\argmin_{\beta}\mathbb{E}_{\mathbb{Q}}[\ell(X,Y;\beta)]\) for any $\tilde\beta$. 
Hence, 
\[
\tag{Equiv.}
\label{eq:set_equiv}
\Omega_n(\delta)=\left\{\beta\in\mathbb{R}^d:R_n(\beta)\le \delta\right\}.
\]

For a fixed \(\Lambda\), this motivates us to choose the Wasserstein radius as
\(\delta=\delta_{\alpha,n}(\Lambda)\), the \((1-\alpha)\)-quantile of \(R_n(\beta_*)\) where $\beta_*$ is the true target parameter. This choice balances robustness and statistical efficiency: the resulting profile region \(\Omega_n(\delta_{\alpha,n}(\Lambda))\) is sufficiently large to cover the truth \(\beta_*\) with the prescribed probability, while avoiding an unnecessarily large uncertainty set in WDRO that would lead to excessive conservatism. Moreover, \(\Omega_n(\delta_{\alpha,n}(\Lambda))\) provides a ``knowledge-$\Theta$-guided'' confidence region for \(\beta_*\), serving as an alternative to conventional M-estimation confidence regions based solely on the target sample. Its geometry is shaped by our representation-aware transport cost and therefore reflects external knowledge \(\Theta\). In addition, the region \(\Omega_n(\delta_{\alpha,n}(\Lambda))\) is compatible with our READ point estimator in the sense that it is ensured to contain \(\hat{\beta}_{\mathrm{READ}}\).

To compute the radius \(\delta_{\alpha,n}(\Lambda)\) and the confidence region, we rely on the large-sample asymptotics of RWP function $R_n(\beta)$. In particular, we show  in \Cref{thm:rwpi} the functional central limit theorem (CLT) $nR_n\bigl(\beta_*+n^{-1/2}\,\cdot\,\bigr)
    \leadsto
    \psi_{p,\Lambda}^*(H-\Sigma\,\cdot\,)$, where 
\begin{align*}
& H\sim N(0,\cov(h(X,Y;\beta_*))), \quad \psi_{p,\Lambda}(z)
\coloneqq
\frac{1}{4}\mathbb{E}_{\mathbb{P}_*}\big[\varphi_{p,\Lambda}^2(\Xi z)\big],
\quad
\psi_{p,\Lambda}^*(h)
\coloneqq
\sup_{\xi\in\mathbb{R}^d}\bigl\{\xi^\top h-\psi_{p,\Lambda}(\xi)\bigr\}, \notag \\
 &\Sigma \coloneqq -\mathbb{E}_{\mathbb{P}_*}[\nabla_\beta h(X,Y;\beta_*)], \quad \Xi \coloneqq [\nabla_x h(X,Y;\beta_*)]^\top.
\end{align*} 
Let \(\eta_\alpha(\Lambda)\) denote the \((1-\alpha)\)-quantile of \(\psi_{p,\Lambda}^*(H)\). Our RWPI procedure therefore selects $\hat\delta_{\alpha,n}(\Lambda)=n^{-1}\eta_\alpha(\Lambda)$ as the \((1-\alpha)\)-quantile of \(R_n(\beta_*)\). Also, the large-sample approximation of the confidence region can be expressed as:
\[
\mathcal{C}_{1-\alpha}(\Lambda) = \hat\beta_{\erm} + n^{-1/2}\left\{ u\in\mathbb{R}^d:\psi_{p,\Lambda}^*(\Sigma u)\le \eta_\alpha(\Lambda) \right\}.
\]
In practice, we replace \(\psi_{p,\Lambda}^*\), \(\Sigma\), and \(\eta_\alpha(\Lambda)\) by their empirical versions \(\hat{\psi}_{p,\Lambda}^*\), \(\hat{\Sigma}\), and \(\hat{\eta}_\alpha(\Lambda)\) defined in \Cref{alg:delta}. Direct evaluation of \(\psi_{p,\Lambda}^*\) is often inconvenient. Therefore, in \Cref{prop:variational_rep} of the Supplementary Material, we introduce a useful variational representation of $\psi_{p,\Lambda}^*(z)$ for its easy computation:
\[
\psi_{p,\Lambda}^*(z) = \inf \left\{ \mathbb{E}[c_{q,\Lambda}(U)] : U\in L^2(\Omega;\mathbb{R}^d),\ \mathbb{E}[\Xi^\top U]=z \right\}.
\]
A complete procedure for the selection of the radius $\delta$ and the construction of the confidence region for $\beta_*$ is given in \cref{alg:delta}.


\begin{algorithm}[htbp]
\caption{Robust Wasserstein Profile Inference (RWPI) in the READ Framework.}
\label{alg:delta}
\begin{algorithmic}
\Require \( \mathbf{X}\in \mathbb{R}^{n\times d},\mathbf{y}\in\mathbb{R}^n,\hat{\beta}\in\mathbb{R}^d,\Theta\in\mathbb{R}^{d\times m},\Lambda\in [0,\infty]^m \).
\State \textbf{Step 1:} For each observation \((x_i,y_i)\), set \(\hat\Xi_i = [\nabla_x h(x_i,y_i;\hat\beta)]^\top\). Compute:
\begin{align*}
\label{obj:variational}
\tag{Quadratic}
\hat{\psi}_{p,\Lambda}^*(v)
= \min_{\{u_i\}} \frac{1}{n}\sum_{i=1}^n c_{q,\Lambda}(u_i) \quad \text{s.t.} \quad \frac{1}{n}\sum_{i=1}^n \hat\Xi_i^\top u_i = v.
\end{align*}
\State \textbf{Step 2:} Compute \(\hat{\Sigma}_h\) as the sample covariance matrix of \(\{h(x_i,y_i;\hat\beta)\}_{i=1,\ldots,n}\) and \(\hat\Sigma = -n^{-1}\sum_{i=1}^n\nabla_\beta h(x_i,y_i;\hat\beta)\). Draw \(l\) random vectors \(\{h_k\}_{k\le l} \overset{\mathrm{iid}}{\sim} N(0,\hat{\Sigma}_h)\).
\State \textbf{Return:}
\begin{enumerate}
    \item Radius \(\hat\delta_{\alpha,n}(\Lambda)=\hat{\eta}_{\alpha}(\Lambda)/n\), where \(\hat{\eta}_{\alpha}(\Lambda)\) is the \((1-\alpha)\)-quantile of \(\{\hat{\psi}_{p,\Lambda}^{*}(h_k)\}_{k\le l}\). This radius $\hat\delta_{\alpha,n}(\Lambda)$ is then used in (\ref{obj:read}) to obtain the point estimator $\hat{\beta}_{\readd}$.
    
    \item The READ confidence region:
    \[
    \hat{\mathcal C}_{1-\alpha}(\Lambda)
    = \hat\beta_{\erm} + n^{-1/2} \left\{
    u\in\mathbb{R}^d: \hat\psi_{p,\Lambda}^*(\hat\Sigma u)\le \hat\eta_\alpha(\Lambda) \right\}.
    \]
\end{enumerate}
\end{algorithmic}
\end{algorithm}


To see how RWPI selects a radius $\delta$ that adapts to the informativeness of the representation $\Theta$, consider the one-dimensional case
\(\Theta=\theta\) with \(\lambda=\infty\). Then, any transport induced by $c_{q,\infty}(x,u)$ must preserve both
\(Y\) and \(\theta^\top X\). Thus, when $\beta_*\in \col(\theta)$, the distribution of the projected score $\theta^\top h(X,Y;\beta_*)$ also remains unchanged. By our definition, $\mathbb{E}_{\mathbb{Q}}[\theta^\top h(X,Y;\beta_*) ]$ needs to be zero for some ${\mathbb{Q}}$ satisfying $d_{c_{q,\Lambda}}(\mathbb{P}_n,\mathbb{Q})= \delta_{\alpha,n}(\Lambda)$.  Therefore, when \(\beta^*\) is approximately aligned with \(\theta\), admissible perturbations preserving \(\theta^\top X\) have a relatively limited ability to alter $\mathbb{E}_{\mathbb{Q}}[\theta^\top h(X,Y;\beta_*) ]$. In this case, a larger Wasserstein radius is required so that \(\beta^*\) can be accommodated by some perturbed distribution $\mathbb{Q}$, leading to a larger RWPI-selected value of \(\delta\). Meanwhile, in the dual regularized regression form defined in Proposition \ref{prop:read_estimator}, $\delta$ controls the degree of shrinkage to $\col(\theta)$. Thus, when external knowledge is informative, our RWPI method can select a large $\delta$ that encourages the subsequent READ estimator to stay close to $\col(\theta)$, which can improve the estimation performance.
 

We propose a data-adaptive selection procedure for \(\Lambda\) based
on balanced \(K\)-fold cross-validation; see \Cref{sec:lambda_selection} of the
Supplementary Material. \Cref{prop:lambda-selection} shows that, for a fixed finite
candidate family converging to deterministic
limits, the cross-validation selector converges to the unique
minimizer of a deterministic prediction-risk criterion.
The confidence region constructed after this selection remains
asymptotically valid. The monomial parameterization used in practice
reduces the search to a low-dimensional finite grid and facilitates
the condition required by this proposition.

\section{Asymptotic Theory}
\label{sec:inference}

This section develops the asymptotic theory for READ. In \Cref{sec:rwpi}, we study the RWP function and justify the validity of our proposed representation-aware confidence region. We then derive the limit distribution and asymptotic bias of the READ estimator in \Cref{sec:asymptotics_read}. 

\subsection{RWPI and Confidence Region Construction}
\label{sec:rwpi}


Throughout, we work in the fixed-dimensional asymptotic regime as in the previous WDRO literature \citep{blanchet2022confidence}, where $d$ and $m$ remain fixed as $n\rightarrow\infty$. The behavior of READ in high-dimensional settings is empirically evaluated through our simulation studies and real-data application.

\begin{assumption}
\label{asm:regularity}
The loss function $\ell(x,y;\beta)$ satisfies the following assumptions:
\begin{enumerate}[label=(\alph*),ref=\theassumption(\alph*)]
    \item \label{asm:regularity_i}
    The loss $\ell(x,y;\beta)$ is nonnegative, convex in $\beta$ and twice differentiable in $x$ and $\beta$.

    \item \label{asm:optimality}
    There exists \(\beta_* \in \mathbb{R}^d\) such that $\mathbb{E}_{\mathbb{P}_*} \left[h(X, Y; \beta_*)\right] = 0$. Moreover, \(\mathbb{E}_{\mathbb{P}_*} \|h(X,Y;\beta_*)\|_2^2 < \infty\), \(\Sigma = -\mathbb{E}_{\mathbb{P}_*}[\nabla_\beta h(X,Y;\beta_*)]\succ 0\) and $\mathbb{E}_{\mathbb{P}_*} [\Xi^\top \Xi ]\succ 0$.
    
    \item \label{asm:regularity_ii}
    For every $\beta\in \mathbb{R}^d$ and $p\geq 1$, $\mathbb{E}_{\mathbb{P}_*} \|\nabla_x\ell(X,Y;\beta)\|_p^2 <\infty$ and $\| \nabla_x^2 \ell(x,\cdot\,,y;\beta)\|_p$ is uniformly continuous in $x$ and bounded by a continuous function $M(\beta)$. 
    \item \label{asm:regularity_iii}
    For any $\beta$ in the neighborhood of $\beta_*$ and $q \geq 1$, there exists $0<M'<\infty$ such that $\|\nabla_xh(x,y;\beta)\|_q\leq M'(1+\|x\|_q)$. In addition, the following Lipschitz conditions hold:
    \begin{align*}
        \| \nabla_xh(x + u,y;\beta_* + \zeta) - \nabla_xh(x,y;\beta_*)\|_q &\leq K(x) \left( \|u\|_q+\|\zeta\|_q\right),\\
        \| \nabla_\beta h(x + u,y;\beta_* + \zeta) - \nabla_\beta h(x,y;\beta_*)\|_q &\leq K'(x) \left( \|u\|_q+\|\zeta\|_q\right),
    \end{align*} for $\|u\|_q+\|\zeta\|_q\leq 1$, where $K(x),K'(x)>0$ and $\mathbb{E}_{\mathbb{P}_*}[K(X)^2], \mathbb{E}_{\mathbb{P}_*}[K'(X)^2]<\infty$.
\end{enumerate}
\end{assumption}

Recall the following function associated with \(\varphi_{p,\Lambda}^2\) defined on \(\mathbb{R}^d\):
\[
\psi_{p,\Lambda}(x)
= \frac{1}{4}\mathbb{E}_{\mathbb{P}_*}\big[\varphi_{p,\Lambda}^2(\Xi x)\big]
= \frac{1}{4}\mathbb{E}_{\mathbb{P}_*}\left[\inf_{\kappa\in\mathbb{R}^m}
\left\{ \|\Xi x - \Theta \kappa\|_p^2 + \|\kappa\|_{\Lambda^{-1}}^2 \right\}\right].
\]
Let \(\psi_{p,\Lambda}^*\) denote its convex conjugate.

\begin{assumption}
\label{asm:pd}
For all \(\Lambda \in \diag([0,\infty]^m)\) and all nonzero \(\xi \in \mathbb{R}^d\), $\mathbb{P}_*\bigl(\varphi_{p,\Lambda}^2(\Xi \xi) > 0\bigr) > 0$.
\end{assumption}

\cref{asm:regularity_i,asm:optimality} are standard smoothness and identification conditions for M-estimation \citep[Chapter 3.2.2]{wellner2023weak}. \cref{asm:regularity_ii,asm:regularity_iii} are also used in the asymptotic analysis of standard WDRO \citep{blanchet2019rwpi,blanchet2022confidence}. In particular, \cref{asm:optimality} imposes the second-moment and non-degeneracy conditions. 
\cref{asm:regularity_ii} ensures that the inner maximum value in the WDRO objective is finite, and \cref{asm:regularity_iii} provides the local smoothness needed for the limit theory. \cref{asm:pd} is a nondegeneracy condition for \(\varphi_{p,\Lambda}\). This implies that \(\psi_{p,\Lambda}^*(z)<\infty\) for every \(z\in\mathbb{R}^d\). We now study the asymptotic behavior of the RWP function at $\beta_*$. 

\begin{theorem}
\label{thm:rwpi}
Suppose \cref{asm:regularity,asm:pd} hold and let $H\sim N(0,\cov_{\mathbb{P}_*}h(X,Y;\beta_*))$. Then the following convergences hold jointly:
\begin{enumerate}[leftmargin=2em]
    \item $nR_n\bigl(\beta_*+n^{-1/2}\,\cdot\,\bigr)
    \leadsto
    \psi_{p,\Lambda}^*(H-\Sigma\,\cdot\,)$ in $C(\mathbb{R}^d)$, uniformly in compact subsets of $\mathbb{R}^d$.

    \item For each fixed $\eta>0$, if $\delta_n=\eta/n$, then
    \[
    \sqrt{n}\bigl(\Omega_n^+(\delta_n)-\beta_*\bigr)
    \leadsto
    \Sigma^{-1}H+\{u\in\mathbb{R}^d:\psi_{p,\Lambda}^*(\Sigma u)\le \eta\},
    \]
    where $\Omega_n^+(\delta)$ denotes the closure of $\cap_{\varepsilon>0}\Omega_n(\delta+\varepsilon)$.
\end{enumerate}
\end{theorem}

The asymptotic limit \(\psi_{p,\Lambda}^*(H)\) introduced in \Cref{thm:rwpi} is a quadratic-form-type functional of a Gaussian vector. Combined with \eqref{eq:set_equiv}, this yields the approximation \(\delta_{\alpha,n}(\Lambda)\approx \eta_\alpha(\Lambda)/n\), where \(\eta_\alpha(\Lambda)\) is the \((1-\alpha)\)-quantile of \(\psi_{p,\Lambda}^*(H)\).  As introduced in \Cref{sec:adaptive_tuning}, this approximation is used in RWPI to obtain the radius parameter. Combining the second point in \cref{thm:rwpi} with the classical convergence result $\sqrt{n}(\hat{\beta}_{\erm}-\beta_*)\leadsto \Sigma^{-1}H$, we can justify the validity of the confidence region obtained by RWPI.


\begin{corollary}
\label{corollary:confidence_region}
Suppose \cref{asm:regularity,asm:pd} hold for a fixed $\Lambda$, the set
\[
\hat{\mathcal C}_{1-\alpha}(\Lambda)=\hat\beta_{\erm} + n^{-1/2}\left\{ u \in \mathbb{R}^d : \hat{\psi}^*_{p,\Lambda}(\hat{\Sigma} u) \leq \hat{\eta}_{\alpha}(\Lambda) \right\}
\]
is an asymptotically valid \((1-\alpha)\)-confidence region for \(\beta_*\). 
\end{corollary}


\begin{remark}
\Cref{corollary:confidence_region} is established for a deterministic
\(\Lambda\). \Cref{sec:lambda_selection} of the Supplement introduces \Cref{alg:lambda-cv}, a \(K\)-fold
cross-validation procedure for selecting \(\Lambda\) in practice. Under a unique limiting minimizer condition on the candidate set of $\Lambda$, \Cref{prop:lambda-selection} establishes both consistency of the selected
alignment parameter and asymptotic validity of the resulting plug-in
confidence region \(\hat{\mathcal{C}}_{1-\alpha}(\hat\Lambda)\) .
\label{remark:valid}
\end{remark}

To instantiate Theorem \ref{thm:rwpi}, we now derive the limiting distribution of \(\psi^*_{p,\Lambda}(H)\) in closed form for the linear regression model with \(p=q=2\). The resulting expression shows that \(\psi^*_{2,\Lambda}\) varies smoothly with \(\Lambda\) and that the corresponding confidence region is an ellipsoid whose shape deforms continuously with \(\Lambda\). 

\begin{proposition}
\label{prop:linearmodel}
Assume that \(\Sigma_X \coloneqq \mathbb{E}[XX^\top] \succ 0\). Consider the linear model $Y = X^\top \beta_* + \varepsilon$ with the risk function \(\ell(x,y;\beta) = (y-x^\top\beta)^2\), where the noise term
satisfies $\mathbb{E}[\varepsilon \mid X] = 0$ and $\mathbb{E}[\varepsilon^2 \mid X] = 1$. Then, for \(p=q=2\) and each \(\Lambda \in \diag([0,\infty]^m)\),
\[
\psi_{2,\Lambda}(z) = \frac{1}{4} z^\top \Gamma z,
\qquad
\Gamma = \Psi_{\Lambda} + \|\beta_*\|_{\Psi_\Lambda}^2 \Sigma_X.
\]
When \(\beta_* \notin \col(\Theta_\infty)\), \(\psi_{2,\Lambda}^*(H)\) follows the generalized \(\chi^2\)-distribution \(H^\top \Gamma^{-1} H\), and the corresponding confidence region becomes
\[
\label{eq:confidence_region}
\tag{CR}
\mathcal{C}_{1-\alpha}(\Lambda) = \hat{\beta}_{\erm}
+
n^{-1/2}\left\{
u\in\mathbb{R}^d:
\|\Sigma_X u\|_{\Gamma^{-1}}^2
\le
\eta_\alpha(\Lambda)
\right\}.
\]
\end{proposition}

\begin{remark}
In the context of Proposition \ref{prop:linearmodel}, Assumption \ref{asm:pd} holds if and only if \(\beta_* \notin \col(\Theta_{\infty})\), where \(\Theta_{\infty}\) is the submatrix of \(\Theta\) corresponding to the entries of \(\Lambda\) equal to \(\infty\). Under a ``perfectly aligned'' scenario with \(\beta_* \in \col(\Theta_{\infty})\), theoretical studies of READ can be further simplified because the RWPI-based selection of $\delta$ degenerates to $\delta=\infty$ and the dual problem of DRO reduces to some constrained optimization problem. This is studied in Section \ref{sec:asymptotics_constrained} of the Supplementary Material, as a complement to our theoretical framework.  
\end{remark}


\subsection{Limiting Distribution of READ}
\label{sec:asymptotics_read}

In this subsection, we characterize the limiting distribution of \(\hat\beta_{\readd}\). We start by introducing a regularity condition.

\begin{assumption}
\label{asm:pd_2}
For all \(\Lambda \in \diag([0,\infty]^m)\), $\mathbb{P}_*\bigl(\varphi_{p,\Lambda}^2(\nabla_x \ell(X,Y;\beta_*))>0\bigr)>0$.
\end{assumption}

Assumption \ref{asm:pd_2} ensures that the function
$
\mathcal{V}(\beta;\Lambda)
\coloneqq
\left(
\mathbb{E}_{\mathbb{P}_*}\bigl[\varphi_{p,\Lambda}^2(\nabla_x\ell(X,Y;\beta))\bigr]
\right)^{1/2}
$
is differentiable in a neighborhood of \(\beta_*\). This function can be viewed as a sensitivity measure as it quantifies the local perturbation effect of \(X\) on the expectation of the loss $\ell(X,Y;\beta)$. Under the linear model considered in \cref{prop:linearmodel}, Assumption \ref{asm:pd_2} is equivalent to \cref{asm:pd}. But in general, one of them does not imply the other. We derive the asymptotic properties of READ in the following theorem.

\begin{theorem}
\label{thm:asymptotic_dro}
Suppose \cref{asm:regularity,asm:pd_2} hold and recall \(\delta_n=\hat{\eta}_{\alpha}(\Lambda)/n\). Then, for any \(p \in (1,\infty)\) and \(\Lambda\in[0,\infty]^m\),
\[
\sqrt{n}(\hat\beta_\readd-\beta_*)
\leadsto
\Sigma^{-1}H-\sqrt{{\eta}_{\alpha}(\Lambda)}\Sigma^{-1}\nabla_\beta \mathcal{V}(\beta_*;\Lambda),
\]
where $H\sim N\left(0,\cov_{\mathbb{P}_*}h(X,Y;\beta_*)\right)$ and $\Sigma=-\mathbb{E}_{\mathbb{P}_*}[\nabla_\beta h(X,Y;\beta_*)]$. 
\end{theorem}

\cref{thm:asymptotic_dro} shows that the READ estimator tuned by RWPI retains the same asymptotic variance as the empirical risk minimization (ERM) estimator through the component \(\Sigma^{-1}H\), while \(\Lambda\) affects the asymptotic bias through the deterministic effect \(\sqrt{\eta_\alpha(\Lambda)}\,\Sigma^{-1}\nabla_\beta \mathcal{V}(\beta_*;\Lambda)\). Under the linear regression model of \cref{prop:linearmodel}, the squared-norm of this bias term scales, up to a multiplicative constant, as \(\eta_{\alpha}(\Lambda)\|\beta_*\|_{\Psi_\Lambda}^{-2}\|\Psi_{\Lambda}(\beta_*)\|_2^2\). The derivation of this form also helps us justify the validity of $\hat{\mathcal C}_{1-\alpha}(\hat\Lambda)$ discussed in Remark \ref{remark:valid}.


Note that \Cref{thm:asymptotic_dro} focuses on scenarios with \(p>1\). When \(p=1\), the limiting distribution of $\hat\beta_\readd$ is generally non-Gaussian. Instead, it is given by the minimizer of a convex objective consisting of a quadratic Gaussian term and an \(\ell_1\)-term. The detailed results are presented in \Cref{sec:asymptotic_sqrt_lasso} of the Supplementary Material.



\section{Properties Under Future Distribution Shift}
\label{sec:robustness}

So far, our analysis has focused on inference for the current target distribution \(\mathbb{P}_*\). We now study how READ behaves on future distributions that may differ from \(\mathbb{P}_*\) but remain similar to it in the sense that their model coefficients share an invariant component in the linear space $\col(\Theta)$. For this purpose, we introduce a $\Theta$-invariant random-coefficient model that accommodates both local and non-vanishing heterogeneity between the current and future populations.

Let \(e\in\{*\}\cup\mathcal E\) index a population, where $*$ denotes the current target population and \(\mathcal E\) is the collection of future populations. For population $e$, suppose
\[
\label{asm:drift_model}
\tag{$\Theta$-Invariant($r_n$)}
\beta_e
=
\Theta\kappa+r_n\varepsilon_e,
\qquad
\varepsilon_e\sim N\left(
0,
\tau^2\Theta(\Theta^\top\Theta)^{-1}\Theta^\top
+
\sigma^2\Theta_\perp\Theta_\perp^\top
\right),
\]
where \(\kappa\neq0\) is fixed, $\Theta_\perp$ is an orthonormal basis for the orthogonal complement of $\col(\Theta)$, and $r_n\in\{n^{-1/2},1\}$ is a scaling parameter. Equivalently, we can write
\[
\varepsilon_e
=
\Theta(\Theta^\top\Theta)^{-1/2}\mu_e
+
\Theta_\perp\gamma_e,
\qquad
\mu_e\sim N(0,\tau^2\mathbb I_m),
\qquad
\gamma_e\sim N(0,\sigma^2\mathbb I_{d-m}),
\]
where $\mu_e$ and $\gamma_e$ are independent. The first component describes variation within $\col(\Theta)$, whereas the second describes variation in orthogonal directions. When $\tau^2\ll\sigma^2$, the shared component $\Theta\kappa$ is relatively stable between populations. The scale $r_n$ distinguishes two complementary regimes. The choice $r_n=n^{-1/2}$ places the heterogeneity between the populations on the same scale as sampling uncertainty in the target sample. Therefore, it produces a nondegenerate coverage analysis for the conventional $n^{-1/2}$-scale RWPI confidence region. The choice $r_n=1$ allows the difference between the current and future population parameters to remain non-vanishing as $n$ grows. In that case, an ordinary $n^{-1/2}$-scale confidence region generally cannot cover an independently drawn future model parameter asymptotically, and instead a robust region with a fixed scale is needed. We first study the local regime $r_n=n^{-1/2}$ and then give a complementary result for $r_n=1$.

For simplicity, we assume $\Sigma_X=\mathbb E(XX^\top)=\mathbb I_d$ throughout this section. For population $e$, suppose that the data is generated from the linear model
\[
Y_e=X_e^\top\beta_e+\epsilon_e,
\qquad
\mathbb E(\epsilon_e\mid X_e)=0,
\qquad
\mathbb E(\epsilon_e^2\mid X_e)=1.
\]
The decomposition in \eqref{asm:drift_model} is related to random-effects models in meta-analysis \citep{borenstein2010basic} and to the multitask learning frameworks of \citet{duan2023adaptive} and \citet{tian2025similar-rep}. 

To operationalize the $\Theta$-invariant regime, $\Theta$ or its stable component $\Theta\kappa$ can be estimated from external data. In multi-source transfer learning, $\Theta\kappa$ may be constructed by aggregating pretrained estimators from external populations. In multitask learning, $\Theta$ may be estimated from the coefficient matrix of auxiliary tasks. Section \ref{sec:theta_invariance} studies future coverage under local shifts, Section \ref{sec:fixed_scale_robustness} gives the complementary result under non-vanishing shifts, and Section \ref{sec:projected_inference} studies inference for the invariant coordinates.

\subsection{Generalization Under Local Representation-Invariant Shift}
\label{sec:theta_invariance}

We first take $r_n=n^{-1/2}$ in \eqref{asm:drift_model} and consider the conventional $n^{-1/2}$-scale confidence region defined in \eqref{eq:confidence_region}. For \Cref{thm:robustness}, we assume that $\Theta^\top\Theta=\mathbb I_m$ and restrict the alignment parameter to the common-scalar form $\Lambda=\lambda\mathbb I_m$, with $\lambda\in[0,\infty]$. The orthonormality of $\Theta$ is primarily a notational simplification, since any full-column-rank representation matrix can be orthonormalized as $Q=\Theta(\Theta^\top\Theta)^{-1/2}$. We discuss the extension to the nonorthonormal case in Section \ref{sec:proof_robustness_theta} of the Supplement.

For an unseen future model $\beta_e$, define
\[
p_\alpha(\lambda)
\coloneqq
\lim_{n\to\infty}
\mathbb P\bigl(\beta_e\in\mathcal C_{1-\alpha}(\lambda)\bigr),
\]
where $\mathcal C_{1-\alpha}(\lambda)$ is the RWPI confidence region in \eqref{eq:confidence_region}, with $\Lambda=\lambda\mathbb I_m$. Recall that $\lambda=0$ recovers the standard WDRO, so $p_\alpha(0)$ is the limiting future-parameter coverage of the standard RWPI region \citep{blanchet2022confidence}.

\begin{theorem}
\label{thm:robustness}
Suppose \eqref{asm:drift_model} holds with $r_n=n^{-1/2}$, $\Theta^\top\Theta=\mathbb I_m$, $\Lambda=\lambda\mathbb I_m$, $d-m\ge1$, and $\sigma^2>0$.
\begin{enumerate}[leftmargin=2em]
\item[(1)] If $\tau^2=0$, then $p_\alpha(\lambda)$ is strictly increasing in $\lambda\in[0,\infty]$. In particular,
$
p_\alpha(\infty)>p_\alpha(0).
$
\item[(2)] If $\tau^2>0$, then there exists a constant $C>0$ such that
$
\tau^2<C
~\Rightarrow~
p_\alpha(\infty)>p_\alpha(0).
$
An explicit choice of $C$ is given in \Cref{sec:proof_of_robustness} of the Supplement.
\end{enumerate}
\end{theorem}

\begin{figure}[htb!]
    \centering
    \includegraphics[width=0.6\textwidth]{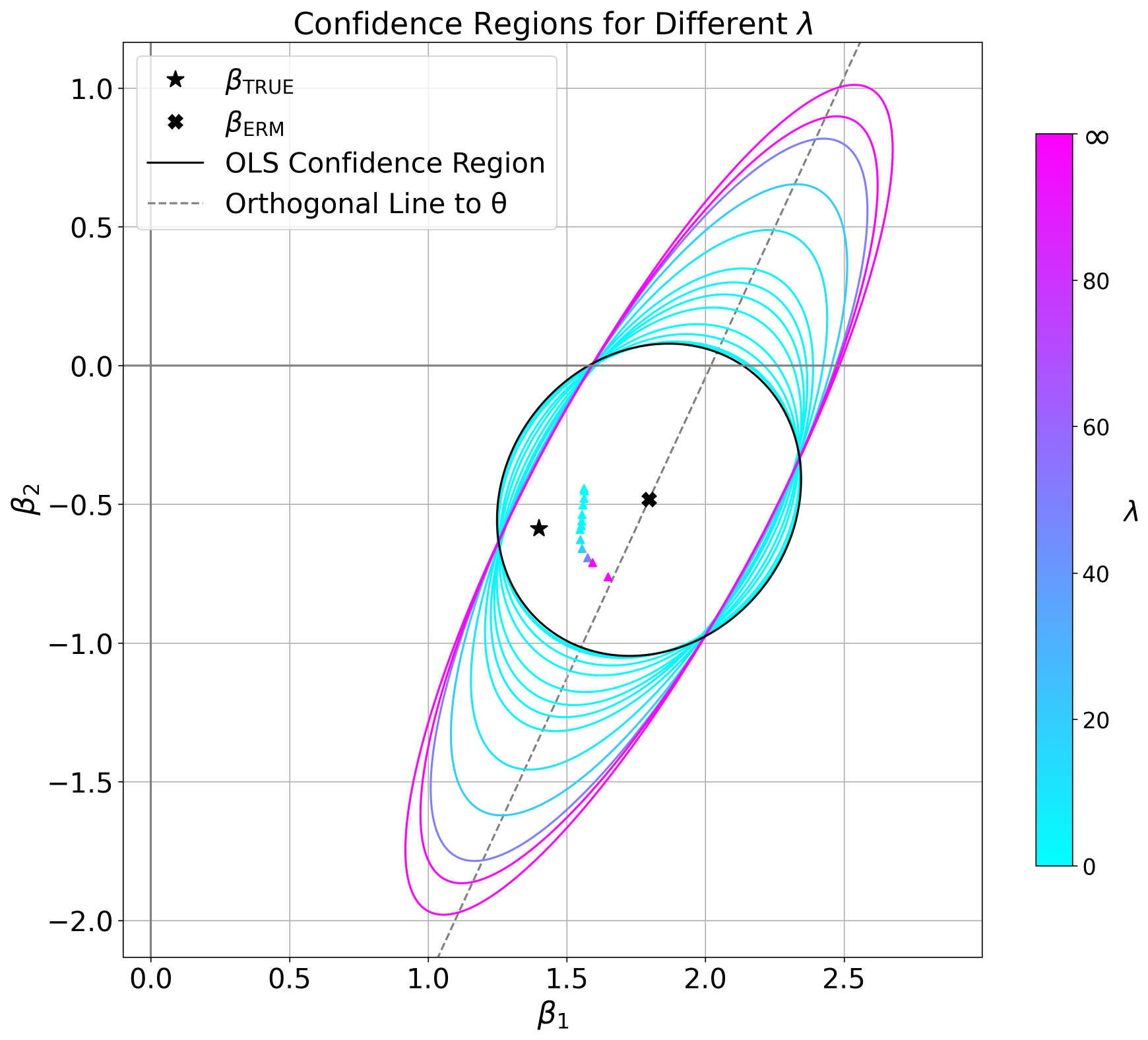}
    \caption{An illustration of the geometric reshaping of the READ confidence region in a linear model with $d=2$ and $m=1$. As the alignment hyperparameter $\lambda$ increases from 0 to $\infty$ (traced by the color bar), the region becomes narrower along the representation subspace $\operatorname{col}(\Theta)$ and wider along the orthogonal subspace $\operatorname{col}(\Theta_\perp)$. This reshaping improves coverage for future population parameters that remain stable in $\operatorname{col}(\Theta)$ but shift primarily in orthogonal directions.}
    \label{fig:confidence_region}
\end{figure}

Whenever $\tau^2$ is zero or small relative to $\sigma^2$, $\beta_e$ remains stable along $\col(\Theta)$ between the current and future populations. The READ confidence region reduces uncertainty within $\col(\Theta)$ by imposing invariance of $\Theta^\top X$ through the transport cost $c_{q,\Lambda}$. Because RWPI calibrates the Wasserstein radius to preserve current-target coverage, the region becomes narrower in $\col(\Theta)$ and wider in $\col(\Theta_\perp)$. This increases the chance of covering future parameters that shift primarily in orthogonal directions. The geometric intuition is illustrated in \Cref{fig:confidence_region}.

While \Cref{thm:robustness} assumes that the true $\Theta$ is known, practical applications rely on an empirical estimate $\hat\Theta\in\mathbb R^{d\times m}$ derived from external data. \Cref{thm:error_in_theta} shows that the local coverage advantage persists when the estimated and true representation subspaces are sufficiently aligned.

\begin{theorem}
\label{thm:error_in_theta}
Suppose \eqref{asm:drift_model} holds with $r_n=n^{-1/2}$ and $d-m\ge2$. Let $\angle_m$ denote the largest principal angle between $\col(\hat\Theta)$ and $\col(\Theta)$; see \cref{def:principal_angles} in the Supplement. Then there exist constants $K_1\in(0,1)$ and $K_2>0$ such that, if $\sin(\angle_m)<K_1$ and $\tau^2<K_2$, then $p_\alpha(\infty)>p_\alpha(0)$. The constants are defined in \Cref{sec:proof_robustness_theta} of the Supplement.
\end{theorem}

\subsection{Generalization Under Fixed-Scale Shifts}
\label{sec:fixed_scale_robustness}

We next take $r_n=1$ in \eqref{asm:drift_model}, so the difference between $\beta_e$ and $\beta_*$ does not vanish with the target sample size. The conventional confidence region $\mathcal C_{1-\alpha}(\lambda)$ has a diameter of the order $n^{-1/2}$ and therefore generally has zero limiting coverage for an independently drawn future parameter when  $r_n=1$. In this regime, we retain the READ geometry but replace the $n^{-1/2}$ scaling with the fixed scale $1$. In specific, define the scale-one robust region
\begin{equation}
\mathcal R_{1-\alpha}(\lambda)
=
\hat\beta_{\erm}
+\left\{
u\in\mathbb{R}^d:
\|u\|_{\Gamma^{-1}}^2
\le
\eta_\alpha(\Lambda)
\right\}.
\label{eq:fixed_scale_region}
\end{equation}
This region is the fixed-scale analogue of \eqref{eq:confidence_region} when $\Sigma_X=\mathbb I_d$. It is not intended to be a $(1-\alpha)$ confidence region for the current parameter $\beta_*$. Instead, it is a robust region whose scale matches the non-vanishing random effect in \eqref{asm:drift_model} with $r_n=1$. Then we define the limiting coverage probability as:
\[
p_\alpha^{\mathrm{fix}}(\lambda)
\coloneqq
\lim_{n\to\infty}
\mathbb P\bigl(\beta_e\in\mathcal R_{1-\alpha}(\lambda)\bigr).
\]

\begin{theorem}
\label{thm:fixed_scale_robustness}
Suppose \eqref{asm:drift_model} holds with $r_n=1$, $\Theta^\top\Theta=\mathbb I_m$, $\Lambda=\lambda\mathbb I_m$, $d-m\ge1$, and $\sigma^2>0$.
\begin{enumerate}[leftmargin=2em]
\item[(1)] If $\tau^2=0$, then $p_\alpha^{\mathrm{fix}}(\lambda)$ is strictly increasing in $\lambda\in[0,\infty]$. 
\item[(2)] There exists a constant $C_{\mathrm{fix}}>0$ such that
$
\tau^2<C_{\mathrm{fix}}
~\Rightarrow~
p_\alpha^{\mathrm{fix}}(\infty)
>
p_\alpha^{\mathrm{fix}}(0).
$
\end{enumerate}
\end{theorem}

\Cref{thm:fixed_scale_robustness} shows that the advantage of representation alignment is not an artifact of the local $n^{-1/2}$ model. Under non-vanishing model heterogeneity between the current and future populations, stronger alignment still improves the coverage of a scale-matched robust region when shifts are absent or sufficiently small within $\col(\Theta)$. Note that \Cref{thm:fixed_scale_robustness} compares robust regions with the same fixed scale and does not assert nominal coverage as conventional inference. 

\subsection{Projected Inference for the Invariant Coordinates}
\label{sec:projected_inference}

We return to the local regime $r_n=n^{-1/2}$ and the conventional confidence region $\mathcal C_{1-\alpha}(\Lambda)$. The previous subsections studied robustness for the full model coefficients. We now examine inference for the invariant coordinates $\kappa$. Suppose $\tau=0$, so the random shifts in \eqref{asm:drift_model} are entirely in $\col(\Theta_\perp)$. Then we have
$
(\Theta^\top\Theta)^{-1}\Theta^\top\beta_e=\kappa,
$
so projection onto $\col(\Theta)$ retains the stable coordinates. These coordinates may be more scientifically meaningful than the full coefficients $\beta_e$; see, for example, the context of causal invariance learning \citep{peters2016invariant}. We study the projected confidence region
$
(\Theta^\top\Theta)^{-1}\Theta^\top\mathcal C_{1-\alpha}(\Lambda)
$
and restrict the alignment parameter to $\Lambda=\lambda(\Theta^\top\Theta)^{-1}$, which applies the same alignment strength in standardized representation coordinates.

\begin{theorem}
\label{thm:interval_estimate}
Suppose $r_n=n^{-1/2}$, $\tau=0$, and $\Lambda=\lambda(\Theta^\top\Theta)^{-1}$, with $\lambda\in[0,\infty]$. Let $\Pi\in\mathbb R^{m\times m}$ be any fixed positive-definite symmetric matrix and define
\[
V_{\Pi,n}(\lambda)
\coloneqq
\vol_\Pi\left(
(\Theta^\top\Theta)^{-1}\Theta^\top
\mathcal C_{1-\alpha}(\Lambda)
\right),
\]
where $\vol_\Pi(C)=\int_C\sqrt{\det(\Pi)}\,dx$ for any $C\subset\mathbb R^m$. Then we have
\begin{enumerate}[leftmargin=2em]
\item[(1)] For any $\lambda>0$,
\[
\liminf_{n\to\infty}
\mathbb P\left\{
\kappa\in
(\Theta^\top\Theta)^{-1}\Theta^\top
\mathcal C_{1-\alpha}(\Lambda)
\right\}
\ge1-\alpha.
\]
\item[(2)] $V_{\Pi,n}(\lambda)$ is strictly decreasing in $\lambda\in[0,\infty]$, and $n^{m/2}V_{\Pi,n}(\lambda)$ converges to a strictly decreasing positive volume function of $\lambda$.
\end{enumerate}
\end{theorem}

\Cref{thm:interval_estimate} shows that stronger representation alignment (e.g., larger $\lambda$) yields sharper inference for $\kappa$ while maintaining asymptotic coverage. Thus, READ can provide more precise inference for the stable representation coordinates $\kappa$ than standard RWPI, which corresponds to $\lambda=0$.

\section{Simulation Studies}
\label{sec:simulation}

We evaluate the finite-sample performance of READ for both point estimation and interval coverage, testing its robustness on the current target distribution and its generalizability to future distributions. Additional configuration details are deferred to Section \ref{sec:details_sim} of the Supplementary Material. Let \(n\), \(m\), and \(d\) denote the sample size, the number of sources, and the dimension of the model coefficients. We generate data from a linear model with the target parameter $\beta_*=\rho\Theta\kappa+\sqrt{1-\rho^2}\varepsilon$, with $\varepsilon\sim N(0,(C/d)\mathbb{I}_d)$. The parameter \(\rho\in[0,1]\) controls the alignment between \(\beta_*\) and the representation \(\Theta\). We test our framework in two distinct settings. First, in terms of estimation and inference of the current target, we explore both low-$m$ and high-$m$ regimes. In the high-$m$ scenario, the invariant coefficient $\kappa$ is sparse, mimicking the practical issue of non-informative knowledge. Second, in terms of out-of-distribution (OOD) generalization, each fitted estimator or interval is tested on future populations with model coefficients generated according to \eqref{asm:drift_model} with $\tau^2 = C/(25d)$ and $\sigma^2 =C/(5d)$.  

For point estimation, we compare \texttt{READ} with four benchmarks, including the standard Wasserstein \texttt{DRO} and three state-of-the-art transfer learning methods: distance-based shrinkage estimation (\texttt{TransLasso}) \citep{li2021translasso}, angle-based method (\texttt{AngleBased}) \citep{gu2025angle-based} and representation-based method (\texttt{RepBased}) \citep{tian2025similar-rep}. For interval coverage, we compare with the standard inference approach based on empirical risk minimization (\texttt{ERM}) and RWPI based on the standard Wasserstein \texttt{DRO} \citep{blanchet2019rwpi}.



\begin{figure}[!htb]
    \centering
    \includegraphics[width=1.0\linewidth]{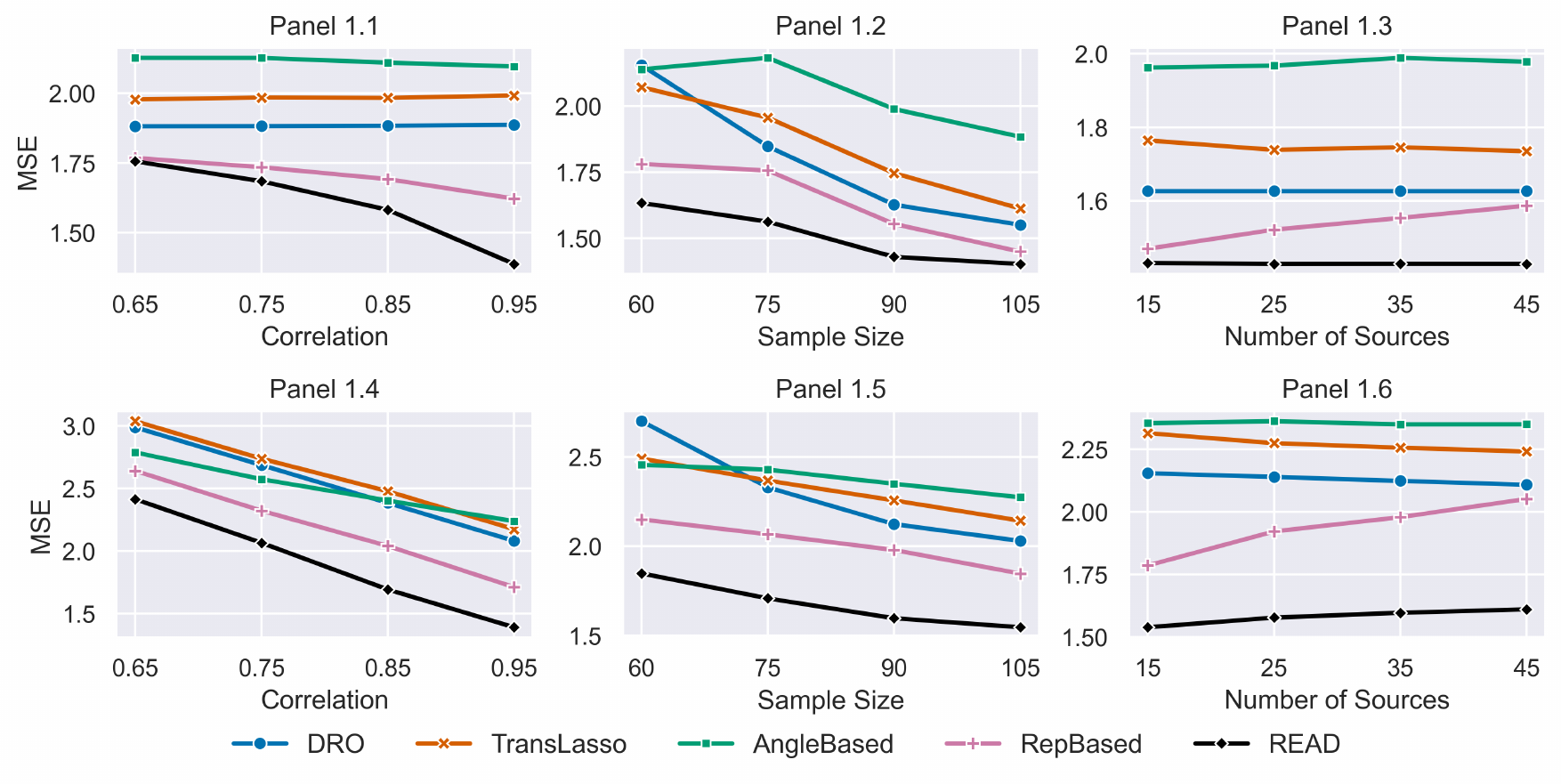}
    \caption{Results for the high-\(m\) regime. Panels 1.1--1.3 report Task 1 point-estimation performance under the target distribution, while Panels 1.4--1.6 report Task 2 OOD performance under future distributions generated according to \eqref{asm:drift_model}.} 
    
    \label{fig:sim_1}
\end{figure}

\textbf{Task 1: Prediction on the current target.} We fix \(C=2\) and \(d=50\). We vary one of \(\rho\), \(n\), and \(m\) while fixing the other two. The results are shown in the upper three panels of \Cref{fig:sim_1}. Across all 12 high-\(m\) settings in \Cref{table:task1_highm}, \texttt{READ} achieves the lowest mean squared error (MSE) in terms of estimating the target $\beta^*$. Its average MSE is 1.513, which is 6.9\% lower than the strongest competitor, \texttt{RepBased}. As \(m\) increases from 15 to 45, the MSE of \texttt{READ} remains stable. This confirms the importance of tuning \(\Lambda\), as using all directions in \(\Theta\) may encounter negative knowledge transfer when many of them are weak or misleading and the target sample size is small. The low-\(m\) results are deferred to \Cref{sec:details_sim}. In this regime, \texttt{READ} has the best performance on average and attains the lowest MSE in 10 of the 12 settings in \Cref{table:task1_lowm}. 

\textbf{Task 2: Prediction on future distributions.} Next, we evaluate the OOD prediction performance under the model shift regime \eqref{asm:drift_model}. We use the same high-\(m\) configurations as in \textbf{Task 1}. Each simulation averages the OOD MSE over 30 future test populations. The results are shown in the lower three panels of \Cref{fig:sim_1}. In the high-\(m\) regime, \texttt{READ} achieves the lowest OOD MSE in all 12 configurations in \Cref{table:task2_highm}. Its average MSE is 1.716, which is 16.0\% lower than the strongest competing method, \texttt{RepBased}, and 25.7\% lower than \texttt{DRO}. Again, the advantage is especially clear as $\Theta$'s dimension \(m\) increases. The low-\(m\) results are reported in \Cref{table:task2_lowm} and \texttt{READ} also achieves the best performance.

\textbf{Task 3: Inference on the current target.} We assess the empirical coverage of the READ confidence region constructed by \Cref{alg:lambda-cv}, with \(\Lambda\) selected by our data-driven procedure. We consider \((d,m)=(5,2),(10,4),(15,7)\) and vary the sample-to-dimension ratio \(n/d\in\{15,30,45\}\). For each configuration, in \Cref{sec:cr_coverage}, we report the coverage rate computed on 1000 replications. Overall, the coverage rate of our method is close to the nominal level. When \(n/d=45\), the coverage rates are all within $2\%$ of the nominal level, in all settings of the other hyperparameters. When \(n/d=15\), the coverage rates decrease slightly while still remaining $\leq 4\%$ below the nominal level, and most deviations from the nominal level are within $2\%$. These results support the validity of the RWPI-based confidence region after data-driven selection of \(\Lambda\) justified in \Cref{sec:lambda_selection}.

\begin{figure}[htb!]
    \centering
    \includegraphics[width=1.0\linewidth]{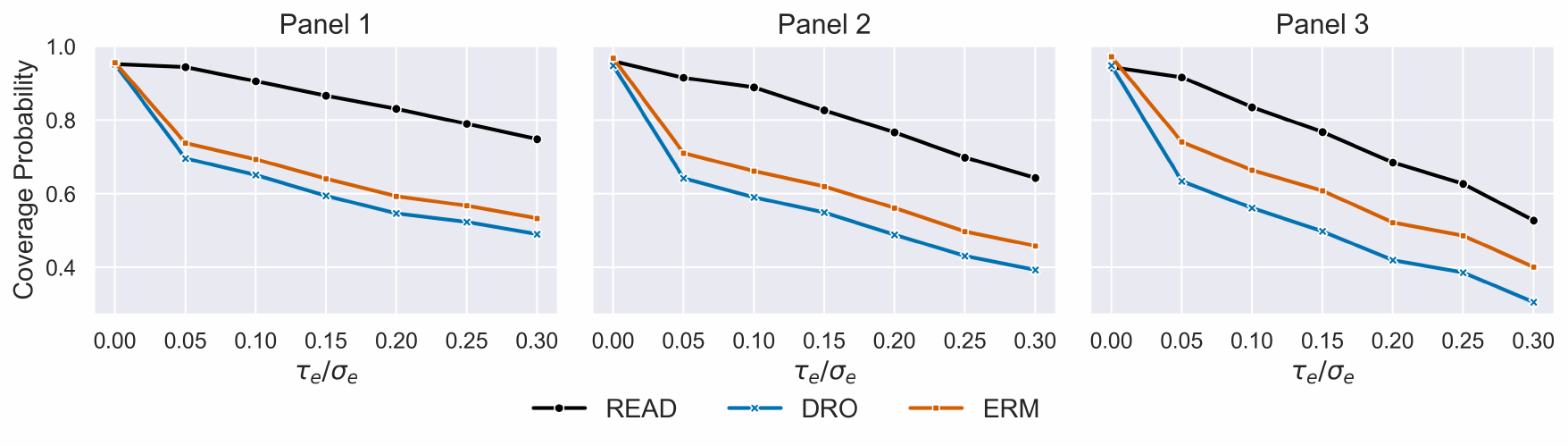}
    \caption{Coverage performance on future distributions in Task 4. The panels correspond to \((d,m)=(5,2)\), \((10,4)\), and \((15,7)\), respectively. Results are averaged over the coverage for future draws of \(\beta_e\) according to \eqref{asm:drift_model}.}
    \label{fig:sim_2}
\end{figure}

\textbf{Task 4: Inference on future distributions.} Finally, we evaluate coverage performance for future parameters under distribution shifts. Both the target parameter \(\beta_*\) and the future model parameters \(\beta_e\) are generated as in \eqref{asm:drift_model}. We set \(n/d=15\) and vary the signal ratio \(\tau/\sigma\in\{0.05,0.10,0.15,0.20,0.25,0.30\}\), where smaller values of $\tau/\sigma$ correspond to a stronger invariant signal along \(\Theta\). We also include the case \(\tau=\sigma=0\), which reduces to inference of the current target model parameter. We use \texttt{READ}, \texttt{DRO} and \texttt{ERM} to construct confidence regions to cover the future \(\beta_e\), with nominal level \(0.95\) on the current target data. We use the same three settings of \((d,m)\) as in Task 3. Each configuration is repeated 250 times, with coverage estimated from 100 draws of \(\beta_e\) in each replication. The results in \Cref{fig:sim_2} show that the coverage decreases as \(\tau/\sigma\) increases, reflecting greater variation within the representation subspace and hence the larger expected discrepancy between the current \(\beta_*\) and the future \(\beta_e\). Excluding the target-inference case \(\tau=\sigma=0\), \texttt{READ} achieves the highest coverage in all configurations. This superior OOD coverage rate empirically supports the theoretical results in Section \ref{sec:robustness}. By geometrically reshaping confidence regions to expand along directions orthogonal to $\Theta$ while remaining tight within the stable representation subspace, READ successfully captures future population shifts that elude standard methods.

\section{Application to Single-Cell RNA Data}
\label{sec:data}

We evaluated \texttt{READ} on a multimodal single-cell dataset \citep{luecken2021a} and compared it with \texttt{TransLasso}, \texttt{AngleBased}, and \texttt{RepBased}. The task is to predict antibody-derived tags (ADTs) from gene expression (GEX) measured by RNA counts. We focus on the annotated CD14$+$monocytes (\texttt{CD14+Mono}) cell type. The data set consists of human bone marrow cells collected from 12 separate batches. The distributional heterogeneity between them naturally arises from technical variations and donor-specific biological differences. However, the fundamental biological mechanism linking GEX to ADT is supposed to be conserved in all environments and tasks. Following the preprocessing detailed in Supplement \Cref{sec:screen}, we retain the top 400 highly variable genes using Scanpy \citep{wolf2018scanpy}. Then using a separate batch of \texttt{CD14+Mono} cells, we identify the 9 most predictable ADTs (Supplement \Cref{table:screen_values}) to serve as our target outcomes. To avoid overfitting, this batch is held out in all the downstream training and evaluation tasks.

We evaluate the practical utility of READ in three scenarios that reflect common single-cell analytical challenges: (1) multi-source (batch) transfer learning, which leverages external batches to improve predictions on a data-scarce target batch; (2) multi-task learning, which transfers shared structural information between different ADT tasks within the same population; and (3) statistical inference, which demonstrates how READ produces sharper confidence regions for stable invariant coordinates. 


\textbf{Scenario 1: Multi-source transfer learning.}
We first study transfer learning for each of the 9 selected ADT outcomes across different data batches. The target data consists of two batches randomly picked from the data set. To create a data-scarce, high-dimensional regression problem, we randomly subsample 360 observations for training and validation for each target, reserving the remainder for testing. The source data comprises four batches randomly selected from the data, each treated as an independent source, with sample sizes ranging from 2,500 to 3,000. To construct the representation matrix $\Theta \in \mathbb{R}^{400 \times 4}$, we fit an elastic-net regression with standardization in each source and stack the resulting coefficient vectors. Due to the randomness in picking the source and target data, we repeat the above-introduced procedure 10 times and report the average results. 

We include the same set of benchmark methods as in Section \ref{sec:simulation}. Their performance is measured by the improvements on the test-set $R^2$ relative to the standard \texttt{DRO} baseline. For \texttt{READ}, we turn to an alternative data-driven tuning strategy for the radius $\delta$ due to high-dimensionality of features, while the RWPI tuning strategy is considered later in Scenario 3 with lower-dimensional covariates. Additional details of the implementation are provided in \Cref{sec:details_task_i}.

\begin{figure}[htb!]
    \centering
    \includegraphics[width=1\linewidth]{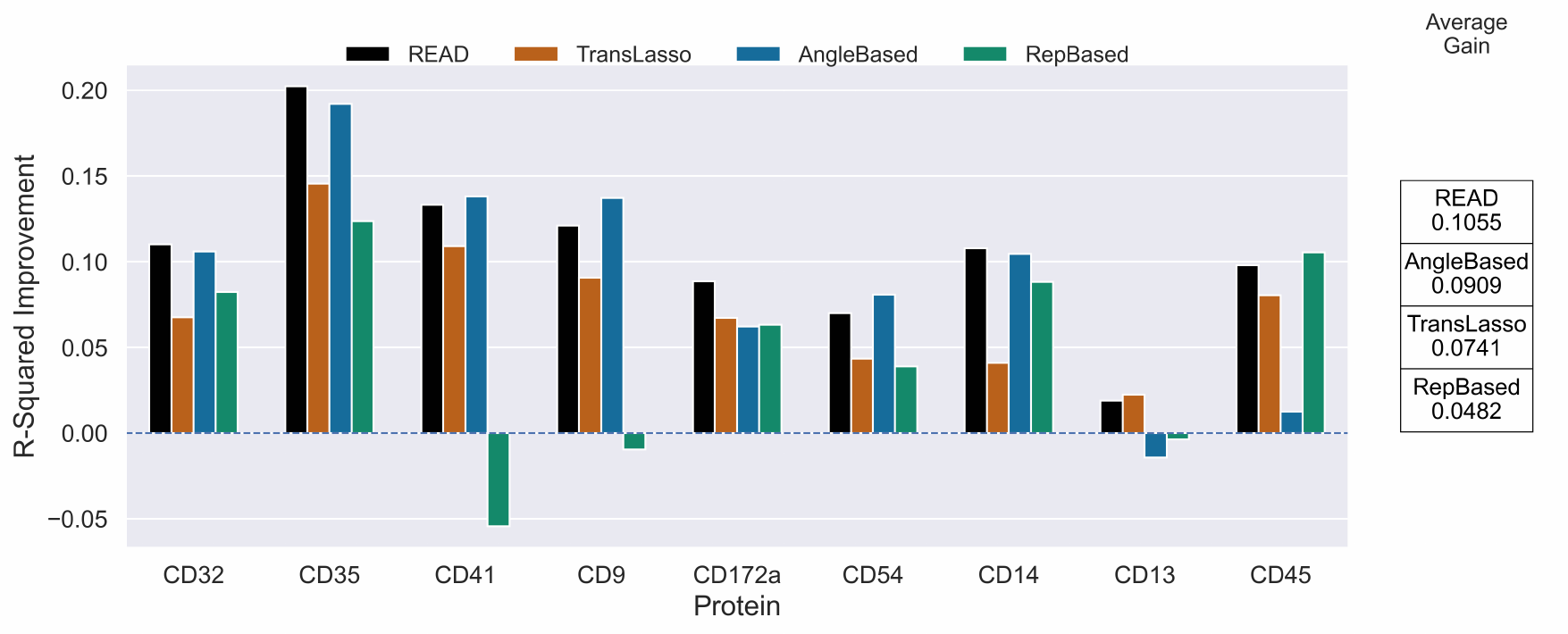}
    \caption{Prediction performance in the high-dimensional multi-source batch transfer scenario (Scenario 1). Bars report the improvement in average test-set $R^{2}$ for predicting nine ADT surface proteins from gene expression features, measured relative to the vanilla \texttt{DRO} baseline.}
    \label{fig:data1}
\end{figure}

\Cref{fig:data1} reports \(R^2\)-improvement of each method relative to the baseline \texttt{DRO}. For more than half of the ADT outcomes, \texttt{READ} improves over \texttt{DRO} by more than \(0.1\). \texttt{READ} achieves the best performance for the largest number of outcomes and the highest average \(R^2\). \texttt{TransLasso} also improves consistently over the baseline, but by smaller margins. By contrast, \texttt{RepBased} falls below the baseline in several tasks and \texttt{AngleBased} does so once. Unlike \texttt{TransLasso} or \texttt{AngleBased}, our method avoids forcing the target estimator to align with a rigid average of the source models. Instead, it adaptively transfers information in the representation matrix $\Theta$ through $\Lambda$ and $\delta$ that are carefully selected in our framework.



\textbf{Scenario 2: Multi-Task Learning.} This scenario focuses on transferring shared structural information across different ADT prediction tasks within the same batch. We frame this as a multi-task learning problem that holds out one of the nine ADT outcomes to serve as the target prediction task and uses the other eight as auxiliary tasks in turn. For each target task, we train regression models for the remaining eight ADTs and stack the learned coefficient vectors to construct $\Theta$. This matrix captures the shared biological structure among the source tasks, which is then transferred to improve the prediction of the target outcome. Using the same training and validation split strategy as in Scenario 1, we repeat this process 10 times per target ADT. Additional details are provided in \Cref{sec:details_task_ii}. 

\begin{figure}[htb!]
    \centering
    \includegraphics[width=1.0\linewidth]{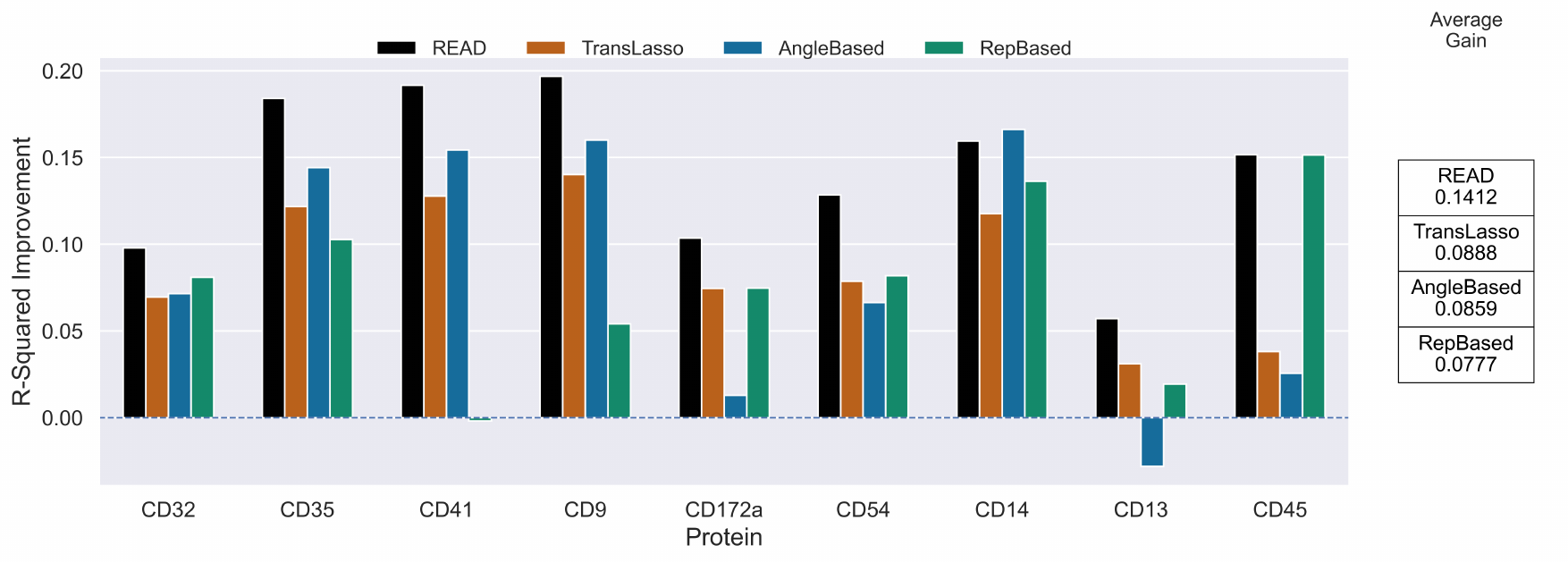}
    \caption{Prediction performance in the multi-task learning scenario (Scenario 2). Bars report the improvement in average test-set $R^{2}$ for predicting a held-out ADT target using shared structural information from eight auxiliary ADT targets, measured relative to the vanilla DRO baseline.}
    \label{fig:data2}
\end{figure}

The results in \Cref{fig:data2} illustrate that \texttt{READ} performs the best overall by a clear margin. It wins 8 of the 9 tasks, outperforms the baseline \texttt{DRO} in all tasks, and achieves the highest average \(R^2\), surpassing the second-best method by \(0.05\). \texttt{AngleBased} and \texttt{RepBased} are competitive but less consistent, whereas \texttt{TransLasso} is more stable but yields only modest gains for some targets. Compared with batch transfer, this multi-task scenario has a more complex structure of $\Theta$, and \texttt{READ} uses this structure more effectively.

\textbf{Scenario 3: Inference for Invariant Coordinates.} As reported in \Cref{sec:additional_task} of the Supplement, we examine whether the representation-aware geometry of READ yields sharper inference on the invariant coordinates as was theoretically studied in Section \ref{sec:projected_inference}. To make RWPI-based inference feasible, we first select 20 important GEX features for the 9 ADT outcomes using elastic-net on the screening batch independent of our training data. For each ADT outcome, we then implement a similar procedure as in Scenario 1 to fit linear regressions with the four source batches and stack the resulting source coefficient vectors to form \(\Theta\in\mathbb R^{20\times 4}\). We construct the \texttt{READ} confidence region \(\hat{\mathcal{C}}_{0.95}(\hat\Lambda)\) on the target batch and compare it with the vanilla \texttt{DRO} region \(\hat{\mathcal{C}}_{0.95}(0)\). We evaluate the projected marginal interval lengths both after projecting onto \(\operatorname{col}(\Theta)\) and after mapping to the invariant coordinates \(\kappa=(\Theta^\top\Theta)^{-1}\Theta^\top\beta\). In these directions, \texttt{READ} generally produces shorter intervals than \texttt{DRO}. It shows positive length reductions over \texttt{DRO} among 8 of the 9 ADT outcomes in the projected GEX-coordinate intervals and positive reductions for all 9 outcomes in terms of the invariant-coordinate $\kappa$. These results support the theoretical message of Section \ref{sec:projected_inference}: representation-aware alignment can sharpen the inference for stable representation-level coordinates $\kappa$ and their corresponding model coefficients $\Theta\kappa$.

\begin{figure}[htb!]
    \centering
    \includegraphics[width=0.9\linewidth]{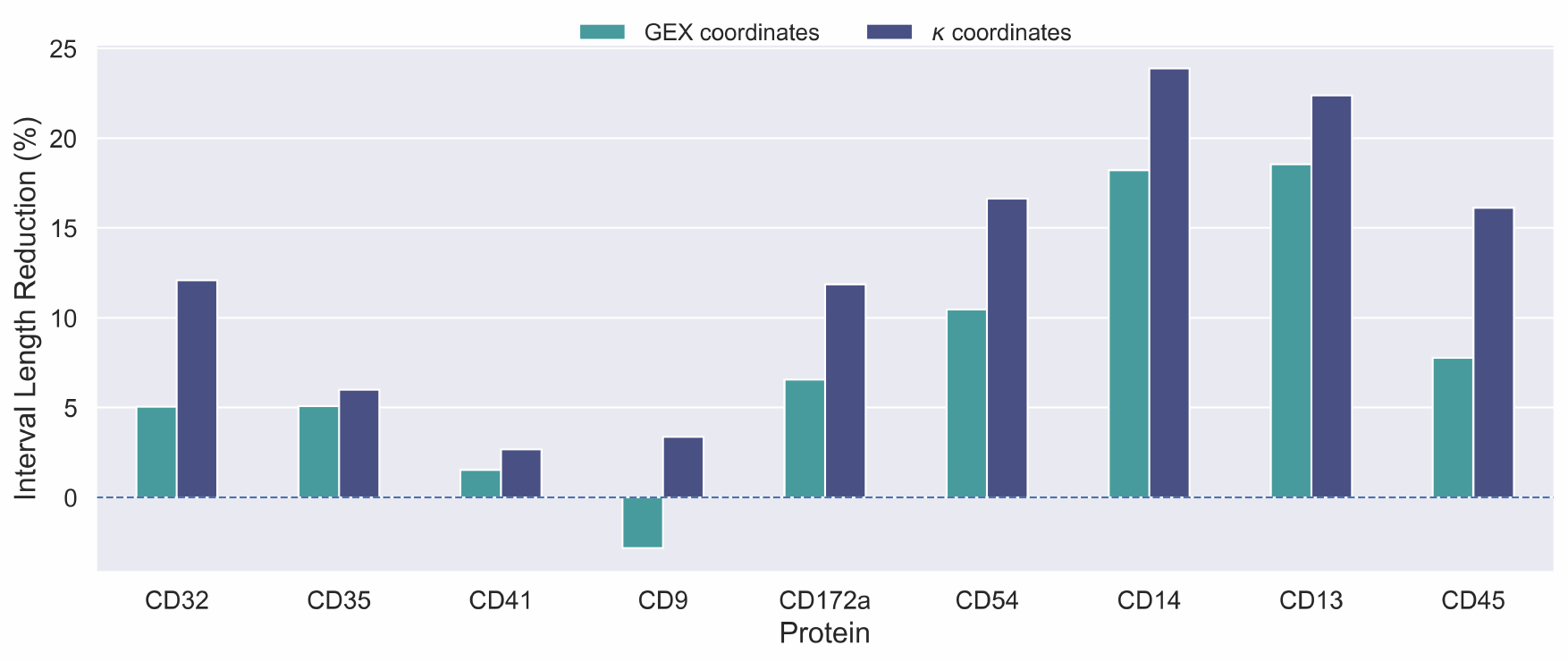}
    \caption{Projected interval-length comparison under the invariant coordinates inference scenario (Scenario 3). Bars display the average percentage reduction in marginal interval lengths of READ relative to the vanilla DRO baseline. Results are shown for projections onto both the invariant coordinates $\kappa$ and their corresponding GEX coordinates $\Theta\kappa$. Positive values indicate relatively sharper confidence intervals by using READ compared to DRO. }
    \label{fig:additional_data}
\end{figure}

\section{Discussion}\label{sec:diss}

This paper develops READ as a representation-aware extension of Wasserstein distributionally robust inference. Beyond its regularized-regression interpretation, a central message of our work is that the representation-aware transport cost reshapes the RWPI confidence region in WDRO. In particular, it contracts uncertainty along the externally informed representation subspace and expands protection in the orthogonal directions where future population shifts are more likely to occur. This geometry provides a mechanism for improved coverage of future population parameters when the model shifts tend to occur in a ``representation-invariant'' way. Importantly, our inferential approach remains valid on the current target, even after the RWPI-based selection of the radius and alignment parameters that helps with adaptive knowledge transfer. The improved future-population coverage of READ should be interpreted together with the size of the robust region. Stronger alignment makes the region narrower along the stable representation subspace but may enlarge it in directions where population shifts are allowed. Thus, higher coverage can come at the cost of greater uncertainty in those directions. 

Several promising directions remain for future work. First, extending our inference method to high-dimensional regimes will require integrating representation-aware geometry with sparsity or low-rank structures. Second, adapting the framework for nonlinear models (e.g., neural networks or tree-based methods) could allow a more flexible and general use of our key idea. Finally, applying our representation-guided uncertainty principle to other DRO formulations (such as $f$-divergence or group DRO) would further expand the scope of knowledge-guided robust inference.  

{\spacingset{1}
\bibliography{references}
}

\newpage
\appendix
\setcounter{page}{1}

\numberwithin{figure}{section}
\numberwithin{table}{section}
\numberwithin{algorithm}{section}
\numberwithin{equation}{section}

\renewcommand{\thefigure}{\thesection\arabic{figure}}
\renewcommand{\thetable}{\thesection\arabic{table}}
\renewcommand{\thealgorithm}{\thesection\arabic{algorithm}}
\renewcommand{\theequation}{\thesection\arabic{equation}}

\renewcommand*{\theHfigure}{\theHsection.\arabic{figure}}
\renewcommand*{\theHtable}{\theHsection.\arabic{table}}
\renewcommand*{\theHalgorithm}{\theHsection.\arabic{algorithm}}
\renewcommand*{\theHequation}{\theHsection.\arabic{equation}}

\part{}
\begin{center}
{\large\bf SUPPLEMENTARY MATERIAL}
\end{center}

{\spacingset{1.0}\parttoc}
\newpage
\spacingset{1.5}
\section{Additional Theoretical Discussion}

This section presents additional theoretical results.

\subsection{Asymptotics of READ for \texorpdfstring{$p=1$}{p = 1}}
\label{sec:asymptotic_sqrt_lasso}
The assumptions of \cref{thm:asymptotic_dro} require $p > 1$. We now study the asymptotic distribution of the READ estimator in the case $p = 1$. Under \cref{asm:regularity}, the loss function $\ell(x,y;\beta)$ is twice differentiable in both $x$ and $\beta$. We therefore define
\[
g(x,y;\beta) = \nabla_x \ell(x,y;\beta) \in \mathbb{R}^d,
\]
and, for simplicity, let $\zeta = (X,Y) \sim \mathbb{P}_*$. We write
\[
g_*(\zeta) = g(X,Y;\beta_*),
\quad
J_*(\zeta) = \nabla_\beta g(X,Y;\beta_*) \in \mathbb{R}^{d \times d}.
\]

For any $z,h \in \mathbb{R}^d$, define the directional derivative of the $\ell_1$ norm at $z$ in the direction $h$ by
\begin{align*}
    \sigma(z;h)
    &= \max_{s \in \partial \|z\|_1} s^\top h \\
    &= \sum_{j:z_j \neq 0}\operatorname{sign}(z_j)h_j + \sum_{j:z_j = 0}|h_j|.
\end{align*}

For any $\Lambda \in [0,\infty]^m$, recall that
\[
\varphi_{1,\Lambda}^2(z)
=
\inf_\kappa \left\{\|z-\Theta\kappa\|_1^2 + \|\kappa\|_{\Lambda^{-1}}^2\right\}.
\]
For each $z$, define
\[
K_\Lambda(z)
=
\argmin_\kappa \left\{\|z-\Theta\kappa\|_1^2 + \|\kappa\|_{\Lambda^{-1}}^2\right\}.
\]
When all entries of $\Lambda$ are positive and finite, $K_\Lambda(z)=\{\kappa_\Lambda^*(z)\}$ is a singleton. By Danskin's theorem \citep[Theorem 9.26]{shapiro2021lectures}, the directional derivative of $\varphi_{1,\Lambda}^2$ is
\[
D\varphi_{1,\Lambda}^2(z;h)
=
\min_{\kappa \in K_\Lambda(z)}
\left\{
2\|z-\Theta\kappa\|_1 \sigma(z-\Theta\kappa;h)
\right\}.
\]
When $K_\Lambda(z)=\{\kappa_\Lambda^*(z)\}$ is a singleton, define $r_*(z)=z-\Theta\kappa_\Lambda^*(z)$. In this case,
\[
D\varphi_{1,\Lambda}^2(z;h)
=
2\|r_*(z)\|_1\sigma(r_*(z);h).
\]

We now state the main theorem. Its proof follows from that of \cref{thm:asymptotic_dro}, after extending \cref{lem:bias_linearization} to the directionally differentiable case $p=1$.

\begin{theorem}
    Suppose \cref{asm:regularity,asm:pd_2} hold, and suppose the directional differentiability conditions stated above hold. Let $\delta_n = \eta/n$ for any fixed $\eta > 0$. Then for $p=1$ and $\Lambda\in[0,\infty]^m$, we have 
    \[
    \sqrt{n}(\hat\beta_\readd-\beta_*)
    \leadsto
    \argmin_{u} \left\{-H^\top u + \frac{1}{2}u^\top\Sigma u + \frac{\sqrt{\eta}}{2\mathcal{V}(\beta_*;\Lambda)}\mathbb{E}_{\mathbb{P}_*}[D\varphi_{1,\Lambda}^2(g_*(\zeta);J_*(\zeta)u)]\right\}
    \]
    where $H\sim N\left(0,\cov_{\mathbb{P}_*}h(X,Y;\beta_*)\right)$, and $\Sigma=-\mathbb{E}_{\mathbb{P}_*}[\nabla_\beta h(X,Y;\beta_*)]$. Suppose that $\lambda_j < \infty$ for all $j\in [m]$, and $r_*(\zeta) = g_*(\zeta) - \Theta \kappa_\Lambda^*(g_*(\zeta))$ almost surely has no components equal to zero, then $\sqrt{n}(\hat\beta_\readd - \beta_*)$ converges to a Gaussian distribution.
\end{theorem}

\subsection{Asymptotics under Perfect Alignment}
\label{sec:asymptotics_constrained}

The conditions in \cref{asm:pd,asm:pd_2} exclude the regime in which the true parameter lies entirely in the prior subspace, that is, \(\beta_* \in \col(\Theta)\), as well as the more general case in which the distance from \(\beta_*\) to \(\col(\Theta)\) vanishes faster than \(n^{-1/2}\) when \(\Theta\) itself depends on the target sample size \(n\). This is a distinct perfect-alignment regime and requires a separate asymptotic analysis. To complement \cref{thm:asymptotic_dro}, we therefore study the constrained estimator
\[
\label{obj:constrained}
\tag{Aligned}
\argmin_{\beta \in \col(\Theta)} \mathbb{E}_{\mathbb{P}_*}[\ell(X,Y;\beta)]
=\argmin_{\beta \in \mathbb{R}^d}
\Bigl\{ \mathbb{E}_{\mathbb{P}_*}[\ell(X,Y;\beta)]
:\varphi_{p,\infty}^2(\beta)=0 \Bigr\},
\]
where
\[
\varphi_{p,\infty}^2(\beta)=\inf_{\kappa}\|\beta-\Theta\kappa\|_p^2.
\]

This constrained formulation is conceptually clean, but we analyze a more general relaxed estimator that contains it as a special case. Specifically, define the Lagrangian estimator
\[
\hat \beta_{\mathcal L}
\coloneqq \argmin_{\beta \in \mathbb{R}^d}
\Bigl\{ \mathbb{E}_{\mathbb{P}_n}[\ell(X,Y;\beta)]
+ \delta_n \varphi_{p,\Lambda_n}^2(\beta;\Theta_n) \Bigr\},
\]
where the prior knowledge \(\Theta_n\) may depend now on \(n\). For each linear subspace \(U \subseteq \mathbb{R}^d\), let \(P_U\) denote the orthogonal projection onto \(U\), and define
\[
d_{\op}(U,V)=\|P_U-P_V\|_{\op}
\coloneqq \sup_{\|x\|_2=1}\|(P_U-P_V)x\|_2,
\]
which is a metric on the collection of linear subspaces of \(\mathbb{R}^d\). We study growth conditions on \(\delta_n\to\infty\) and \(\lambda_{j,n}\to\infty\), \(j=1,\ldots,m\), under which the relaxed estimator is asymptotically equivalent to the constrained problem. \textit{Suppose that \(\Theta_n\) and \(\Theta\) have orthonormal columns}.

\begin{assumption}
\label{asm:Theta}
Assume that \(d_{\op}(\col(\Theta_n),\col(\Theta))=O_p(n^{-\gamma})\) for some \(\gamma>1/2\).
\end{assumption}

This assumption requires the source subspace to be estimated at a rate faster than \(n^{-1/2}\), the scale of the target estimator.

\begin{theorem}
\label{thm:lagrangian_asymptotics}
Suppose \cref{asm:regularity_i,asm:regularity_iii,asm:Theta} hold and \cref{asm:optimality} holds with \(\beta_*\in\col(\Theta)\). If \(\delta_n^{-1}=o(1)\) and \(\delta_n n^{1/2-\gamma}=o(1)\), and if \(\lambda_{j,n}^{-1}=o(1)\) and \(\delta_n n^{1/2}/\lambda_{j,n}=o(1)\) for all \(j\in[m]\), then
\[
\sqrt{n}(\hat\beta_{\mathcal L}-\beta_*)
\leadsto
\Theta(\Theta^\top \Sigma \Theta)^{-1}\Theta^\top H,
\]
where \(H \sim N(0,\cov_{\mathbb{P}_*}(h(X,Y;\beta_*)))\) and \(\Sigma=-\mathbb{E}_{\mathbb{P}_*}[\nabla_\beta h(X,Y;\beta_*)]\).
\end{theorem}

The upper bound on the divergence rate of \(\delta_n\) and the lower bound on the divergence rate of \(\min_j\lambda_{j,n}\) ensure that the relaxation asymptotically recovers the hard constraint without introducing additional bias. The first prevents the penalty term from dominating the stochastic fluctuations of the estimator, while the second ensures that directions orthogonal to \(\Theta\) are penalized strongly enough. Moreover, by \cref{lem:schurcomplement},
\[
\Theta(\Theta^\top \Sigma \Theta)^{-1}\Theta^\top \preceq \Sigma^{-1}.
\]
Hence the limiting covariance of the constrained estimator is smaller than that of the ERM estimator. This agrees with the classical asymptotic distribution of \(M\)-estimators under linear constraints.

\subsection{Variational Dual Form of \texorpdfstring{$\psi_{p,\Lambda}^*$}{Psi-p-Lambda}}
\label{sec:selection}
In this section, we study the simulation of the quantile $\eta_\alpha(\Lambda)$ under the target distribution \(\mathbb{P}_*\), where \(\eta_\alpha(\Lambda)\) denotes the \((1-\alpha)\)-quantile of \(\psi_{p,\Lambda}^*(H)\). We let $\Lambda \in [0,\infty]^m$ be fixed. In general, \(\psi_{p,\Lambda}^*(H)\) does not admit a closed form:
\[
\psi_{p,\Lambda}^*(H)
=
\sup_{\xi \in \mathbb{R}^d}
\Bigl\{
\xi^\top H
-
\mathbb{E}_{\mathbb{P}_*}\bigl[\tfrac{1}{4}\varphi_{p,\Lambda}^2(\Xi \xi)\bigr]
\Bigr\},
\qquad
\Xi^\top \coloneqq \nabla_x h(X,Y;\beta_*).
\]

A direct approximation of \(\eta_\alpha(\Lambda)\) is to repeatedly evaluate the convex conjugate of the empirical criterion \(\mathbb{E}_{\mathbb{P}_n}[4^{-1}\varphi_{p,\Lambda}^2(\Xi\xi)]\) via the conic program
\begin{align*}
\label{obj:cone}
\tag{Conic}
\hat{\psi}_{p,\Lambda}^{\,*}(h)
&= \max_{\xi,\{t_i\},\{\kappa_i\}}
\left\{
h^\top \xi - \frac{1}{4n}\sum_{i=1}^n t_i
\right\} \\
&\text{s.t.} \quad
t_i \geq \bigl\Vert \Xi_i \xi - \Theta \kappa_i \bigr\Vert_p^2 + \|\kappa_i\|_{\Lambda^{-1}}^2,
\quad \forall i \leq n.
\end{align*}
However, because \(\varphi_{p,\Lambda}^2\) is itself defined through an inner optimization, this formulation introduces \(n\) conic constraints and may be computationally demanding. We therefore derive next a variational representation of \(\psi_{p,\Lambda}^*\) that is more convenient for simulation. Let \((X,Y)\in\mathbb{R}^{d+1}\) be defined on an underlying probability space \((\Omega,\mathcal{F},P_*)\), and let \(\Xi\in\mathbb{R}^{d\times d}\) be the corresponding \(\mathcal{F}\)-measurable random matrix. Recall that the convex conjugate of \(\frac{1}{4}\varphi_{p,\Lambda}^2\) is \(c_{q,\Lambda}\).

\begin{proposition}
\label{prop:variational_rep}
Assume that \(\mathbb{E}_{P_*}[\|\Xi\|_{\op}^2] < \infty\). Define
\[
\psi_{p,\Lambda}(x)
=
\frac{1}{4}\,\mathbb{E}_{P_*}\bigl[\varphi_{p,\Lambda}^2(\Xi x)\bigr]
=
\mathbb{E}_{P_*}\bigl[c_{q,\Lambda}^*(\Xi x)\bigr],
\qquad x \in \mathbb{R}^d.
\]
Then, for every \(z\in\mathbb{R}^d\),
\[
\psi_{p,\Lambda}^*(z)
=
\inf\left\{
\mathbb{E}_{P_*}\bigl[c_{q,\Lambda}(U)\bigr]
:
U \in L^2(\Omega;\mathbb{R}^d),\;
\mathbb{E}_{P_*}[\Xi^\top U] = z
\right\},
\]
with the convention that the infimum over the empty set is \(+\infty\).
\end{proposition}

Given samples \(\{\Xi_1,\ldots,\Xi_n\}\), constructed using an estimator of \(\beta_*\), we may then compute
\begin{align*}
\tag{Quadratic}
\hat{\psi}_{p,\Lambda}^*(h)
&= \min_{\{u_i\}} \frac{1}{n}\sum_{i=1}^n c_{q,\Lambda}(u_i) \\
&\text{s.t.} \quad \frac{1}{n}\sum_{i=1}^n \Xi_i^\top u_i = h.
\end{align*}
When \(\Lambda\) has no infinite entries, \eqref{obj:variational} involves only a single constraint and is often easier to solve in practice. Moreover, \eqref{obj:variational} and \eqref{obj:cone} are Fenchel--Rockafellar dual; see \citet[Theorem 31.1]{rockafellar1970convex}. Let \(\hat{\beta}\) denote any consistent estimator of \(\beta_*\). The RWPI-based \(\delta\)-selection algorithm is given in \Cref{alg:delta}.

\begin{proposition}
\label{prop:alg_delta_consistency}
\Cref{alg:delta} consistently estimates \(\eta_{\alpha}(\Lambda)\), the \((1-\alpha)\)-quantile of \(\psi_{p,\Lambda}^*(H)\), as \(n\) and \(l\) both diverge.
\end{proposition}

\section{Additional Methodological Discussion}

This section presents additional methodological descriptions and results.

\subsection{Selecting the Alignment Parameters}
\label{sec:lambda_selection}
We first turn to the selection strategy of the alignment parameter
\(\Lambda \in [0,\infty]^m\) that preserves the validity of RWPI based on it. Our goal is to choose \(\Lambda\) in a representation-aware manner that adapts to the target distribution \(\mathbb{P}_*\). In particular, we seek a choice of \(\Lambda\) that yields the best READ estimator under \(\mathbb{P}_*\) where the Wasserstein
radius \(\delta\) is determined by RWPI given \(\Lambda\). For a fixed \(\Lambda\), let
\[
\hat\beta_\Lambda
\in
\argmin_{\beta}
\sup_{\mathbb{P}\in \mathcal{B}_{\delta_n}(\mathbb{P}_n;c_{q,\Lambda})}
\mathbb{E}_{\mathbb{P}}[\ell(X,Y;\beta)],
\qquad
\delta_n=\eta_\alpha(\Lambda)/n.
\]
A natural criterion is to choose \(\Lambda\) by minimizing the
cross-validated prediction risk. We use a fixed and finite candidate
family for $\Lambda$ to ensure the consistency of the resulting selector.

\begin{algorithm}[htb!]
\caption{Cross-validated tuning of \(\Lambda\) for linear regression}
\label{alg:lambda-cv}
\textbf{Require:}
\(X\in\mathbb R^{n\times d}\), \(y\in\mathbb R^n\),
\(\Theta\in\mathbb R^{d\times m}\), a fixed integer \(K\geq 2\),
and a finite index set \(\mathcal F\).

\begin{enumerate}
    \item Partition the observations into \(K\) disjoint folds
    \(I_1,\ldots,I_K\) of equal size \(n_h=n/K\), and let
    \(n_t=n-n_h\).

    \item Construct a finite indexed family of candidate alignment
    parameters
    $
    \left\{\Lambda_n(f):f\in\mathcal F\right\}.
    $

    \item For each \(f\in\mathcal F\) and \(k\in[K]\), use the
    observations outside \(I_k\) to compute
    \[
    \hat\delta^{(-k)}_f
    =
    \frac{\hat\eta^{(-k)}_\alpha
    \{\Lambda_n(f)\}}{n_t},
    \]
    and fit the corresponding READ estimator
    \(\hat\beta_f^{(-k)}\).

    \item Define the \(K\)-fold cross-validation criterion
    \[
    \hat{\mathrm{CV}}_n(f)
    =
    \frac{1}{K}
    \sum_{k=1}^K
    \frac{1}{n_h}
    \sum_{i\in I_k}
    \left\{
    y_i-x_i^\top\hat\beta_f^{(-k)}
    \right\}^2.
    \]

    \item Select
    $
    \hat f
    \in
    \arg\min_{f\in\mathcal F}
    \hat{\mathrm{CV}}_n(f),
    $
    and
    $
    \hat\Lambda=\Lambda_n(\hat f).
    $

    \item Using the full target sample, compute
    $
    \hat\delta
    =
    {\hat\eta_\alpha(\hat\Lambda)}/{n},
    $
    and refit READ with \((\hat\Lambda,\hat\delta)\).
\end{enumerate}

\textbf{Return:}
\(\hat\Lambda\), \(\hat\delta\), and the final READ estimator.
\end{algorithm}

To construct the candidate family in practice, we restrict attention
to a low-dimensional parameterization. Let
\[
\hat\kappa_{\mathrm{init}}
=
\arg\min_{\kappa\in\mathbb R^m}
\left\|
\hat\beta-\Theta\kappa
\right\|_p,
\]
where \(\hat\beta\) is a consistent initial estimator, and define
\[
\Lambda_n(a,b)
=
a\,
\operatorname{diag}
\left(
\left|
\hat\kappa_{\mathrm{init}}
\right|^b
\right).
\]
We evaluate this family over a fixed finite grid of \((a,b)\). The
candidate matrices are allowed to be data dependent, but Assumption \ref{asm:alg_lambda}
below requires each indexed candidate to converge to a deterministic
limit. This condition is satisfied when
\(\hat\kappa_{\mathrm{init}}\) is consistent and the
parameterization is continuous at its population limit.


We next establish the validity of the confidence region after
data-driven selection of \(\Lambda\) in the linear regression model.
The argument has two components. First, the balanced \(K\)-fold
criterion consistently selects the minimizer of a deterministic
second-order prediction-risk criterion. Second, once the selected
alignment parameter converges to a deterministic limit, the empirical
RWPI quantities may be evaluated at the selected value without
affecting their first-order limit. An analogous result for logistic
regression would require a separate second-order expansion of its
cross-validation criterion and is not pursued here.
\begin{assumption}\label{asm:alg_lambda}
Let \(\mathcal F=\{1,\ldots,J\}\) be a fixed finite index set, and let
\(K\geq2\) be fixed. For every \(f\in\mathcal F\), the empirical candidate set
\[
\Lambda_n(f)
\overset{p}{\rightarrow}
\Lambda(f),
\]
where the limiting matrices belong to a compact set
\(\mathcal L\subset\operatorname{diag}([0,\infty]^m)\). Moreover,
\[
\inf_{\Lambda\in\mathcal L}
\lambda_{\min}(\Gamma_\Lambda)>0,
\qquad
\inf_{\Lambda\in\mathcal L}
\|\beta_*\|_{\Psi_\Lambda}^2>0,
\]
where \(\Gamma_\Lambda\) and \(\Psi_\Lambda\) are defined in
Proposition~\ref{prop:linearmodel}. Also, assume that
\[
f_*
=
\arg\min_{f\in\mathcal F}
\mathcal Q\{\Lambda(f)\}
\]
is unique, where
\[
\mathcal Q(\Lambda)
=
\eta_\alpha(\Lambda)
\frac{
\|\Psi_\Lambda\beta_*\|_{\Sigma_X^{-1}}^2
}{
\|\beta_*\|_{\Psi_\Lambda}^2
},
\qquad
\Sigma_X=\mathbb{E}_{\mathbb{P}_*}(XX^\top).
\]

\end{assumption}

\begin{proposition}\label{prop:lambda-selection}
Suppose Assumption~\ref{asm:alg_lambda} and the
assumptions of Proposition~4.1 hold. Then the selector in Algorithm~\ref{alg:lambda-cv} satisfies
\[
\Pr(\hat f=f_*)\rightarrow1,
\qquad
\hat\Lambda
\overset{p}{\rightarrow}
\Lambda_*:=\Lambda(f_*),
\]
as $n\rightarrow\infty$. Consequently, the plug-in confidence region
\[
\hat C_{1-\alpha}(\hat\Lambda)
=
\hat\beta_{\mathrm{ERM}}
+
n^{-1/2}
\left\{
u\in\mathbb R^d:
\hat\psi^*_{2,\hat\Lambda}
(\hat\Sigma u)
\leq
\hat\eta_\alpha(\hat\Lambda)
\right\}
\]
satisfies that
\[
\lim_{n\rightarrow\infty}\Pr\left\{
\beta_*\in
\hat C_{1-\alpha}(\hat\Lambda)
\right\}
=
1-\alpha.
\]
\end{proposition}

\subsection{Polyhedral Approximation of READ Confidence Region}
\label{sec:region_approx}
When \(\psi^*_{p,\Lambda}\) is not available in closed form, the confidence region can be approximated geometrically. Writing $\mathcal C_\eta \coloneqq \{u\in\mathbb{R}^d : \psi^*_{p,\Lambda}(\Sigma u)\le \eta\}$, its support function is $\sigma_{\mathcal C_\eta}(x)=2\sqrt{\eta\,\psi_{p,\Lambda}(\Sigma^{-1}x)}.$ Hence, if \(\{v_1,\ldots,v_k\}\) are sampled on the unit sphere, then
\[
\bigcap_{l=1}^k \left\{u\in\mathbb{R}^d : v_l^\top u \le \sigma_{\mathcal C_\eta}(v_l)\right\}
\]
is a polyhedral outer approximation of \(\mathcal C_\eta\); see \citet[Proposition 5]{blanchet2022confidence}.

\section{Proofs of Statements}
\label{app:proofs}

\subsection{Duality of Seminorms and Extended Norms}

Let \(g : \mathbb{R}^d \to [0,\infty]\) be an extended norm. We show that taking the dual norm establishes a one-to-one correspondence between the collection \(\mathcal{G}\) of extended norms on \(\mathbb{R}^d\) and the collection \(\mathcal{R}\) of seminorms on \(\mathbb{R}^d\). For \(V_g \coloneqq \dom{g}\), it is straightforward to verify that \(V_g\) is a linear subspace of \(\mathbb{R}^d\). Denote by \(V_g^\perp\) its orthogonal complement.

\begin{lemma}
\label{lem:seminorm_duality}
The following statements hold:
\begin{enumerate}[leftmargin=2em]
    \item The dual norm of \(g\), defined by \(g_*(x) = \sup_{y : g(y) \leq 1} y^\top x\), is a seminorm on \(\mathbb{R}^d\) with null space exactly \(V_g^\perp\).
    \item The dualization operation is an involution on \(\mathcal{G}\); that is, \(g_{**}(x) = g(x)\).
    \item There is a one-to-one correspondence between \(\mathcal{G}\) and \(\mathcal{R}\).
    \item The convex conjugate of \(g^2\) satisfies
    \[
    (g^2)^*(x) = \frac{1}{4} \sup_{g(y) \leq 1} (y^\top x)^2 = \frac{1}{4} g_*^2(x),
    \]
    and therefore \((g^2)^*\) is quadratically homogeneous with domain \(\mathbb{R}^d\).
    \item The convex conjugate of \(g\) is the convex indicator of the unit ball of the dual seminorm:
    \[
    g^*(x) = I_{\{g_*\le 1\}}(x).
    \]
\end{enumerate}
\end{lemma}

\begin{proof}
To prove $(1)$, by definition we have $g_*(x) = \sup_{y:g(y)\leq 1}y^\top x$. We first show that the null space $\mathcal{N}$ of $g_*$ is $V_g^\perp$. Choose $w\in V_g^\perp$, then clearly $g_*(w)=\sup_{y:g(y)\leq 1} \langle y,w\rangle = 0$. Conversely choose $w\notin V_g^\perp$, with $w\neq0$ then define $w^{\#} = P_{V_g} (w)$ the Euclidean projection of $w$ onto $V_g$, since $w\notin V_g^{\perp}$, $w^{\#}\neq 0$ and $g(w^\#)<\infty$. By Pythagoras Theorem $\langle w,w^{\#}\rangle = \Vert w^\#\Vert_2^2>0$, we have $g_*(w) = \sup_{y:g(y)\leq 1}\langle y,w\rangle \geq\langle w^\#/g(w^\#),w\rangle>0$. This shows that $\mathcal{N}=V_g^\perp$. The positive homogeneity and triangle inequality follows easily from definition of $g_*$. The finiteness of $g_*(x)$ follows from the compactness of the unit ball $B_g\coloneqq \{y\in \mathbb{R}^d:g(y)\leq1\}$.

Secondly we show that the involution $g_{**}(x) = \sup_{y:g_*(y)\leq 1} \langle y,x\rangle = g(x)$ holds. It follows that the unit ball of $g_*$ is \[
B_{g_*} = \{x:g_*(x)\leq 1\} = \{x:\sigma_{B_g}(x)\leq 1\} = B_g^\circ,
\]where $\sigma_{B_g}(x) = \sup_{y:y\in B_g}\langle y,x\rangle$ is the support function of the set $B_g$, and $B_g^\circ = \{x\in \mathbb{R}^d:\langle y,x\rangle \leq 1, \forall y\in B_g\}$ is the \textit{polar} of $B_g$. Because $B_g$ is a closed convex set of $\mathbb{R}^d$ containing the origin, by \citet[Corollary 15.1.2]{rockafellar1970convex}, we have \[
g_{**}(x) = \sigma_{B_{g_*}}(x) = \sigma_{B_g^\circ}(x) = \iota_{B_g}(x),
\]where $\iota_{B_g}(x) = \inf\{t\in (0,\infty]:x\in tB_g\}$ is the \textit{Minkowski functional} of the set $B_g$. It follows from the definition that \[
\iota_{B_g}(x) = \begin{cases}
    g(x) & \text{if } x \in V_g,\\
    \infty & \text{otherwise.}
\end{cases}
\]Completing the proof of $(2)$.

Claim $(3)$ follows from the similar line of argument in $(1)$ to show that $r_*\in\mathcal{G}$ for all $r\in\mathcal{R}$ with $\dom{r_*} = \mathcal{N}_r^\perp$, where $\mathcal{N}_r$ is the null space of $r$. Thus bijectivity follows from $(2)$.

For claim $(4)$, choose any $x\in \dom{g}\setminus\{0\}$, then there exists $t\in \mathbb{R}$ and some $y\in \mathbb{R}^d$ such that $x = ty$ with $g^2(y)=1$. To see this, we can always set $t = g(x)$ and let $y = t^{- 1} x$. Then by definition of $(g^2)^*(x^*)$ we may substitute $x=ty$ and get \begin{align*}
   (g^2)^*(x^*) &= \sup_{x\in\dom{g}} \left\{\langle x,x^*\rangle-g^2(x) \right\}\\
   & = \sup_{y:g^2(y)=1}\sup_{t\geq0}\left\{t\langle y,x^*\rangle-t^2  \right\}\\
   & = \dfrac{1}{4}\sup_{y:g(y)=1} \langle y,x^*\rangle^2 = \frac{1}{4}g_*^2(x^*),
\end{align*}
where the second last equality solves the global supremum over $t$ because the set $\{y:g^2(y)=1\}$ is centrally symmetric. It is straightforward to see that $(g^2)^*(x^*)$ is quadratically-homogeneous on $\dom{(g^2)^*}=\mathbb{R}^d$ where the latter follows from $g_*$ being finite on $\mathbb{R}^d$. 

For the final claim, note by definition of convex conjugate we have \[
    g^*(x^*) = \sup_{x\in V_g}\left\{\langle x,x^*\rangle - g(x)\right\},
    \]set $t = g(x)$ and $z = x/t\in B_g$, for which $x \neq 0$, then by a change of variable, we have \[
    g^*(x^*) = \sup_{t\geq 0}\sup_{z\in B_g} \Big\{t(\langle z,x^*\rangle - 1 )\Big\} = \sup_{t\geq 0}\Big\{t(g_*(x^*) - 1) \Big\} = \begin{cases}0 &\text{if } g_*(x^*)\leq 1,\\
        \infty &\text{otherwise.}
    \end{cases}
    \]
\end{proof}

\subsection{Proof of \texorpdfstring{\cref{prop:linear_regression}}{Proposition: LinearRegression}}
\begin{proof}
We use \cref{prop:duality} (Strong Duality) to reformulate the primal problem. From \cref{prop:duality}, we need to solve the maximization problem \[
\phi_{\gamma} (x,y;\beta) = \sup_{(u,v)}\left\{ \ell(u,v;\beta) - \gamma c\big((u,v),(x,y)\big)\right\}.
\]Because $\phi_{\gamma}$ appears in a minimization problem, we look for finite values of $\phi_\gamma$ and by definition of $c$, $\phi_{\gamma}$ is finite then necessarily $y=v$. Let $\epsilon(\beta) \coloneqq y - x^\top \beta$ and $\Delta \coloneqq u-x$, expanding the objective in $\phi_\gamma$ we get
\begin{align*}
    \ell\left( u,v;\beta \right) -\gamma c\big( (u,v) ,(x,y)\big) 
    &=\epsilon(\beta)^2 + \big\{(\beta^\top \Delta)^2 - 2\epsilon(\beta)\beta^\top\Delta - \gamma g^2(\Delta) \big\}\\
    &=\epsilon(\beta)^2 + \big\{f_\beta(\Delta) - \gamma g^2(\Delta) \big\},
\end{align*}
where we let $f_\beta(\Delta) \coloneqq (\beta^\top \Delta)^2 - 2\epsilon(\beta)\beta^\top\Delta$. Now the maximization becomes \[
\phi_{\gamma} (x,y;\beta) = \epsilon(\beta)^2 + \sup_{\Delta\in \mathbb{R}^d} \big\{f_\beta(\Delta) - \gamma g^2(\Delta)\big\}.
\]This is a difference of convex optimization, and we try to express $f_\beta(\Delta)$ as a supremum of infinitely many affine functions so the difference of convex function becomes an affine function plus a concave function. By \cref{lem:1} the (convex) biconjugate function of $f_\beta$ is \[f^{**}_\beta(\Delta) = \sup_{\alpha\in \mathbb{R}} \left\{\alpha (\beta^\top\Delta) - \frac{1}{4}{(\alpha+2\epsilon(\beta))^2}\right\}.\]By Fenchel-Moreau we have $f^{**}_\beta(\Delta) = f_\beta(\Delta)$. Then \begin{align*}
    \phi_\gamma(x,y;\beta)&= \epsilon(\beta)^2+\sup_{\Delta} \big\{f_\beta(\Delta) -\gamma g^2(\Delta) \big\}\\
    &=\epsilon(\beta)^2+\sup_{\Delta} \left\{ \sup_{\alpha\in \mathbb{R}} \left(\alpha (\beta^\top\Delta) - \frac{1}{4}{(\alpha+2\epsilon(\beta))^2}\right) - \gamma g^2(\Delta)\right\}\\
    &=\epsilon(\beta)^2+\sup_\alpha \left\{ \sup_\Delta \big(\Delta ^\top(\alpha \beta)-\gamma g^2(\Delta)\big) - \frac{1}{4}{(\alpha+2\epsilon(\beta))^2} \right\}\\
    (1)&=\epsilon(\beta)^2+\sup_\alpha \left\{ \frac{1}{\gamma}(g^2)^*(\alpha\beta) - \frac{1}{4}{(\alpha+2\epsilon(\beta))^2}\right\}\\
    (2)&=\epsilon(\beta)^2 + \frac{1}{4}\sup_{\alpha}\left\{\frac{1}{\gamma}\alpha^2 r^2(\beta) -(\alpha+2\epsilon(\beta))^2\right\},
\end{align*}
where in (1) we used the convex conjugate property $(\gamma g^2)^*(x^*) = \gamma (g^2)^*\left( \frac{x^*}{\gamma}\right) = \frac{1}{\gamma}(g^2)^*(x^*)$ because $(g^2)^*$ is quadratically-homogeneous and also $(g^2)^*(\alpha\beta)$ is finitely valued and in (2) we used the dual norm representation of $(g^2)^*$ in \cref{lem:seminorm_duality} with $r=g_*$. Therefore, \begin{align*}
    \phi_\gamma(x,y;\beta) 
    &=\dfrac{1}{4}\sup_{\alpha}\left\{\left(\dfrac{r^2(\beta)}{\gamma}- 1\right)\alpha^2 - 4\epsilon(\beta)\alpha\right\}\\
    &= \begin{cases}
        \dfrac{\epsilon(\beta)^2\gamma}{\gamma-r^2(\beta)} &\text{if }r^2(\beta) < \gamma,\\
        \infty&\text{otherwise.}
    \end{cases}
    \end{align*}
Then the dual objective can be simplified as \begin{align*}
  &\inf_{\beta \in \mathbb{R}^d} \min_{\gamma \geq 0} 
\left\{
\gamma \delta + \dfrac{1}{n} \sum_{i=1}^n \phi_\gamma(x_i, y_i; \beta)
\right\}\\
= &\inf_{\beta} \inf_{\gamma > r^2(\beta)} 
\left\{
\gamma \delta + \dfrac{1}{n} \sum_{i=1}^n \dfrac{\epsilon_i(\beta)^2 \gamma}{\gamma - r^2(\beta)}
\right\}\\
=&\inf_\beta \inf_{\gamma > r^2(\beta)} 
\left\{
\gamma \delta + \text{MSE}(\beta) \dfrac{\gamma}{\gamma - r^2(\beta)}
\right\}\\
=&\inf_\beta \left(\sqrt{\text{MSE}(\beta)} + \sqrt{\delta}r(\beta)\right)^2.
\end{align*}
This completes the proof.
\end{proof}

\subsection{Proof of \texorpdfstring{\cref{prop:binary_classification}}{Proposition: BinaryClassification}}
\begin{proof}
We first prove the logistic-loss case. By Proposition~\ref{prop:duality},
\[
\inf_{\gamma\ge 0}\left\{
\gamma\delta+\frac1n\sum_{i=1}^n
\sup_{u\in\mathbb R^d}
\left(\log\left(1+e^{-y_i\beta^\top u}\right)-\gamma g(x_i-u)\right)\right\}.
\]
For each $i$, let
\[
\phi_\gamma(x_i,y_i;\beta)
\coloneqq\sup_{u\in\mathbb R^d}
\left\{\log\left(1+e^{-y_i\beta^\top u}\right)-\gamma g(x_i-u)\right\}.
\]
Then a standard extension of \citet[Lemma 1]{abadeh2015drologistic} yields
\[
\phi_\gamma(x_i,y_i;\beta)=
\begin{cases}
\log\left(1+e^{-y_i x_i^\top\beta}\right),& g_*(\beta)\le \gamma,\\
+\infty,& \text{otherwise}.
\end{cases}
\]
Hence
\begin{align*}
\sup_{\mathbb P\in\mathcal B_\delta(\mathbb P_n;c)}
\mathbb E_{\mathbb P}\left[\log\left(1+e^{-YX^\top\beta}\right)\right]
&=
\inf_{\gamma\ge g_*(\beta)}
\left\{\gamma\delta+\frac1n\sum_{i=1}^n\log\left(1+e^{-y_i x_i^\top\beta}\right)\right\}\\
&=
\frac1n\sum_{i=1}^n
\log\left(1+e^{-y_i x_i^\top\beta}\right)+\delta g_*(\beta).
\end{align*}Now consider the hinge-loss case. Again by Proposition~\ref{prop:duality},
\[
\inf_{\gamma\ge 0}\left\{\gamma\delta+\frac1n\sum_{i=1}^n
\sup_{u\in\mathbb R^d}
\left((1-y_i\beta^\top u)^+-\gamma g(x_i-u)\right)\right\}.
\]
Let $\Delta\coloneqq u-x_i$. Then
\begin{align*}
&\sup_{u}
\left((1-y_i\beta^\top u)^+-\gamma g(x_i-u)\right)\\
&=
\sup_{\Delta}
\left((1-y_i\beta^\top(x_i+\Delta))^+-\gamma g(\Delta)\right)\\
&=
\sup_{\Delta}\sup_{0\le\alpha\le 1}
\left(\alpha(1-y_i\beta^\top(x_i+\Delta))-\gamma g(\Delta)\right)\\
&=
\sup_{0\le\alpha\le 1}\left\{\alpha(1-y_i x_i^\top\beta)+
\sup_{\Delta}\left(-\alpha y_i\beta^\top\Delta-\gamma g(\Delta)\right)\right\},
\end{align*}
where we used $t^+=\sup_{0\le\alpha\le 1}\alpha t$. For fixed $\alpha$,
\[
\sup_{\Delta}
\left(-\alpha y_i\beta^\top\Delta-\gamma g(\Delta)\right)=(\gamma g)^*(-\alpha y_i\beta)
=\begin{cases}
0,& \alpha g_*(\beta)\le \gamma,\\
\infty,& \text{otherwise}.
\end{cases}
\]
Therefore, if $\gamma<g_*(\beta)$, taking $\alpha=1$ yields $+\infty$. If $\gamma\ge g_*(\beta)$, then the inner supremum is zero for every $\alpha\in[0,1]$, and thus
\[
\sup_{0\le\alpha\le 1}\alpha(1-y_i x_i^\top\beta)
=
(1-y_i x_i^\top\beta)^+.
\]
Hence
\[
\sup_{u}
\left(
(1-y_i\beta^\top u)^+-\gamma g(x_i-u)
\right)
=
\begin{cases}
(1-y_i x_i^\top\beta)^+,& g_*(\beta)\le \gamma,\\
+\infty,& \text{otherwise}.
\end{cases}
\]
So
\begin{align*}
&\inf_{\beta}\inf_{\gamma\ge 0}
\left\{
\gamma\delta+\frac1n\sum_{i=1}^n
\sup_u\left((1-y_i\beta^\top u)^+-\gamma g(x_i-u)\right)
\right\}\\
&=
\inf_{\beta}\inf_{\gamma\ge g_*(\beta)}
\left\{
\gamma\delta+\frac1n\sum_{i=1}^n(1-y_i x_i^\top\beta)^+
\right\}\\
&=
\inf_{\beta}
\left\{
\frac1n\sum_{i=1}^n(1-y_i x_i^\top\beta)^+
+\delta g_*(\beta)
\right\}.
\end{align*}
This completes the proof.
\end{proof}

\subsection{Convex Conjugate of the READ Cost Function}
\label{sec:read_reformulation}

Fix $\Lambda \in [0,\infty]^m$, and define 
\[
\mathbf{1}_\infty \coloneqq \{ j \in [m]:\lambda_j = \infty\},
\]
then the read cost function is of the form
\[
c_{q,\Lambda}(x) = \|x\|_q^2 + \sum_{j \not\in \mathbf{1}_\infty}\lambda_j(\theta_j^\top x)^2 + \sum_{j \in \mathbf{1}_\infty}I_{\{0\}}(\theta_j^\top x),
\]where we recall $I_{\{0\}}$ is the convex indicator of the singleton $\{0\}$. It is evident that \(c_{q,\Lambda}\) is proper, lower semicontinuous, convex, and satisfies
\[
c_{q,\Lambda}(x)\ge \|x\|_q^2, \qquad x\in\mathbb{R}^d.
\]
For simplicity, we hide the convex indicator and implicitly assume that $\infty \cdot a = 0$ if and only if $a = 0$ for all $a \geq 0$ and $\infty$ otherwise.
\begin{lemma}
\label{lem:regularizer}
Fix \(\Theta \in \mathbb{R}^{d\times m}\) and \(\Lambda \in [0,\infty]^m\). Define $c_{q,\Lambda}(x) = \|x\|_q^2 + \|\Theta^\top x\|_{\Lambda}^2$, whose convex conjugate is denoted by \(c^*_{q,\Lambda}(x)\). Then for $z \in \mathbb{R}^d$
\[
c_{q,\Lambda}^*(z) =\frac{1}{4}\inf_{\kappa \in \mathbb{R}^m} 
\big\{ \|z - \Theta \kappa\|_p^2 + \|\kappa\|_{\Lambda^{-1}}^2 \big\},
\]
where \(p^{-1} + q^{-1} = 1\), with the convention that \(0^{-1} = \infty\) and \(\infty^{-1} = 0\) on $[0,\infty]$, and $0 \cdot \infty = 0$.
\end{lemma}

\begin{proof}
Write $\Lambda=\diag(\lambda_1,\dots,\lambda_m)$ and interpret
\[
\|u\|_{\Lambda}^2=\sum_{j=1}^m \lambda_j u_j^2,
\]
with the conventions
\[
0\cdot a^2=0,
\quad
\infty\cdot a^2=
\begin{cases}
0,& a=0,\\
\infty,& a\neq 0.
\end{cases}
\]
Likewise,
\[
\|v\|_{\Lambda^{-1}}^2=\sum_{j=1}^m \lambda_j^{-1} v_j^2,
\quad
0^{-1}=\infty,\ \ \infty^{-1}=0.
\]Define
\[
F(z)\coloneqq \frac14 \inf_{\kappa\in\mathbb{R}^m}
\Big\{\|z-\Theta\kappa\|_p^2+\|\kappa\|_{\Lambda^{-1}}^2\Big\},
\quad z\in\mathbb{R}^d.
\]
Since $(z,\kappa)\mapsto \|z-\Theta\kappa\|_p^2+\|\kappa\|_{\Lambda^{-1}}^2$ is jointly convex,
$F$ is convex as a partial infimum of a jointly convex function. Moreover,
\[
F(z)\le \frac14\|z\|_p^2<\infty
\]
by choosing $\kappa=0$, so $F$ is finite everywhere on $\mathbb{R}^d$. Hence $F$ is continuous, and therefore closed and proper. We now compute its convex conjugate. For any $x\in\mathbb{R}^d$,
\begin{align*}
F^*(x)
&=
\sup_{z\in\mathbb{R}^d}
\left\{
\langle x,z\rangle
-\frac14\inf_{\kappa\in\mathbb{R}^m}
\big(\|z-\Theta\kappa\|_p^2+\|\kappa\|_{\Lambda^{-1}}^2\big)
\right\} \\
&=
\sup_{z\in\mathbb{R}^d,\ \kappa\in\mathbb{R}^m}
\left\{
\langle x,z\rangle
-\frac14\|z-\Theta\kappa\|_p^2
-\frac14\|\kappa\|_{\Lambda^{-1}}^2
\right\}.
\end{align*}
Set $y=z-\Theta\kappa$, so that $z=y+\Theta\kappa$. Then
\begin{align*}
F^*(x)
&=
\sup_{y\in\mathbb{R}^d,\ \kappa\in\mathbb{R}^m}
\left\{
\langle x,y+\Theta\kappa\rangle
-\frac14\|y\|_p^2
-\frac14\|\kappa\|_{\Lambda^{-1}}^2
\right\} \\
&=
\sup_{y\in\mathbb{R}^d}
\left\{
\langle x,y\rangle-\frac14\|y\|_p^2
\right\}
+
\sup_{\kappa\in\mathbb{R}^m}
\left\{
\langle \Theta^\top x,\kappa\rangle
-\frac14\|\kappa\|_{\Lambda^{-1}}^2
\right\}.
\end{align*}
The first term is the standard conjugacy relation for dual norms:
\[
\sup_{y\in\mathbb{R}^d}
\left\{
\langle x,y\rangle-\frac14\|y\|_p^2
\right\}
=\|x\|_q^2.
\]
For the second term, writing $u=\Theta^\top x\in\mathbb{R}^m$, separability gives
\[
\sup_{\kappa\in\mathbb{R}^m}
\left\{
\langle u,\kappa\rangle-\frac14\|\kappa\|_{\Lambda^{-1}}^2
\right\}
=
\sum_{j=1}^m
\sup_{\kappa_j\in\mathbb{R}}
\left\{
u_j\kappa_j-\frac14\lambda_j^{-1}\kappa_j^2
\right\}.
\]
For each coordinate:
\[
\sup_{\kappa_j\in\mathbb{R}}
\left\{
u_j\kappa_j-\frac14\lambda_j^{-1}\kappa_j^2
\right\}
=
\begin{cases}
\lambda_j u_j^2, & 0<\lambda_j<\infty,\\[1mm]
0, & \lambda_j=0,\\[1mm]
0 \text{ if } u_j=0,\ \infty \text{ otherwise}, & \lambda_j=\infty.
\end{cases}
\]
This is exactly $\lambda_j u_j^2$ under the above conventions, so
\[
\sup_{\kappa\in\mathbb{R}^m}
\left\{
\langle u,\kappa\rangle-\frac14\|\kappa\|_{\Lambda^{-1}}^2
\right\}
=
\|u\|_{\Lambda}^2
=
\|\Theta^\top x\|_{\Lambda}^2.
\]
Therefore,
\[
F^*(x)=\|x\|_q^2+\|\Theta^\top x\|_{\Lambda}^2=c_{q,\Lambda}(x).
\]
Since $F$ is closed proper convex, Fenchel--Moreau gives $F=c_{q,\Lambda}^*$. That is,
\[
c_{q,\Lambda}^*(x)
=
\frac14\inf_{\kappa\in\mathbb{R}^m}
\Big\{
\|x-\Theta\kappa\|_p^2+\|\kappa\|_{\Lambda^{-1}}^2
\Big\}.
\]
\end{proof}
The following two matrices are equivalent when $\Lambda$ is invertible, 
\[\mathbb I_d - \Theta(\Theta^\top\Theta + \Lambda^{-1})^{-1}\Theta^\top = (\mathbb{I}_d + \Theta\Lambda\Theta^\top)^{-1},\] and when $\Theta$ is orthonormal, it further can be written as $P_{S^\perp} + \Theta(\mathbb{I}_m + \Lambda)^{-1}\Theta^\top$, where $S = \col(\Theta)$ and $P_S$ denotes the projection onto $S$. The convex objective function\[
\inf_{\beta\in\mathbb{R}^d}  \left\{\sqrt{{\rm MSE}_n(\beta)} + \sqrt{\delta} \min_{\kappa\in \mathbb{R}^m} \left\{\sqrt{\| \beta - \Theta\kappa\|_p^2 + \|\kappa\|_{\Lambda^{-1}}^2}\right\} \right\},
\]can be reformulated into the following \emph{convex conic programming} for efficient optimization,
\begin{align*}
    \inf_{\beta,\kappa,t,s_1,s_2} &\left\{ \frac{1}{\sqrt{n}}\Vert \mathbf{y}-\mathbf{X}\beta\Vert_2 + \sqrt{\delta}t\right\}\\
    \text{s.t }& s_1\geq \Vert \beta - \Theta\kappa\Vert_p,\quad s_2\geq \Vert \Lambda^{-1/2} \kappa\Vert_2,\\
    & t\geq \Vert[s_1,s_2]^\top\Vert_2,\\
    &s_1,s_2,t\geq0.
\end{align*}

Let $\Omega \succeq 0$ be PSD, and $\Omega = Q \Lambda Q^\top$, where $Q^\top Q=\mathbb{I}_m$, and $\Lambda$ is diagonal with nonnegative entries. Then the READ cost function: \[
c_{q,\Omega}(x) = \|x\|_q^2 + \|\Theta^\top x\|_\Omega^2 = \|x\|_{q}^2 + \|(\tilde \Theta)^\top x\|_\Lambda^2,
\]where $\tilde \Theta = \Theta Q$.
\subsection{Proof of \texorpdfstring{\cref{prop:linearmodel}}{Proposition: LinearModels}}
\begin{proof}[Proof of \cref{prop:linearmodel}]
To avoid confusion with the matrix \(\Sigma=-\mathbb E_{\mathbb P_*}[\nabla_\beta h(X,Y;\beta_*)]\) used in \Cref{sec:rwpi}, write $\Sigma_X \coloneqq \mathbb E_{\mathbb P_*}[XX^\top]\succ 0$.
Suppose
\[
Y=X^\top\beta_*+\varepsilon,
\quad
\mathbb E[\varepsilon\mid X]=0,
\quad
\mathbb E[\varepsilon^2\mid X]=1.
\]
Under the definition \(h(x,y;\beta)=(y-x^\top\beta)x\), we have
\[
h(X,Y;\beta_*)=\varepsilon X,
\quad
\cov_{\mathbb P_*}(h(X,Y;\beta_*))=\Sigma_X,
\quad
\Xi^\top=\varepsilon \mathbb I_d-X\beta_*^\top.
\]For \(p=q=2\), \(\varphi_{2,\Lambda}^2(z)=\|z\|_{\Psi_\Lambda}^2\), where $\Psi_\Lambda=(\mathbb I_d+\Theta\Lambda\Theta^\top)^{-1}$ in the extended sense described after \Cref{lem:regularizer}. Therefore
\[
\psi_{2,\Lambda}(\xi)
=
\frac14\mathbb E_{\mathbb P_*}\|\Xi\xi\|_{\Psi_\Lambda}^2
=
\frac14\xi^\top \Gamma \xi,
\]
where
\[
\Gamma
\coloneqq
\Psi_\Lambda+\|\beta_*\|_{\Psi_\Lambda}^2\Sigma_X.
\]
Consequently, whenever \(\Gamma\) is positive definite, $\psi_{2,\Lambda}^*(H)=H^\top \Gamma^{-1}H$. Let
\[
\bar{\Gamma}
\coloneqq
\|\beta_*\|_{\Psi_\Lambda}^2 \mathbb I_d
+
\Sigma_X^{-1/2}\Psi_\Lambda\Sigma_X^{-1/2},
\]
so that \(\Gamma=\Sigma_X^{1/2}\bar\Gamma\Sigma_X^{1/2}\). Since $H\sim N(0,\Sigma_X)$, we may write \(\Sigma_X^{-1/2}H= Z\) with \(Z\sim  N(0, \mathbb I_d)\), and hence
\[
\label{equ:psi_distribution}
\tag{Psi}
\psi_{2,\Lambda}^*(H)
=
 Z^\top\bar\Gamma^{-1}Z
=
\sum_{i=1}^d \frac{1}{\gamma_i}W_i,
\]
where \(\gamma_1,\ldots,\gamma_d\) are the eigenvalues of \(\bar \Gamma\), and \(W_1,\ldots,W_d\) are independent \(\chi_1^2\) random variables. Thus \(\psi_{2,\Lambda}^*(H)\) has a generalized chi-square distribution. Again, we see that we can raise $\psi^*_{2,\Lambda}(H)$ in the stochastic order by lowering the eigenvalues of $\bar \Gamma$, i.e. raising the entries of $\Lambda$.

Let \(\eta_\alpha(\Lambda)\) denote the \((1-\alpha)\)-quantile of \(\psi_{2,\Lambda}^*(H)\). Since \(\Sigma=\Sigma_X\) under the present convention for \(h\), and \(\psi_{2,\Lambda}^*\) is quadratically homogeneous, an asymptotically valid \((1-\alpha)\)-confidence region for \(\beta_*\) is
\[
\hat \beta_{\erm} +
\left\{ u\in\mathbb R^d: \|\Sigma_Xu\|_{\Gamma^{-1}}^2
\le \frac{\eta_\alpha(\Lambda)}{n} \right\}.
\]

We now verify the positive-definiteness condition in \Cref{asm:pd}. For any nonzero \(\xi\in\mathbb R^d\),
\[
\mathbb E_{\mathbb P_*}\!\left[\varphi_{2,\Lambda}^2(\Xi\xi)\right]
=
4\psi_{2,\Lambda}(\xi)
=
\xi^\top\Gamma\xi.
\]
Since \(\varphi_{2,\Lambda}^2(\Xi\xi)\ge0\), we have
\[
\mathbb P_*\left(\varphi_{2,\Lambda}^2(\Xi\xi)>0\right)>0
\iff
\mathbb E_{\mathbb P_*}\left[\varphi_{2,\Lambda}^2(\Xi\xi)\right]>0
\iff
\xi^\top\Gamma\xi>0.
\]
Thus \Cref{asm:pd} holds for every nonzero direction if and only if \(\Gamma\succ0\). If all entries of \(\Lambda\) are finite, then \(\Psi_\Lambda\succ0\), so \(\Gamma\succ0\) automatically. Now suppose that some entries of \(\Lambda\) are equal to \(\infty\), and let \(\Theta_\infty\) denote the submatrix of \(\Theta\) formed by the corresponding columns. By construction,
\[
\ker(\Psi_\Lambda)=\col(\Theta_\infty),
\]
so
\[
\|\beta_*\|_{\Psi_\Lambda}^2=0
\iff
\beta_*\in\col(\Theta_\infty).
\]
If \(\beta_*\notin\col(\Theta_\infty)\), then \(\|\beta_*\|_{\Psi_\Lambda}^2>0\). Since \(\Sigma_X\succ0\), for every nonzero \(\xi\),
\[
\xi^\top\Gamma\xi
=
\xi^\top\Psi_\Lambda\xi
+
\|\beta_*\|_{\Psi_\Lambda}^2\xi^\top\Sigma_X\xi
>0.
\]
Hence \(\Gamma\succ0\), and \Cref{asm:pd} holds. Conversely, if \(\beta_*\in\col(\Theta_\infty)\), then \(\|\beta_*\|_{\Psi_\Lambda}^2=0\), so
\[
\Gamma=\Psi_\Lambda.
\]
Because \(\Psi_\Lambda\) is singular whenever at least one entry of \(\Lambda\) is equal to \(\infty\), there exists a nonzero \(\xi\in\ker(\Psi_\Lambda)\), and then
\[
\xi^\top\Gamma\xi=0.
\]
Therefore \(\Gamma\) is not positive definite, and \Cref{asm:pd} fails. This proves the stated equivalence when \(\Lambda\) has at least one infinite entry.
\end{proof}

\subsection{Proof of \texorpdfstring{\cref{thm:rwpi}}{Theorem: RWPI}}

\begin{proof}[Proof of (1)]
    Fix \(\Lambda \in[0,\infty]^m\). For \(R>0\), let
\[
\mathcal{K}_R\coloneqq \{u\in\mathbb{R}^d:\|u\|_2\le R\},
\qquad
G_{n}(u)\coloneqq nR_n(\beta_*+n^{-1/2}u),
\qquad
\Gamma(z,u)\coloneqq \psi_{p,\Lambda}^*(z-\Sigma u).
\]
Let \(\varepsilon>0\). Since \(\mathbb{E}_{\mathbb{P}_*}\|h(X,Y;\beta_*)\|_2^2<\infty\) by assumption, we have
\[
\sup_{n\ge 1}\mathbb{E}_{\mathbb{P}_*}\|H_n\|_2^2<\infty.
\]
Hence, by Markov's inequality, there exists \(M<\infty\) such that
\[
\sup_{n\ge 1}\mathbb{P}_*\bigl(\|H_n\|_2>M\bigr)\le \varepsilon.
\]
By \cref{lem:coercivity_psi}, for this \(M\) and the present \(R\), there exists \(b=b(M,R)\) such that
\[
\Gamma(z,u)=\Phi_{b}(z,u)
\qquad
\text{for all }\|z\|_2\le M,\ u\in\mathcal{K}_R,
\]with $\Phi_b$ defined in \cref{lem:local_dual}. Therefore,
\[
\mathbb{P}_*\left(
\sup_{u\in\mathcal{K}_R}|G_{n}(u)-\Gamma(H_n,u)|>\varepsilon
\right)
\le
\mathbb{P}_*\left(
\sup_{u\in\mathcal{K}_R}|G_{n}(u)-\Phi_{b}(H_n,u)|>\varepsilon
\right)
+
\mathbb{P}_*(\|H_n\|_2>M).
\]
The second term is at most \(\varepsilon\) by construction of \(M\), and the first term tends to \(0\) by \cref{lem:local_dual}. Hence
\begin{equation}
\label{eq:slutsky}
\sup_{u\in\mathcal{K}_R}|G_{n}(u)-\Gamma(H_n,u)|\to 0
\qquad\text{in probability.}
\end{equation}
The map $z\mapsto \Gamma(z,\cdot)\in C(\mathcal{K}_R)$ is continuous, because \(\psi_{p,\Lambda}^*\) is continuous and \((z,u)\mapsto z-\Sigma u\) is continuous on \(\mathbb{R}^d\times \mathcal{K}_R\). Since \(H_n\leadsto H\), the continuous mapping theorem yields
\[
\Gamma(H_n,\cdot)\leadsto \Gamma(H,\cdot)=\psi_{p,\Lambda}^*(H-\Sigma\cdot)
\]
in \(C(\mathcal{K}_R)\). Combining this with \eqref{eq:slutsky} and Slutsky's lemma gives
\[
G_{n}\leadsto \psi_{p,\Lambda}^*(H-\Sigma\cdot)
\]
in \(C(\mathcal{K}_R)\). Since every compact subset of \(\mathbb{R}^d\) is contained in some \(\mathcal{K}_R\), the convergence holds in \(C(\mathbb{R}^d,\mathbb{R})\) endowed with the topology of uniform convergence on compact sets. This proves the theorem.
\end{proof}

\begin{proof}[Proof of (2)]
Define $\delta_n = \eta/n$, where $\eta > 0$ is fixed. From the equivalence \eqref{eq:set_equiv}, we have 
\[
\Omega_n^+(\delta_n) = \{\beta: R_n(\beta) \leq \eta/n\}.
\]
Now define the bijective mapping
\[
T_n(u) = \beta_* + n^{-1/2}u, \qquad T_n^{-1}(\beta) = n^{1/2}(\beta - \beta_*).
\]
Then,
\[
G_n(u) \leq \eta \iff R_n(\beta_*+n^{-1/2}u) \leq \eta /n \iff R_n(\beta) \leq \eta / n,
\]where in the last equivalence we write $\beta = T_n(u)$. Therefore, under the bijection of $T_n$, we have 
\[
\{\beta: R_n(\beta) \leq \eta/n\} = T_n(\{u:G_n(u) \leq \eta\}) = \beta_* + n^{-1/2}\{u:G_n(u)\leq \eta\}.
\]Let $\operatorname{lev}{(f,k)} = \{x : f(x) \leq k\}$ denote the level set of a function $f:\mathbb{R}^d \to \mathbb{R}$, then by \citet[Proposition 14]{blanchet2021statistical}, we have the set convergence 
\[
\{u:G_n(u)\leq \eta\} \to \operatorname{lev}{(G,\eta)},
\]
where $G(u) = \psi_{p,\Lambda}^*(H-\Sigma u)$. Define the reparameterization $u = \Sigma^{-1}H + v$, then
\begin{align*}
    \operatorname{lev}{(G,\eta)}
    &= \{u:\psi_{p,\Lambda}^*(H-\Sigma u) \leq \eta\} \\
    &= \Sigma^{-1}H + \{v: \psi_{p,\Lambda}^*(-\Sigma v)\leq \eta\}\\
    &= \Sigma^{-1}H + \{v:\psi_{p,\Lambda}^*(\Sigma v) \leq \eta\},
\end{align*}
where in the last equality we used that $\psi^*_{p,\Lambda}$ is even. This complete the proof.
\end{proof}

\subsection{Proof of \texorpdfstring{\cref{thm:asymptotic_dro}}{Theorem: Asymptotics}}

Define
\[
f_\eta(x)=x-\eta^{1/2}\nabla_\beta \mathcal{V}(\beta_*), \qquad \mathcal{V}(\beta_*;\Lambda)=\left(\mathbb{E}_{\mathbb{P}_*}\big[\varphi_{p,\Lambda}^2(\nabla_x \ell(X,Y;\beta_*))\big]\right)^{1/2},
\]as well as the worst-case objective,
\[
\Psi_n(\beta;\delta)\coloneqq \sup_{\mathbb{P}\in \mathcal{B}_\delta(\mathbb{P}_n;c_{q,\Lambda})}\mathbb{E}_{\mathbb{P}}[\ell(X,Y;\beta)].
\]

\begin{proposition}
\label{prop:V^DRO}
Suppose the assumptions of \cref{thm:asymptotic_dro} hold. Let \(\delta_n=\eta/n\), where \(\eta>0\) is fixed, and define
\[
V_n^{\dro}(u)\coloneqq n\bigl(\Psi_n(\beta_*+n^{-1/2}u;\delta_n)-\Psi_n(\beta_*;\delta_n)\bigr).
\]
Then, for any \(\varepsilon,\varepsilon',K>0\), there exists \(n_0>0\) such that
\[
\mathbb{P}\left(\big|V_n^{\dro}(u)-f_\eta(H_n)^\top u+\frac{1}{2}u^\top\Sigma u\big|\le \varepsilon'\right)\ge 1-\varepsilon
\]
for every \(n\ge n_0\) and every \(\|u\|_2\le K\), where
\[
H_n=\sqrt{n}\mathbb{E}_{\mathbb{P}_n}[h(X,Y;\beta_*)],
\qquad
\Sigma=-\mathbb{E}_{\mathbb{P}_*}[\nabla_\beta h(X,Y;\beta_*)].
\]
\end{proposition}
\begin{proof}
Fix \(K>0\), and write
\[
\beta_{n,u}\coloneqq \beta_*+n^{-1/2}u, \qquad \|u\|_2\le K.
\]
By \cref{lem:worst_case_expansion}, applied at \(\beta=\beta_{n,u}\) and \(\beta=\beta_*\), we have
\[
V_n^{\dro}(u)
=
n\Bigl( \mathbb E_{\mathbb P_n}[\ell(X,Y;\beta_{n,u})]
- \mathbb E_{\mathbb P_n}[\ell(X,Y;\beta_*)] \Bigr)
+
\eta^{1/2}\sqrt n
\Bigl( \mathcal V_n(\beta_{n,u})-\mathcal V_n(\beta_*) \Bigr) + o_{\mathbb P}(1),
\]
uniformly over \(\|u\|_2\le K\).

The first term is the empirical-loss increment that converges to $H_n^\top u + 2^{-1}u^\top \Sigma u$. The second term is handled by \cref{lem:bias_linearization}. Since the remainder terms are uniform on \(\|u\|_2\le K\), this proves the proposition.
\end{proof}

\begin{proof}[Proof of \cref{thm:asymptotic_dro}]
The proof follows the same general argument as in \citet{blanchet2022confidence}. By the central limit theorem, \(H_n \leadsto H\), where \(H_n = \sqrt{n}\mathbb{E}_{\mathbb{P}_n}[h(X,Y;\beta_*)]\), \(H \sim N(0,\cov_{\mathbb{P}_*}(h(X,Y;\beta_*)))\). Then \cref{prop:V^DRO} and the continuous mapping theorem imply that \(V_n^\dro(\cdot)\Rightarrow V(-f_\eta(H),\cdot)\) in \(C(\mathbb{R}^d,\mathbb{R})\) equipped with the topology of uniform convergence on compact sets, where \[V(x,u)=x^\top u+\frac{1}{2}u^\top\Sigma u.\]

As shown in \citet[Proposition 13]{blanchet2022confidence}, \(V_n^\dro\) is convex, and \(V_n^\dro(u)\leq 0\) implies that the family \(\{\argmin_u V_n^\dro(u)\}_n\) is contained in a compact set with arbitrarily high probability. In particular, the sequence of minimizers is tight. It therefore follows that
\[
\sqrt{n}(\hat\beta_\readd-\beta_*)=\argmin_u V_n^\dro(u)\leadsto \argmin_u V^\dro(u)=\Sigma^{-1}f_\eta(H).
\]
This completes the proof.
\end{proof}

\subsection{Proof of \texorpdfstring{\cref{thm:robustness}}{Theorem: Robustness}}
\label{sec:proof_of_robustness}

Throughout this section, suppose that
\(\Theta^\top\Theta=\mathbb I_m\) and that the alignment parameter has the
common-scalar form
\[
\Lambda=\lambda\mathbb I_m,\qquad \lambda\in[0,\infty].
\]
Let
\[
P\coloneqq \Theta\Theta^\top,
\qquad
P_\perp\coloneqq \Theta_\perp\Theta_\perp^\top,
\qquad
r\coloneqq d-m,
\qquad
K\coloneqq \|\kappa\|_2^2.
\]
To make the local asymptotic regime explicit, write
\[
\beta_{e}
=
\Theta\kappa+n^{-1/2}
\bigl(\Theta\mu_e+\Theta_\perp\gamma_e\bigr),
\qquad e\in\{*\}\cup\mathcal E.
\]
We assume that the hyperpopulation random effects are mutually independent
across populations and are independent of the target sample conditional on
\(\beta_{*}\). In particular,
\[
\mu_e,\mu_*\stackrel{\mathrm{iid}}{\sim}
N(0,\tau^2\mathbb I_m),
\qquad
\gamma_e,\gamma_*\stackrel{\mathrm{iid}}{\sim}
N(0,\sigma^2\mathbb I_r).
\]

Define that
$
G_n\coloneqq \sqrt n\bigl(\hat\beta_{\erm}-\beta_{*}\bigr).
$
Under the linear model and \(\mathbb E(XX^\top)=\mathbb I_d\), the usual
conditional central limit theorem gives the joint convergence
\[
G_n\rightsquigarrow G,
\qquad
G\sim N(0,\mathbb I_d),
\qquad
G\ \perp\ (\mu_e,\mu_*,\gamma_e,\gamma_*).
\]
Consequently,
\begin{equation}
\label{eq:future_local_difference}
\sqrt n\bigl(\beta_{e}-\hat\beta_{\erm}\bigr)
=
\Theta(\mu_e-\mu_*)
+
\Theta_\perp(\gamma_e-\gamma_*)
-G_n
\rightsquigarrow D,
\end{equation}
where
$
D
=
\Theta(\mu_e-\mu_*-G_\parallel)
+
\Theta_\perp(\gamma_e-\gamma_*-G_\perp),
$
with
$
G_\parallel\sim N(0,\mathbb I_m)
$
and
$
G_\perp\sim N(0,\mathbb I_r)
$
independent of one another and of the random effects.

We first derive the limiting coverage for a fixed finite \(\lambda\). Since
\[
\Psi_\lambda
=
(\mathbb I_d+\lambda\Theta\Theta^\top)^{-1}
=
\frac{1}{1+\lambda}P+P_\perp,
\]
we have
\begin{align*}
 c_{n,\lambda}
 &\coloneqq
 \|\beta_{*}\|_{\Psi_\lambda}^2 \\
 &=
 \frac{1}{1+\lambda}
 \left\|\kappa+n^{-1/2}\mu_*\right\|_2^2
 +n^{-1}\|\gamma_*\|_2^2
 \rightarrow
 c_\lambda
 \coloneqq
 \frac{K}{1+\lambda}
\end{align*}
almost surely. Therefore,
\[
\Gamma_{n,\lambda}
\coloneqq
\Psi_\lambda+c_{n,\lambda}\mathbb I_d
\rightarrow
\Gamma_\lambda
=
 d_\lambda P+e_\lambda P_\perp,
\]
where
\[
 d_\lambda
 =\frac{1+K}{1+\lambda},
 \qquad
 e_\lambda
 =1+\frac{K}{1+\lambda}.
\]
Define
\begin{equation}
\label{eq:a_lambda_scalar}
 a_\lambda
 \coloneqq
 \frac{e_\lambda}{d_\lambda}
 =
 1+\frac{1}{1+K}\lambda.
\end{equation}
Thus \(a_\lambda\) is strictly increasing in \(\lambda\).

Let \(X\sim\chi_m^2\) and \(Y\sim\chi_r^2\) be independent. The Gaussian
quadratic form used to calibrate the RWPI radius satisfies
\[
H^\top\Gamma_\lambda^{-1}H
\overset{d}{=}
\frac{1}{d_\lambda}X
+
\frac{1}{e_\lambda}Y,
\qquad
H\sim N(0,\mathbb I_d).
\]
The continuity of the generalized chi-squared distribution implies
$\eta_{n,\alpha}(\lambda)\to\eta_\alpha(\lambda)$, and therefore
\begin{equation}
\label{eq:eta_scalar_lambda}
 e_\lambda\eta_\alpha(\lambda)
 =
q_{1-\alpha}\bigl(a_\lambda X+Y\bigr).
\end{equation}
Moreover, the two orthogonal components of \(D\) in
\eqref{eq:future_local_difference} are independent and satisfy
\[
\frac{\|PD\|_2^2}{1+2\tau^2}
\sim\chi_m^2,
\qquad
\frac{\|P_\perp D\|_2^2}{1+2\sigma^2}
\sim\chi_r^2.
\]
Combining this with \eqref{eq:eta_scalar_lambda}, for every finite
\(\lambda\),
\begin{equation}
\label{eq:coverage_scalar_lambda}
 p_\alpha(\lambda)
 =
 \mathbb P\left\{
 c_\parallel a_\lambda X+c_\perp Y
 \le
 q_{1-\alpha}(a_\lambda X+Y)
 \right\},
\end{equation}
where
\[
 c_\parallel
 \coloneqq
 1+2\tau^2,
 \qquad
 c_\perp
 \coloneqq
 1+{2\sigma^2}>1.
\]

\begin{proof}[Proof of Part \textnormal{(1)}]
Suppose \(\tau^2=0\), so that \(c_\parallel=1\). We first establish that
\[
p(a)
\coloneqq
\mathbb P\left\{
 aX+c_\perp Y\le q_{1-\alpha}(aX+Y)
\right\}
\]
is strictly increasing in \(a\in[1,\infty)\).

Let \(\mathbb P_0\) denote the joint law of \((X,Y)\), and let
\(\mathbb P_c\) denote the joint law of \((X,c_\perp Y)\). The likelihood
ratio of \(\mathbb P_c\) with respect to \(\mathbb P_0\) is
\begin{equation}
\label{eq:lr_orthogonal_inflation}
L(y)
=
 c_\perp^{-r/2}
 \exp\left\{\frac{1-c_\perp^{-1}}{2}y\right\},
\end{equation}
which is strictly increasing in \(y\). For \(a\ge1\), set
\[
q(a)\coloneqq q_{1-\alpha}(aX+Y),
\qquad
A_a\coloneqq\{(x,y)\in\mathbb R_+^2:ax+y\le q(a)\}.
\]
Then
\[
\mathbb P_0(A_a)=1-\alpha,
\qquad
p(a)=\mathbb P_c(A_a).
\]

Take \(a_2>a_1\). Since \(a_2X+Y\) strictly stochastically dominates
\(a_1X+Y\), its continuous quantile satisfies \(q(a_2)>q(a_1)\). The two
boundary lines intersect at
\[
x_0
=\frac{q(a_2)-q(a_1)}{a_2-a_1},
\qquad
y_0=q(a_1)-a_1x_0=q(a_2)-a_2x_0.
\]
Because \(\mathbb P_0(A_{a_1})=\mathbb P_0(A_{a_2})\) and the two sets are
distinct, neither set contains the other; hence \(x_0>0\) and \(y_0>0\).
Except for boundaries of probability zero,
\[
A_{a_2}\setminus A_{a_1}\subset\{y>y_0\},
\qquad
A_{a_1}\setminus A_{a_2}\subset\{y<y_0\}.
\]
Also,
\[
\mathbb P_0(A_{a_2}\setminus A_{a_1})
=
\mathbb P_0(A_{a_1}\setminus A_{a_2})>0.
\]
Using the strict monotonicity of \(L\) in \eqref{eq:lr_orthogonal_inflation},
we obtain
\begin{align*}
\mathbb P_c(A_{a_2}\setminus A_{a_1})
&>
L(y_0)\mathbb P_0(A_{a_2}\setminus A_{a_1}) \\
&=
L(y_0)\mathbb P_0(A_{a_1}\setminus A_{a_2}) \\
&>
\mathbb P_c(A_{a_1}\setminus A_{a_2}).
\end{align*}
Therefore \(p(a_2)>p(a_1)\). Since \(a_\lambda\) in
\eqref{eq:a_lambda_scalar} is strictly increasing, \eqref{eq:coverage_scalar_lambda}
shows that \(p_\alpha(\lambda)\) is strictly increasing for finite
\(\lambda\).

It remains to treat \(\lambda=\infty\), which cannot be obtained by simply
substituting \(\lambda=\infty\) into the finite-\(\lambda\) expression for
\(\Gamma_\lambda\). At infinite alignment,
\[
\Psi_\infty=P_\perp,
\qquad
c_{n,\infty}
=\|\beta_{*}\|_{P_\perp}^2
=n^{-1}\|\gamma_*\|_2^2,
\]
and hence
\[
\Gamma_{n,\infty}
=
 n^{-1}\|\gamma_*\|_2^2P
+
\left(1+n^{-1}\|\gamma_*\|_2^2\right)P_\perp.
\]
Let \(q_m\coloneqq q_{1-\alpha}(\chi_m^2)\). Conditional on
\(\gamma_*\), whose norm is nonzero almost surely,
\[
\frac{\eta_{n,\alpha}(\infty)}{n}
\rightarrow
\frac{q_m}{\|\gamma_*\|_2^2}.
\]
Dividing the confidence-region inequality by \(n\) therefore shows that its
limiting constraint is
\[
\|PD\|_2^2\le q_m;
\]
the component in \(\col(\Theta_\perp)\) becomes asymptotically unrestricted.
When \(\tau^2=0\), \(PD=-\Theta G_\parallel\), and consequently
\[
p_\alpha(\infty)
=
\mathbb P(\chi_m^2\le q_m)
=1-\alpha.
\]
For every finite \(\lambda\), \(c_\perp>1\) and \(Y>0\) almost surely, so
\[
p_\alpha(\lambda)
<
\mathbb P\left\{
 a_\lambda X+Y\le q_{1-\alpha}(a_\lambda X+Y)
\right\}
=1-\alpha.
\]
This proves strict monotonicity on \([0,\infty]\), including the endpoint.
\end{proof}

\begin{proof}[Proof of Part \textnormal{(2)}]
For \(\lambda=0\), \(a_0=1\), so \eqref{eq:coverage_scalar_lambda} gives
\begin{equation}
\label{eq:p_zero_local}
p_\alpha(0)
=
\mathbb P\left\{
 c_\parallel X+c_\perp Y\le q_d
\right\},
\qquad
q_d\coloneqq q_{1-\alpha}(\chi_d^2).
\end{equation}
The infinite-alignment calculation above remains valid when \(\tau^2>0\)
and yields
\begin{equation}
\label{eq:p_infinity_local}
p_\alpha(\infty)
=
F_{\chi_m^2}\left(\frac{q_m}{c_\parallel}\right),
\qquad
q_m\coloneqq q_{1-\alpha}(\chi_m^2).
\end{equation}

Define
\[
\pi_0
\coloneqq
\mathbb P\left\{X+c_\perp Y\le q_d\right\}.
\]
Because \(c_\perp>1\) and \(Y>0\) almost surely,
\[
0<\pi_0<
\mathbb P(X+Y\le q_d)
=1-\alpha.
\]
Furthermore, \(c_\parallel\ge1\), and therefore \eqref{eq:p_zero_local}
implies
\[
p_\alpha(0)\le \pi_0.
\]
Let \(q_{\pi_0}(\chi_m^2)\) denote the \(\pi_0\)-quantile of
\(\chi_m^2\), and define
\begin{equation}
\label{eq:C_local_scalar}
C
\coloneqq
\frac{1}{2}
\left\{
\frac{q_{1-\alpha}(\chi_m^2)}{q_{\pi_0}(\chi_m^2)}-1
\right\}>0.
\end{equation}
If \(\tau^2<C\), then
\[
c_\parallel
=1+{2\tau^2}
<
\frac{q_{1-\alpha}(\chi_m^2)}{q_{\pi_0}(\chi_m^2)}.
\]
Using \eqref{eq:p_infinity_local},
\[
\begin{aligned}
p_\alpha(\infty)
&=
F_{\chi_m^2}\left(
\frac{q_{1-\alpha}(\chi_m^2)}{c_\parallel}
\right) \\
&>
F_{\chi_m^2}\left(q_{\pi_0}(\chi_m^2)\right)
=
\pi_0
\ge
p_\alpha(0).
\end{aligned}
\]
This proves the second claim. Notice that the local argument only requires
\(d-m\ge1\), rather than \(d-m>2\).
\end{proof}

\begin{remark}
\label{rem:conditional_endpoint_local}
The endpoint inequality used later also holds after conditioning on the current
orthogonal random effect. In fact, when \(\tau^2=0\), infinite alignment
constrains only the \(\col(\Theta)\) component, so
\[
p_{\alpha,\gamma_*}(\infty)=1-\alpha.
\]
At \(\lambda=0\), conditional on \(\gamma_*\), the limiting displacement is
Gaussian with mean \(-\Theta_\perp\gamma_*\) and covariance
\[
P+(1+\sigma^2)P_\perp.
\]
The standard RWPI region is the centered Euclidean ball with probability
\(1-\alpha\) under \(N(0,\mathbb I_d)\). Anderson's inequality,
first for the mean shift and then for the covariance inflation in
\(\col(\Theta_\perp)\), gives
\[
p_{\alpha,\gamma_*}(0)<1-\alpha
=p_{\alpha,\gamma_*}(\infty)
\]
for almost every \(\gamma_*\), because \(r\ge1\) and \(\sigma^2>0\).
\end{remark}

\subsection{Proof of \texorpdfstring{\cref{thm:error_in_theta}}{Theorem: ErrorTheta}}
\label{sec:proof_robustness_theta}

Let
$
\mathcal S=\col(\Theta),
$
$
\hat{\mathcal S}=\col(\hat\Theta),
$
and write \(P\) and \(\hat P\) for the orthogonal projections onto
\(\mathcal S\) and \(\hat{\mathcal S}\), respectively. Since
\(\hat\Theta\) is obtained from external data, we condition on
\(\hat\Theta\) throughout the proof. Thus, \(\hat{\mathcal S}\) is
regarded as fixed. Replacing \(\Theta\) and \(\hat\Theta\) by
orthonormal bases of their column spaces does not change either the
principal angles or the two endpoint geometries \(\Lambda=0\) and
\(\Lambda=\infty\).

\begin{definition}[Principal angles]
\label{def:principal_angles}
Let \(U\) and \(V\) be two \(m\)-dimensional subspaces of
\(\mathbb R^d\). The principal angles
\[
0\leq \angle_1\leq\cdots\leq\angle_m\leq \frac{\pi}{2}
\]
are defined by
\[
\cos(\angle_k)=\sigma_k(Q_U^\top Q_V),
\qquad k=1,\ldots,m,
\]
where \(Q_U\) and \(Q_V\) have orthonormal columns spanning \(U\)
and \(V\), and the singular values are arranged in nonincreasing
order. In particular,
\[
\|P_U-P_V\|_{\mathrm{op}}=\sin(\angle_m).
\]
\end{definition}

Set
\[
Q=\Theta(\Theta^\top\Theta)^{-1/2},
\qquad
\tilde\kappa=(\Theta^\top\Theta)^{1/2}\kappa,
\qquad
\theta_0=Q\tilde\kappa=\Theta\kappa,
\]
and choose \(Q_\perp\) so that \([Q,Q_\perp]\) is orthogonal. Under
\eqref{asm:drift_model}, the current and future parameters can be
written as
\[
\beta_{e}
=
\theta_0+n^{-1/2}\{Q\mu_e+Q_\perp\gamma_e\},
\qquad e\in\{*\}\cup\mathcal E.
\]
As in \Cref{sec:proof_of_robustness}, the ERM expansion gives
\[
\sqrt n\bigl(\beta_{e}-\hat\beta_{\erm}\bigr)
\rightsquigarrow D_\tau,
\]
where
$
D_\tau
=
Q(\mu_e-\mu_*-G_\parallel)
+
Q_\perp(\gamma_e-\gamma_*-G_\perp),
$
with
$
G_\parallel\sim N(0,\mathbb I_m)
$
and
$
G_\perp\sim N(0,\mathbb I_{d-m})
$
independent of one another and of the random effects. We couple the
within-subspace effects as
\[
\mu_e=\tau z_e,
\qquad
\mu_*=\tau z_*,
\qquad
z_e,z_*\stackrel{\mathrm{i.i.d.}}{\sim}N(0,\mathbb I_m),
\]
so that \(D_\tau\) varies continuously in \(\tau\) almost surely.

We display the dependence of the endpoint (at $\lambda=0$ or $\lambda=\infty$) coverage probabilities on
\(\hat{\mathcal S}\) and \(\tau\). For \(\lambda\in\{0,\infty\}\), let
\[
p_\alpha(\lambda;\hat{\mathcal S},\tau)
:=
\lim_{n\to\infty}
\mathbb P\left\{
\beta_{e,n}
\in
\mathcal C_{1-\alpha}^{\,\hat{\mathcal S}}(\lambda)
\,\middle|\,
\hat{\mathcal S}
\right\},
\]
where
\(\mathcal C_{1-\alpha}^{\,\hat{\mathcal S}}(\lambda)\)
denotes the READ confidence region constructed using the estimated
representation subspace \(\hat{\mathcal S}\), and \(\tau\) denotes
the standard deviation of the within-representation random effect.
Since we condition on \(\hat{\mathcal S}\) throughout this proof, we
suppress the conditioning in the subsequent probability expressions. At \(\Lambda=0\), the READ region
is the standard RWPI region and does not depend on
\(\hat{\mathcal S}\). Thus, we write
$
p_\alpha(0;\tau)
:=
p_\alpha(0;\hat{\mathcal S},\tau) 
$
and have
\[
p_\alpha(0;\tau)
=
\mathbb P\left\{
\|D_\tau\|_2^2\leq q_{1-\alpha}(\chi_d^2)
\right\}.
\]
We next consider \(\Lambda=\infty\). Define
\[
c(\hat{\mathcal S})
=
\|(\mathbb I_d-\hat P)\theta_0\|_2^2,
\qquad
b(\hat{\mathcal S})
=
\frac{c(\hat{\mathcal S})}{1+c(\hat{\mathcal S})}.
\]
When \(c(\hat{\mathcal S})>0\), the limiting RWPI matrix at infinite
alignment is
\[
\Gamma_\infty(\hat{\mathcal S})
=
\hat P_\perp+c(\hat{\mathcal S})\mathbb I_d
=
c(\hat{\mathcal S})\hat P
+\{1+c(\hat{\mathcal S})\}\hat P_\perp,
\]
where \(\hat P_\perp=\mathbb I_d-\hat P\). Multiplying both the
coverage statistic and its calibration variable by
\(c(\hat{\mathcal S})\) gives
\begin{equation}
\begin{split}
p_\alpha(\infty;\hat{\mathcal S},\tau)
=\mathbb P\Bigl[
&\|\hat P D_\tau\|_2^2
+b(\hat{\mathcal S})\|\hat P_\perp D_\tau\|_2^2\leq
q_{1-\alpha}\left\{
\chi_m^2+b(\hat{\mathcal S})\chi_{d-m}^2
\right\}
\Bigr].
\end{split}
\label{equ:c9:14}
\end{equation}
The two chi-squared variables in (\ref{equ:c9:14}) are independent.
If \(c(\hat{\mathcal S})=0\), the same expression follows directly
from the local calculation used for the case of $\lambda=\infty$ considered in \Cref{sec:proof_of_robustness}, or equivalently by
continuous extension of (\ref{equ:c9:14}) as
\(c(\hat{\mathcal S})\downarrow0\).

At the true subspace \(\hat{\mathcal S}=\mathcal S\), we have
\(c(\mathcal S)=b(\mathcal S)=0\), and hence
\[
p_\alpha(\infty;\mathcal S,\tau)
=
\mathbb P\left\{
\|P D_\tau\|_2^2\leq q_{1-\alpha}(\chi_m^2)
\right\}
=
F_{\chi_m^2}\left
\{
\frac{q_{1-\alpha}(\chi_m^2)}{1+2\tau^2}
\right\}.
\]
In particular,
\[
p_\alpha(\infty;\mathcal S,0)=1-\alpha.
\]
By \Cref{thm:robustness}, and because \(d-m\geq1\) and
\(\sigma^2>0\),
\begin{equation}
p_\alpha(0;0)<1-\alpha
=p_\alpha(\infty;\mathcal S,0).
\label{equ:c9:16}
\end{equation}
It remains to show that the strict gap in Equation (\ref{equ:c9:16}) persists
under a small subspace error and a small within-subspace shift. The
map
\[
\hat P\mapsto c(\hat{\mathcal S})
\mapsto b(\hat{\mathcal S})
\]
is continuous in the operator norm. The random variable on the
left-hand side of the inequality in (\ref{equ:c9:14}) is also
continuous in \((\hat P,\tau)\) under the coupling above. Moreover,
\(q_{1-\alpha}(\chi_m^2+b\chi_{d-m}^2)\) is continuous in
\(b\in[0,1)\), and the corresponding Gaussian quadratic forms have
continuous distributions. It follows from dominated convergence that
\[
(\hat{\mathcal S},\tau)
\mapsto
p_\alpha(\infty;\hat{\mathcal S},\tau)
-p_\alpha(0;\tau)
\]
is continuous at \((\mathcal S,0)\).

The value of this map at \((\mathcal S,0)\) is strictly positive by \Cref{equ:c9:16}. Consequently, there exist
\(K_1\in(0,1)\) and \(\bar\tau>0\) such that
\[
\|\hat P-P\|_{\mathrm{op}}<K_1,
\qquad
0\leq\tau<\bar\tau
\]
imply
\[
p_\alpha(\infty;\hat{\mathcal S},\tau)
>
p_\alpha(0;\tau).
\]
By \Cref{def:principal_angles},
\(\|\hat P-P\|_{\mathrm{op}}=\sin(\angle_m)\). Setting
\(K_2=\bar\tau^2\) proves \Cref{thm:error_in_theta}.

\subsection{Proof of \texorpdfstring{\cref{thm:fixed_scale_robustness}}{Theorem: Fixed-Scale Robustness}}
\label{sec:proof_fixed_scale_robustness}

Throughout this section, suppose that $\Theta^\top\Theta=\mathbb I_m$ and $\Lambda=\lambda\mathbb I_m$, with $\lambda\in[0,\infty]$. Let
\[
P=\Theta\Theta^\top,
\qquad
P_\perp=\Theta_\perp\Theta_\perp^\top,
\qquad
s=d-m,
\qquad
K=\|\kappa\|_2^2.
\]
Under the fixed-scale regime $r_n=1$,
\[
\beta_e
=
\Theta(\kappa+\mu_e)+\Theta_\perp\gamma_e,
\qquad
e\in\{*\}\cup\mathcal E,
\]
where
\[
\mu_e,\mu_*\stackrel{\mathrm{iid}}{\sim}N(0,\tau^2\mathbb I_m),
\qquad
\gamma_e,\gamma_*\stackrel{\mathrm{iid}}{\sim}N(0,\sigma^2\mathbb I_s).
\]
The empirical risk minimizer satisfies
$
\hat\beta_{\erm}\xrightarrow{\mathbb P}\beta_*
$
conditionally on $\beta_*$. Consequently, for every fixed $\lambda$, the plug-in matrix and quantile in \eqref{eq:fixed_scale_region} converge to
\[
\Gamma_\lambda
=
\Psi_\lambda+c_\lambda\mathbb I_d,
\qquad
c_\lambda
=
\|\beta_*\|_{\Psi_\lambda}^2,
\qquad
\eta_\alpha^{\mathrm{fix}}(\lambda)
=
q_{1-\alpha}\left(Z^\top\Gamma_\lambda^{-1}Z\right),
\]
where $Z\sim N(0,\mathbb I_d)$. Thus,
\begin{equation}
\label{eq:fixed_scale_coverage_limit}
p_\alpha^{\mathrm{fix}}(\lambda)
=
\mathbb P\left\{
(\beta_e-\beta_*)^\top\Gamma_\lambda^{-1}(\beta_e-\beta_*)
\le
\eta_\alpha^{\mathrm{fix}}(\lambda)
\right\}.
\end{equation}

\begin{proof}[Proof of Part \textnormal{(1)}]
Suppose $\tau^2=0$. Then
\[
\beta_*
=
\Theta\kappa+\Theta_\perp\gamma_*,
\qquad
\beta_e-\beta_*
=
\Theta_\perp(\gamma_e-\gamma_*).
\]
Condition on $\gamma_*$. Since
\[
\Psi_\lambda
=
\frac{1}{1+\lambda}P+P_\perp,
\]
we have, for finite $\lambda$,
\[
c_\lambda
=
\frac{K}{1+\lambda}+\|\gamma_*\|_2^2.
\]
Write $B=\|\gamma_*\|_2^2$, which is positive almost surely because $s\ge1$ and $\sigma^2>0$. Then
\[
\Gamma_\lambda
=
d_\lambda P+e_\lambda P_\perp,
\qquad
 d_\lambda
 =B+\frac{1+K}{1+\lambda},
\qquad
 e_\lambda
 =1+B+\frac{K}{1+\lambda}.
\]
Define that
$
a_\lambda
={e_\lambda}/{d_\lambda}.
$
With $t=1+\lambda$,
\[
a_\lambda
=
\frac{(1+B)t+K}{Bt+1+K},
\qquad
\frac{d a_\lambda}{dt}
=
\frac{1+B+K}{(Bt+1+K)^2}>0.
\]
Hence $a_\lambda$ is strictly increasing on $[0,\infty]$, with
\[
a_0=1,
\qquad
 a_\infty=\frac{1+B}{B}.
\]

Let $X\sim\chi_m^2$ and $Y\sim\chi_s^2$ be independent. Since
\[
Z^\top\Gamma_\lambda^{-1}Z
\overset d=
 d_\lambda^{-1}X+e_\lambda^{-1}Y,
\]
quadratic homogeneity of quantiles gives
\begin{equation}
\label{eq:fixed_scale_threshold}
t_\alpha(\lambda)
\coloneqq
e_\lambda\eta_\alpha^{\mathrm{fix}}(\lambda)
=
q_{1-\alpha}(a_\lambda X+Y).
\end{equation}
Because $a_\lambda X+Y$ is strictly stochastically increasing in $a_\lambda$, $t_\alpha(\lambda)$ is strictly increasing in $\lambda$.

The displacement $\beta_e-\beta_*$ lies in $\col(\Theta_\perp)$. Therefore, conditional on $\gamma_*$, the event in \eqref{eq:fixed_scale_coverage_limit} is equivalent to
\[
\|\gamma_e-\gamma_*\|_2^2
\le
t_\alpha(\lambda).
\]
The conditional distribution of $\|\gamma_e-\gamma_*\|_2^2$ is continuous and strictly increasing on $(0,\infty)$. Hence the conditional coverage is strictly increasing in $\lambda$. Integrating over $\gamma_*$ proves that $p_\alpha^{\mathrm{fix}}(\lambda)$ is strictly increasing on $[0,\infty]$, and in particular
\[
p_\alpha^{\mathrm{fix}}(\infty)
>
p_\alpha^{\mathrm{fix}}(0).
\]
\end{proof}

\begin{proof}[Proof of Part \textnormal{(2)}]
First, we couple the within-representation random effects for all $\tau\ge0$ by writing
\[
\mu_e=\tau z_e,
\qquad
\mu_*=\tau z_*,
\qquad
z_e,z_*\stackrel{\mathrm{iid}}{\sim}N(0,\mathbb I_m).
\]
At $\lambda=0$, $\Psi_0=\mathbb I_d$ and $\Gamma_0=(1+\|\beta_*\|_2^2)\mathbb I_d$. Hence the scale factor cancels from the calibrated region, and
\begin{equation}
\label{eq:fixed_scale_endpoint_zero}
p_\alpha^{\mathrm{fix}}(0;\tau)
=
\mathbb P\left\{
\tau^2\|z_e-z_*\|_2^2
+
\|\gamma_e-\gamma_*\|_2^2
\le
q_{1-\alpha}(\chi_d^2)
\right\}.
\end{equation}
At $\lambda=\infty$, $\Psi_\infty=P_\perp$ and
\[
c_\infty=\|\gamma_*\|_2^2=B.
\]
Using $a_\infty=(1+B)/B$ and the threshold in \eqref{eq:fixed_scale_threshold},
\begin{equation}
\label{eq:fixed_scale_endpoint_infinity}
p_\alpha^{\mathrm{fix}}(\infty;\tau)
=
\mathbb P\left\{
 a_\infty\tau^2\|z_e-z_*\|_2^2
+
\|\gamma_e-\gamma_*\|_2^2
\le
 t_\alpha(\infty)
\right\}.
\end{equation}

Under the above coupling, the indicators in \eqref{eq:fixed_scale_endpoint_zero} and \eqref{eq:fixed_scale_endpoint_infinity} converge almost surely to their counterparts at $\tau=0$ as $\tau\downarrow0$. The boundary probabilities are zero: the first follows from continuity of $\|\gamma_e-\gamma_*\|_2^2$, and the second follows by conditioning on $\gamma_*$ and using the continuity of the conditional distribution. Then, we have
\[
p_\alpha^{\mathrm{fix}}(0;\tau)
\rightarrow
p_\alpha^{\mathrm{fix}}(0;0),
\qquad
p_\alpha^{\mathrm{fix}}(\infty;\tau)
\rightarrow
p_\alpha^{\mathrm{fix}}(\infty;0).
\]
Part (1) implies
\[
p_\alpha^{\mathrm{fix}}(\infty;0)
-
p_\alpha^{\mathrm{fix}}(0;0)
>0.
\]
Hence there exists $\bar\tau>0$ such that
\[
0<\tau<\bar\tau
\quad\Rightarrow\quad
p_\alpha^{\mathrm{fix}}(\infty;\tau)
>
p_\alpha^{\mathrm{fix}}(0;\tau).
\]
Setting $C_{\mathrm{fix}}=\bar\tau^2$ proves the claim.
\end{proof}

\subsection{Proof of \texorpdfstring{\cref{thm:interval_estimate}}{Theorem: IntervalEstimate}}
\label{sec:proof_projected_inference}

Let
\[
L_\Theta=(\Theta^\top\Theta)^{-1}\Theta^\top,
\qquad
Q=\Theta(\Theta^\top\Theta)^{-1/2},
\qquad
P=QQ^\top,
\qquad
P_\perp=\mathbb I_d-P,
\]
and write \(r=d-m\). Under the assumption \(\tau=0\),
\begin{equation}
\beta_{*,n}
=
\Theta\kappa+n^{-1/2}Q_\perp\gamma_*,
\qquad
L_\Theta\beta_{e,n}=\kappa
\quad\text{for every }e.
\label{equ:c10:17}
\end{equation}
Set
\[
K_\Theta=\|\Theta\kappa\|_2^2,
\qquad
B_n=n^{-1}\|\gamma_*\|_2^2.
\]
For
\(\Lambda=\lambda(\Theta^\top\Theta)^{-1}\),
\[
\Theta\Lambda\Theta^\top=\lambda P,
\qquad
\Psi_\lambda
=(\mathbb I_d+\lambda P)^{-1}
=\frac{1}{1+\lambda}P+P_\perp.
\]
Consequently,
\[
c_{n,\lambda}
:=
\|\beta_{*,n}\|_{\Psi_\lambda}^2
=
\frac{K_\Theta}{1+\lambda}+B_n,
\]
and
\begin{equation}
\Gamma_{n,\lambda}
=
\Psi_\lambda+c_{n,\lambda}\mathbb I_d
=
d_{n,\lambda}P+e_{n,\lambda}P_\perp,
\label{equ:c10:18}
\end{equation}
where
\[
d_{n,\lambda}
=
\frac{1+K_\Theta}{1+\lambda}+B_n,
\qquad
 e_{n,\lambda}
=
1+\frac{K_\Theta}{1+\lambda}+B_n.
\]
The formulas are understood by continuity at \(\lambda=\infty\), so
that
\[
d_{n,\infty}=B_n,
\qquad
 e_{n,\infty}=1+B_n.
\]

Let \(Z\sim N(0,\mathbb I_d)\). From (\ref{equ:c10:18}),
\[
Z^\top\Gamma_{n,\lambda}^{-1}Z
\stackrel{d}{=}
 d_{n,\lambda}^{-1}\chi_m^2
+e_{n,\lambda}^{-1}\chi_r^2,
\]
where the chi-squared variables are independent. Hence
\begin{equation}
\eta_{n,\alpha}(\lambda)
=
 d_{n,\lambda}^{-1}
q_{1-\alpha}\left\{
\chi_m^2+b_{n,\lambda}\chi_r^2
\right\},
\qquad
b_{n,\lambda}
=
\frac{d_{n,\lambda}}{e_{n,\lambda}}.
\label{equ:c10:19}
\end{equation}

For a local displacement \(u=Qx+Q_\perp z\), we have
\[
u^\top\Gamma_{n,\lambda}^{-1}u
=
 d_{n,\lambda}^{-1}\|x\|_2^2
+e_{n,\lambda}^{-1}\|z\|_2^2,
\]
and
\[
L_\Theta u=(\Theta^\top\Theta)^{-1/2}x.
\]
Therefore, the projection of the confidence region in
\eqref{eq:confidence_region} is
\begin{equation}
\begin{split}
L_\Theta\mathcal C_{1-\alpha}(\Lambda)
=
L_\Theta\hat\beta_{\erm}
+n^{-1/2}
\Bigl\{
 w\in\mathbb R^m:
 w^\top(\Theta^\top\Theta)w
\leq
 d_{n,\lambda}\eta_{n,\alpha}(\lambda)
\Bigr\}.
\end{split}
\label{equ:c10:20}
\end{equation}
Indeed, for a fixed projected coordinate \(w\), the left-hand side of
the ellipsoidal constraint is minimized by taking \(z=0\).

Let \(v_m\) denote the volume of the unit ball in \(\mathbb R^m\), and
write \(V_{\Pi,n}(\lambda)\) for the volume in
\Cref{thm:interval_estimate}, making its dependence on \(n\) explicit.
Equations (\ref{equ:c10:19})--(\ref{equ:c10:20}) give
\begin{equation}
V_{\Pi,n}(\lambda)
=
 n^{-m/2}
\frac{v_m\sqrt{\det(\Pi)}}
{\sqrt{\det(\Theta^\top\Theta)}}
\left[
q_{1-\alpha}\left\{
\chi_m^2+b_{n,\lambda}\chi_r^2
\right\}
\right]^{m/2}.
\label{equ:c10:21}
\end{equation}
To verify monotonicity, set \(t=1+\lambda\). Then
\[
b_{n,\lambda}
=
\frac{1+K_\Theta+B_nt}
{K_\Theta+(1+B_n)t},
\]
and
\begin{equation}
\frac{d}{dt}b_{n,\lambda}
=
-
\frac{1+K_\Theta+B_n}
{\{K_\Theta+(1+B_n)t\}^2}
<0.
\label{equ:c10:22}
\end{equation}
Since \(r=d-m\geq1\), the random variable
\(\chi_m^2+b\chi_r^2\) is strictly stochastically increasing in
\(b\). Thus, Equations (\ref{equ:c10:21})--(\ref{equ:c10:22}) show that
\(V_{\Pi,n}(\lambda)\) is strictly decreasing in
\(\lambda\in[0,\infty]\).

The local-volume limit is also explicit. Since \(B_n\to0\) almost
surely,
\[
b_{n,\lambda}
\rightarrow
b_\lambda
:=
\frac{1+K_\Theta}{1+K_\Theta+\lambda},
\]
with \(b_\infty=0\). Therefore,
\[
\begin{split}
n^{m/2}V_{\Pi,n}(\lambda)
\rightarrow
\frac{v_m\sqrt{\det(\Pi)}}
{\sqrt{\det(\Theta^\top\Theta)}}
\left[
q_{1-\alpha}\left\{
\chi_m^2+b_\lambda\chi_r^2
\right\}
\right]^{m/2}.
\end{split}
\]
The right-hand side is strictly decreasing in \(\lambda\), and its
limit at infinite alignment is
\[
\frac{v_m\sqrt{\det(\Pi)}}
{\sqrt{\det(\Theta^\top\Theta)}}
\left\{q_{1-\alpha}(\chi_m^2)\right\}^{m/2}>0.
\]

It remains to establish coverage of the invariant coordinate. By \Cref{equ:c10:17},
\begin{equation}
\{\beta_{*,n}\in\mathcal C_{1-\alpha}(\Lambda)\}
\subseteq
\left\{
\kappa\in
L_\Theta\mathcal C_{1-\alpha}(\Lambda)
\right\}.
\label{equ:c10:24}
\end{equation}
For every finite \(\lambda\), the RWPI validity result, applied to the
local sequence \(\beta_{*,n}\), gives
\[
\mathbb P\{\beta_{*,n}\in\mathcal C_{1-\alpha}(\Lambda)\}
\rightarrow 1-\alpha.
\]
Combining this with (\ref{equ:c10:24}), we have
\[
\liminf_{n\to\infty}
\mathbb P\left\{
\kappa\in
L_\Theta\mathcal C_{1-\alpha}(\Lambda)
\right\}
\geq1-\alpha.
\]
At \(\lambda=\infty\),
\(d_{n,\infty}\eta_{n,\alpha}(\infty)\to
q_{1-\alpha}(\chi_m^2)\). Moreover,
\[
\sqrt n\left(
L_\Theta\hat\beta_{\erm}-\kappa
\right)
\rightsquigarrow
N\left(0,(\Theta^\top\Theta)^{-1}\right).
\]
Thus the quadratic form induced by \(\Theta^\top\Theta\) converges to
\(\chi_m^2\), and the projected region has limiting coverage exactly
\(1-\alpha\). This proves the coverage claim and completes the proof.

\subsection{Proof of \texorpdfstring{\cref{thm:lagrangian_asymptotics}}{Theorem: Lagrangian}}
\begin{proof}[Proof of $p=2$]
    Let $J(x,y;\beta) = \nabla_\beta h(x,y;\beta)$, and recall the notation $\Sigma = -\mathbb{E}_{\mathbb{P}_*} [J(X,Y;\beta_*)]$. The local Lipschitz regularity of $\ell$ using \cref{asm:regularity_iii} implies that for every compact $K\subset\mathbb{R}^d$, we have the local empirical Hessian stability \[
    \sup_{u\in K}\sup_{t \in [0,1]}\left\|-\mathbb{E}_{\mathbb{P}_n}[J(X,Y;\beta_* + tn^{-\frac{1}{2}}u)] - \Sigma \right\|_{\mathrm{op}} \stackrel{p}{\to}0. 
    \]Recall that $\varphi_{p,\Lambda_n}^2(\beta;\Theta_n) = \inf_{\kappa} \{ |\beta - \Theta_n \kappa|_p^2+|\kappa|_{\Lambda_n^{-1}}^2\}$, where $\Lambda_n$ is a diagonal matrix whose entries are nonnegative. We first prove the theorem for the case of $p=2$, in which \[
    \varphi_{p,\Lambda_n}^2(\beta) = \|\beta\|_{\Psi_{\Lambda_n}}^2,\quad \Psi_{\Lambda_n} = \mathbb{I}_d - \Theta_n(\mathbb{I}_m+\Lambda_n^{-1})^{- 1}\Theta_n^\top,
    \]where we assumed that $\Theta_n^\top\Theta_n = \mathbb{I}_m$. Define the objective \[
    R_n(\beta) = \mathbb{E}_{\mathbb{P}_n}[\ell(X,Y;\beta)] + \delta_n \|\beta\|_{\Psi_{n}}^2,\quad V_n(u) = n(R_n(\beta_*+n^{-\frac{1}{2}}u) - R_n(\beta_*))
    \] and $\hat{\beta}_\mathcal{L} = \argmin_\beta R_n(\beta),$ where we used $\Psi_n$ for $\Psi_{\Lambda_n}$ for short. It is easy to see that \[
    \sqrt{n}(\hat{\beta}_\mathcal{L} - \beta_*) = \argmin_{u} V_n(u).
    \] Define $H_n = \sqrt{n}\mathbb{E}_{\mathbb{P}_n} [h(X,Y;\beta_*)]$, central limit theorem yields $H_n \leadsto H$ weakly, with $H \sim N(0,\cov{h(X,Y;\beta_*)})$. Next, by the second-order Taylor expansion with integral remainder and the assumptions, for every compact $K\subset \mathbb{R}^d$, \[
    \sup_{u\in K} \left|n\mathbb{E}_{\mathbb{P}_n}\left[\ell(X,Y;\beta_* + n^{-1/2}u)-\ell(X,Y;\beta_*)\right] + H_n^\top u-\frac{1}{2}u^\top \Sigma u\right| \stackrel{p}{\to}0
    \]Hence, uniformly on every compact \(K\),
    \[
    n\mathbb{E}_{\mathbb{P}_n} \left[\ell(X,Y;\beta_* + n^{-1/2}u) - \ell(X,Y;\beta_*)\right] = -H_n^\top u + \frac{1}{2}u^\top \Sigma u + o_p(1).
    \]For the penalty part of $R_n(\beta)$,
    \[
    n\delta_n\left[
    (\beta_* + n^{-1/2}u)^\top \Psi_n (\beta_* + n^{-1/2}u) - \beta_*^\top \Psi_n \beta_*\right]
    = 2\delta_n \sqrt{n} u^\top \Psi_n \beta_* + \delta_n u^\top \Psi_n u.
    \]We now compare \(\Psi_n\) with \(P_{(\col{\Theta})^\perp}\), let $S = \col{\Theta}$ and $S_n = \col{\Theta_n}$. Since
    \[
    \Psi_n - P_{S^\perp} =(P_{S_n^\perp} - P_{S^\perp}) + \Theta_n (\mathbb{I}_m + \Lambda_n)^{-1}\Theta_n^\top,
    \]
    we have
    \[
    \|\Psi_n - P_{S^\perp}\|_{\mathrm{op}}
    \leq
    \|P_{S_n} - P_S\|_{\mathrm{op}}
    + \max_{1 \leq j \leq m}(1+\lambda_{j,n})^{-1}
    = O_p(n^{-\gamma}) + \max_j (1+\lambda_{j,n})^{-1}.
    \]Because \(\beta_* \in S\), we have \(P_{S^\perp}\beta_* = 0\). Therefore, for every compact \(K \subset \mathbb{R}^d\) with diameter $0< C_K <\infty$,
    \[
    \sup_{u \in K}
    \left| 2\delta_n \sqrt{n} u^\top \Psi_n \beta_* \right|
    \leq 2C_K \delta_n \sqrt{n}\|\Psi_n - P_{S^\perp}\|_{\mathrm{op}} \|\beta_*\|_2 = o_p(1),
    \]Likewise,
    \[
    \sup_{u\in K}\left|\delta_n u^\top(\Psi_n-P_{S^\perp})u\right|
    \leq C_K\delta_n\|\Psi_n-P_{S^\perp}\|_{\mathrm{op}}
    = o_p(1).
    \]Combining the loss expansion and the penalty bounds yields, uniformly on every compact \(K\subset\mathbb{R}^d\),
    \[
    V_n(u)
    =-H_n^\top u+\frac12 u^\top\Sigma u+\delta_n u^\top P_{S^\perp}u+o_p(1).
    \]Now fix a compact \(K\subset S\). Since \(P_{S^\perp}u=0\) for \(u\in S\),
    \[
    \sup_{u\in K}
    \left|
    V_n(u)-\left(-H_n^\top u+\frac12 u^\top\Sigma u\right)
    \right| \stackrel{p}{\to}0.
    \]On the other hand, if \(K\subset\mathbb{R}^d\) is compact and \(\operatorname{dist}(K,S)>0\), then
    \[
    c_K:=\inf_{u\in K}u^\top P_{S^\perp}u
    = \inf_{u\in K}\|P_{S^\perp}u\|_2^2 >0.
    \]Therefore, $\inf_{u\in K}V_n(u) \geq \delta_n c_K - O_p(1) \overset{p}{\to} +\infty$, since \(\delta_n\to\infty\). Thus \(V_n\) converges, in the stochastic epi-sense, to the extended-real convex function
    \[
    V(u):=-H^\top u+\frac12 u^\top \Sigma u+I_S(u),
    \]
    where \(I_S(u)=0\) for \(u\in S\) and \(I_S(u)=+\infty\) otherwise. It remains to identify the unique minimizer of \(V\). Writing \(u=\Theta a\) with \(a\in\mathbb{R}^m\), we get
    \[
    V(\Theta a)=-H^\top \Theta a+\frac12 a^\top (\Theta^\top \Sigma \Theta)a.
    \] Since \(\Theta^\top \Sigma \Theta \succ 0\), this quadratic function has the unique minimizer $a^*=(\Theta^\top \Sigma \Theta)^{-1}\Theta^\top H$. Hence $u^*=\Theta(\Theta^\top \Sigma \Theta)^{-1}\Theta^\top H$. By the convex argmin theorem for stochastic epi-convergence,
    \[
    \argmin_u V_n(u)\Rightarrow \argmin_u V(u)=u^*.
    \]
    Since \(\argmin_u V_n(u)=\sqrt{n}(\hat{\beta}_{\mathcal L}-\beta_*)\), we conclude that
    \[
    \sqrt{n}(\hat{\beta}_{\mathcal L}-\beta_*)\Rightarrow \Theta(\Theta^\top \Sigma \Theta)^{-1}\Theta^\top H.
    \]
\end{proof}

\begin{proof}[Proof of General $p\geq 1$]
Let \(a_p,b_p\) be as in Lemma~\ref{lem:penalty_comparison}, and set
\[
R_{n}^{(p)}(\beta) \coloneqq \mathbb{E}_{\mathbb{P}_n}\ell(X,Y;\beta)+\delta_n\varphi_{p,\Lambda_n}^2(\beta;\Theta_n).
\]
Also define the lower and upper comparison criteria
\begin{align*}
    R_n^-(\beta) &\coloneqq \mathbb{E}_{\mathbb{P}_n}\ell(X,Y;\beta)+a_p\delta_n\varphi_{2,a_p\Lambda_n}^2(\beta;\Theta_n),\\
    R_n^+(\beta)&\coloneqq\mathbb{E}_{\mathbb{P}_n}\ell(X,Y;\beta)+b_p\delta_n\varphi_{2,b_p\Lambda_n}^2(\beta;\Theta_n).
\end{align*}
By Lemma~\ref{lem:penalty_comparison},
\[
R_n^-(\beta)\le R_{n}^{(p)}(\beta)\le R_n^+(\beta),
\quad \beta\in\mathbb{R}^d .
\]For \(\circ\in\{-,(p),+\}\), define the local quadratic
\[
V_n^\circ(u) \coloneqq n\Bigl(R_n^\circ(\beta_*+n^{-1/2}u)-R_n^\circ(\beta_*)\Bigr).
\]
Then, for every \(u\in\mathbb{R}^d\),
\[
V_n^-(u)-C_n \le V_n^{(p)}(u)\le V_n^+(u)+C_n,
\]
where
\[
C_n \coloneqq n\bigl(R_n^+(\beta_*)-R_n^-(\beta_*)\bigr)\ge 0.
\]We now show that \(C_n=o_p(1)\). Since \(\beta_*\in S\coloneqq\col(\Theta)\), the closed-form expression for the Euclidean penalty gives, for each fixed \(c>0\),
\[
c\varphi_{2,c\Lambda_n}^2(\beta_*)=\beta_*^\top M_{n,c}\beta_*,
\]
with
\[
M_{n,c} = cP_{S_n^\perp} +\Theta_n \diag\Bigl(\frac{c}{1+c\lambda_{1,n}},\dots,\frac{c}{1+c\lambda_{m,n}}\Bigr)\Theta_n^\top.
\]
Hence
\[
\frac{C_n}{n\delta_n} = \beta_*^\top(M_{n,b_p}-M_{n,a_p})\beta_*,
\]
so
\[
\frac{C_n}{n\delta_n} \le |b_p-a_p|\|P_{S_n^\perp}\beta_*\|_2^2 + \max_{1\le j\le m}
\left| \frac{b_p}{1+b_p\lambda_{j,n}} - \frac{a_p}{1+a_p\lambda_{j,n}} \right| \|\beta_*\|_2^2 .
\]
Because \(\beta_*\in S\),
\[
\|P_{S_n^\perp}\beta_*\|_2
= \|(P_S-P_{S_n})\beta_*\|_2
\le \|P_S-P_{S_n}\|_{\mathrm{op}}\|\beta_*\|_2
= O_p(n^{-\gamma}),
\]
and
\[
\frac{b_p}{1+b_p\lambda}-\frac{a_p}{1+a_p\lambda}
= \frac{b_p-a_p}{(1+a_p\lambda)(1+b_p\lambda)}
= O(\lambda^{-2}).
\]
Therefore
\[
C_n
= O_p(\delta_n n^{1-2\gamma})
+ O\Bigl(\delta_n n \max_{1\le j\le m}\lambda_{j,n}^{-2}\Bigr).
\]
The first term is \(o_p(1)\) since \(\delta_n n^{1/2-\gamma}\to 0\). The second is also \(o_p(1)\), because
\[
\delta_n n \lambda_{j,n}^{-2}
= \Bigl(\delta_n n^{1/2}\lambda_{j,n}^{-1}\Bigr) \Bigl(n^{1/2}\lambda_{j,n}^{-1}\Bigr),
\]
the first factor tends to \(0\) by assumption, and the second tends to \(0\) even faster for all $j \in [m]$. Applying the \(p=2\) theorem to both \(R_n^-\) and \(R_n^+\), both \(V_n^-\) and \(V_n^+\) epi-converge to
\[
V(u)
=
-H^\top u+\frac12 u^\top\Sigma u + I_S(u),
\]
where \(I_S(u)=0\) if \(u\in S\) and \(I_S(u)=+\infty\) otherwise. Using the comparison
\[
V_n^-(u)-C_n \le V_n^{(p)}(u) \le V_n^+(u)+C_n
\]
together with \(C_n=o_p(1)\), it follows that \(V_n^{(p)}\) epi-converges to the same limit \(V\). This proves the claim.
\end{proof}

\subsection{Proof of \texorpdfstring{\cref{prop:variational_rep}}{Proposition: VariationalDual}}
\begin{proof}
Set
\[
H:=L^2(\Omega;\mathbb R^d),\quad\langle U,V\rangle_H:=\mathbb E_{P_*}[U^\top V].
\]
Let
\[
c:=c_{q,\Lambda}, \quad F:H\to\bar{\mathbb R}, \quad F(U):=\mathbb E_{P_*}[c(U)].
\]Since $c$ is proper, lower semicontinuous, and convex on $\mathbb R^d$, and $c(0)=0$, the functional $F$ is proper and convex on $H$. Moreover, because
\[
c(u)\ge \|u\|_q^2 \ge a\|u\|_2^2, \quad\text{for some }a>0 \text{ and all }u\in\mathbb R^d,
\]
we have $F(U)\ge a\|U\|_H^2$, so $F$ is coercive on $H$.

We also claim that $F$ is lower semicontinuous on $H$. Indeed, if $U_n\to U$ in $H$, then, after passing to a subsequence if necessary, $U_n(\omega)\to U(\omega)$ for $P_*$-a.e.\ $\omega$. By lower semicontinuity of $c$ and Fatou's lemma,
\[
F(U) = \mathbb E_{P_*}[c(U)]
\le \mathbb E_{P_*}\left[\liminf_{n\to\infty} c(U_n)\right]
\le \liminf_{n\to\infty}\mathbb E_{P_*}[c(U_n)]
=\liminf_{n\to\infty}F(U_n).
\]Hence $F$ is strongly lower semicontinuous, and therefore also weakly lower semicontinuous. Now define the linear map
\[
A:H\to\mathbb R^d, \quad AU:=\mathbb E_{P_*}[\Xi^\top U].
\]
Because $\Xi\in L^2(\Omega;\mathbb R^{d\times d})$, this map is continuous:
\[
\|AU\|_2 \le \mathbb E_{P_*}\!\left[\|\Xi\|_{\mathrm{op}}\|U\|_2\right] \le \|\Xi\|_{L^2}\|U\|_H.
\]
Its adjoint $A^*:\mathbb R^d\to H$ is $A^*x=\Xi x$, since for every $x\in\mathbb R^d$ and $U\in H$,
\[
\langle x,AU\rangle_{\mathbb R^d}
= x^\top \mathbb E_{P_*}[\Xi^\top U]
= \mathbb E_{P_*}\left[(\Xi x)^\top U\right]
= \langle \Xi x,U\rangle_H.
\]Fix $x\in\mathbb R^d$ and consider the integrand
\[
h_x(\omega,u)\coloneqq c(u)-\langle \Xi(\omega)x,u\rangle,
\quad (\omega,u)\in\Omega\times\mathbb R^d.
\]
Since $u\mapsto c(u)$ is proper lower semicontinuous convex, and
$(\omega,u)\mapsto \langle \Xi(\omega)x,u\rangle$ is measurable in $\omega$
and continuous in $u$, the function $h_x$ is a convex normal integrand.
Because $H=L^2(\Omega;\mathbb R^d)$ is decomposable, \citet[Theorem 14.60]{rockafellarwets1998} applies and yields
\[
\inf_{U\in H}\mathbb E_{P_*}[h_x(\omega,U(\omega))]
=\mathbb E_{P_*}\left[\inf_{u\in\mathbb R^d} h_x(\omega,u)\right].
\]
Equivalently, after multiplying by $-1$,
\begin{align*}
F^*(A^*x)
&= \sup_{U\in H}\Bigl\{\langle A^*x,U\rangle_H-F(U)\Bigr\}\\
&= \sup_{U\in H}\mathbb E_{P_*}\left[\langle \Xi x,U\rangle-c(U)\right]\\
&= \mathbb E_{P_*}\left[\sup_{u\in\mathbb R^d}\bigl\{\langle \Xi x,u\rangle-c(u)\bigr\}\right]\\
&= \mathbb E_{P_*}[c^*(\Xi x)]\\
&= \frac14\mathbb E_{P_*}\!\left[\varphi_{p,\Lambda}^2(\Xi x)\right]\\
&= \psi_{p,\Lambda}(x).
\end{align*}
Thus $\psi_{p,\Lambda}=F^*\circ A^*$. Next define the value function
\[
v:\mathbb R^d\to\bar{\mathbb R},\quad v(z):=\inf\{F(U): U\in H,\ AU=z\}.
\]Then, for every $x\in\mathbb R^d$,
\begin{align*}
v^*(x)
&= \sup_{z\in\mathbb R^d}\{\langle x,z\rangle-v(z)\}\\
&= \sup_{z\in\mathbb R^d}\sup_{U:AU=z}\{\langle x,z\rangle-F(U)\}\\
&= \sup_{U\in H}\{\langle x,AU\rangle-F(U)\}\\
&= \sup_{U\in H}\{\langle A^*x,U\rangle_H-F(U)\}\\
&= F^*(A^*x)\\
&= \psi_{p,\Lambda}(x).
\end{align*}
Hence $v^*=\psi_{p,\Lambda}.$ It remains to show that $v$ is lower semicontinuous. Let $z_n\to z$ in $\mathbb R^d$, and choose $U_n\in H$ such that
\[
AU_n=z_n, \quad F(U_n)\le v(z_n)+\frac1n.
\]
If $\liminf_n v(z_n)=+\infty$, there is nothing to prove. Otherwise, after passing
to a subsequence, we may suppose that $\sup_n F(U_n)<\infty$. Since $F$ is coercive,
$\{U_n\}$ is bounded in $H$. Therefore, after passing to a further subsequence,
\[
U_n\rightharpoonup U \quad\text{weakly in }H
\]
for some $U\in H$. Because $A$ is continuous linear and the codomain is finite-dimensional,
\[
AU_n \to AU \quad\text{in }\mathbb R^d.
\]
But $AU_n=z_n\to z$, so necessarily $AU=z$. By weak lower semicontinuity of $F$,
\[
v(z)\le F(U)\le \liminf_{n\to\infty}F(U_n)=\liminf_{n\to\infty}v(z_n).
\]
Thus $v$ is lower semicontinuous. Since $v$ is proper, convex, and lower semicontinuous, the Fenchel-Moreau theorem gives
\[
v=v^{**}=(v^*)^*=\psi_{p,\Lambda}^*.
\]
Therefore,
\[
\psi_{p,\Lambda}^*(z)=v(z)=\inf\left\{\mathbb E_{P_*}[c_{q,\Lambda}(U)]:U\in L^2(\Omega;\mathbb R^d),\mathbb E_{P_*}[\Xi^\top U]=z\right\}.
\]This completes the proof.
\end{proof}

\subsection{Proof of \texorpdfstring{\cref{prop:alg_delta_consistency}}{Proposition: Delta Algorithm}}
\begin{proof}
Fix \(\Lambda\). Define
\[
\hat\psi_{n,\Lambda}(\xi)
\coloneqq \frac{1}{4n}\sum_{i=1}^n \varphi_{p,\Lambda}^2\bigl(\Xi_i(\hat\beta)\xi\bigr),
\qquad \psi_{p,\Lambda}(\xi) = \frac14 \mathbb{E}\left[\varphi_{p,\Lambda}^2(\Xi\xi)\right].
\]
By Fenchel--Rockafellar duality, the convex program in Step 2 of \Cref{alg:delta} computes the convex conjugate \(\hat\psi_{n,\Lambda}^*\).
Since \(\hat\beta\overset{p}{\rightarrow}\beta_*\), we have \(\Xi_i(\hat\beta)-\Xi_i(\beta_*)\overset{p}{\rightarrow}0\) for each \(i\). Moreover,
\(\varphi_{p,\Lambda}(z)\le \|z\|_p\), so \(\varphi_{p,\Lambda}^2(\Xi_i(\hat\beta)\xi)\) admits an integrable envelope on every compact \(\xi\)-set. Hence, by a uniform law of large numbers plus dominated continuity, for every compact \(K\subset\mathbb{R}^d\),
\[
\sup_{\xi\in K}
\bigl|
\hat\psi_{n,\Lambda}(\xi)-\psi_{p,\Lambda}(\xi)
\bigr| =o_p(1).
\]
Now \(\psi_{p,\Lambda}\) is finite, convex, continuous, and positive definite by \cref{asm:pd}, hence coercive. Therefore, local uniform convergence of finite convex functions implies local uniform convergence of the Fenchel conjugates: for every compact \(K\subset\mathbb{R}^d\),
\[
\sup_{h\in K}
\bigl| \hat\psi_{n,\Lambda}^*(h)-\psi_{p,\Lambda}^*(h) \bigr| =o_p(1).
\]
Let \(V\coloneqq \cov(h(X,Y;\beta_*))\), so \(H\sim N(0,V)\). By consistency of \(\hat\Sigma_h\), we have $\hat\Sigma_h \stackrel{p}{\to} V$. Therefore, letting \(G_n\sim N(0,\hat\Sigma_h)\), we have \(G_n \leadsto H\). Combining this with the uniform convergence of $\hat \psi_{n,\Lambda}^*$ on compact sets and the continuous mapping theorem gives
\[
\hat\psi_{n,\Lambda}^*(G_n)\leadsto \psi_{p,\Lambda}^*(H).
\]
Let \(q_{n,\alpha}\) be the conditional \((1-\alpha)\)-quantile of \(\hat\psi_{n,\Lambda}^*(G_n)\). Since the cdf of \(\psi_{p,\Lambda}^*(H)\) is continuous at \(\eta_\alpha(\Lambda)\), we have
\[
q_{n,\alpha}\overset{p}{\rightarrow} \eta_\alpha(\Lambda).
\]Finally, \[
\hat\eta_\alpha(\Lambda)\stackrel{p}{\to}\eta_\alpha(\Lambda),
\]
by the law of large number for sample quantiles.
\end{proof}

\subsection{Proof of Proposition~\ref{prop:lambda-selection}}

\begin{proof}
Write
\[
Y_i=X_i^\top\beta_*+\varepsilon_i,
\qquad
E(\varepsilon_i\mid X_i)=0,
\qquad
E(\varepsilon_i^2\mid X_i)=1.
\]
Let \(I_1,\ldots,I_K\) denote the balanced folds in
Algorithm~\ref{alg:lambda-cv}, with
\[
n_h=\frac{n}{K},
\qquad
n_t=n-n_h=\frac{K-1}{K}n.
\]
For each fold \(k\), define
\[
H_k
=
\frac{1}{\sqrt{n_h}}
\sum_{i\in I_k}X_i\varepsilon_i,
\qquad
\hat\Sigma_{X,k}
=
\frac{1}{n_h}
\sum_{i\in I_k}X_iX_i^\top,
\]
and
\[
Z_{-k}
=
\Sigma_X^{-1}
\frac{1}{\sqrt{n_t}}
\sum_{i\notin I_k}X_i\varepsilon_i.
\]

For a deterministic alignment matrix \(\Lambda\), define
\[
b_\Lambda
=
\sqrt{\eta_\alpha(\Lambda)}
\frac{
\Sigma_X^{-1}\Psi_\Lambda\beta_*
}{
\|\beta_*\|_{\Psi_\Lambda}
}.
\]
Then we have
\[
b_\Lambda^\top\Sigma_Xb_\Lambda
=
\eta_\alpha(\Lambda)
\frac{
\|\Psi_\Lambda\beta_*\|_{\Sigma_X^{-1}}^2
}{
\|\beta_*\|_{\Psi_\Lambda}^2
}
=
\mathcal Q(\Lambda).
\]

By the expansion in Theorem \ref{thm:asymptotic_dro}, Proposition~\ref{prop:alg_delta_consistency}, and
Assumption~\ref{asm:alg_lambda}, for every
\(f\in\mathcal F\) and \(k\in[K]\),
\begin{equation}
\sqrt{n_t}
\left(
\hat\beta_f^{(-k)}-\beta_*
\right)
=
Z_{-k}-b_f+o_p(1),
\qquad
b_f:=b_{\Lambda(f)}.
\label{equ:c19}
\end{equation}
Because both \(K\) and \(|\mathcal F|\) are fixed, the remainder in \Cref{equ:c19} is uniform over \(k\in[K]\) and
\(f\in\mathcal F\). The possible data dependence of
\(\Lambda_n(f)\) is asymptotically negligible because
\(\Lambda_n(f)\overset{p}{\rightarrow}\Lambda(f)\) and
\(\Lambda\mapsto b_\Lambda\) is continuous on \(\mathcal L\). For fold \(k\), write
\[
\hat R_k(\beta)
=
\frac{1}{n_h}
\sum_{i\in I_k}
(Y_i-X_i^\top\beta)^2.
\]
The squared-loss expansion around \(\beta_*\) is:
\begin{equation}
\hat R_k(\beta)-\hat R_k(\beta_*)
=
-\frac{2}{\sqrt{n_h}}
H_k^\top(\beta-\beta_*)
+
(\beta-\beta_*)^\top
\hat\Sigma_{X,k}
(\beta-\beta_*).
\label{equ:c20}
\end{equation}
Fix \(f,g\in\mathcal F\). Substituting \Cref{equ:c19} into \Cref{equ:c20}, using
\(\hat\Sigma_{X,k}\overset{p}{\rightarrow}\Sigma_X\), and averaging over the folds
gives
\begin{equation}
\begin{split}
n_t
\left\{
\hat{\mathrm{CV}}_n(f)
-
\hat{\mathrm{CV}}_n(g)
\right\}
={}&
\frac{2}{K}
\sqrt{\frac{n_t}{n_h}}
\sum_{k=1}^K
H_k^\top(b_f-b_g)
\\
&-
\frac{2}{K}
\sum_{k=1}^K
Z_{-k}^\top\Sigma_X(b_f-b_g)
\\
&+
b_f^\top\Sigma_Xb_f
-
b_g^\top\Sigma_Xb_g
+
o_p(1).
\end{split}
\label{equ:c21}
\end{equation}
The first two terms in \Cref{equ:c21} cancel because
\[
\Sigma_XZ_{-k}
=
\sqrt{\frac{n_h}{n_t}}
\sum_{\ell\neq k}H_\ell.
\]
Therefore, we have
\begin{equation}
n_t
\left\{
\hat{\mathrm{CV}}_n(f)
-
\hat{\mathrm{CV}}_n(g)
\right\}
=
\mathcal Q\{\Lambda(f)\}
-
\mathcal Q\{\Lambda(g)\}
+
o_p(1),
\label{equ:c22}
\end{equation}
uniformly over \(f,g\in\mathcal F\). Because \(\mathcal F\) is finite and \(f_*\) is the unique
minimizer, there exists
\[
\Delta
:=
\min_{f\neq f_*}
\left[
\mathcal Q\{\Lambda(f)\}
-
\mathcal Q\{\Lambda(f_*)\}
\right]
>0.
\]
\Cref{equ:c22} then implies
\[
\Pr(\hat f=f_*)\rightarrow1.
\]
Together with
\(\Lambda_n(f_*)\overset{p}{\rightarrow}\Lambda(f_*)\), this yields
\begin{equation}
\hat\Lambda
=
\Lambda_n(\hat f)
\overset{p}{\rightarrow}
\Lambda_*.
\label{equ:c23}
\end{equation}

We now prove the post-selection coverage result. The proof of
Proposition~A.2 extends uniformly over the compact and uniformly
nondegenerate set \(\mathcal L\). In particular, for every fixed
\(M<\infty\),
\begin{equation}
\sup_{\Lambda\in\mathcal L}
\sup_{\|z\|_2\leq M}
\left|
\hat\psi^*_{2,\Lambda}(z)
-
\psi^*_{2,\Lambda}(z)
\right|
=o_p(1),
\label{equ:c24}
\end{equation}
and
\begin{equation}
\sup_{\Lambda\in\mathcal L}
\left|
\hat\eta_\alpha(\Lambda)
-
\eta_\alpha(\Lambda)
\right|
=o_p(1).
\label{equ:c25}
\end{equation}
The uniform nondegeneracy condition in
Assumption~\ref{asm:alg_lambda} ensures that the
coercivity constants used in the proof of Proposition~\ref{prop:alg_delta_consistency} can be
chosen uniformly over \(\mathcal L\).

Recall the asymptotic normality of the ERM estimator,
\[
\sqrt n
\left(
\hat\beta_{\mathrm{ERM}}-\beta_*
\right)
\rightsquigarrow
\Sigma^{-1}H,
\]
where
\[
H\sim
N\left(
0,
\operatorname{Cov}_{P_*}
\{h(X,Y;\beta_*)\}
\right).
\]
Combining this result with
\(\hat\Sigma\overset{p}{\rightarrow}\Sigma\), (\ref{equ:c23}), and (\ref{equ:c24}), we obtain
\[
\hat\psi^*_{2,\hat\Lambda}
\left\{
\hat\Sigma
\sqrt n
(\beta_*-\hat\beta_{\mathrm{ERM}})
\right\}
\rightsquigarrow
\psi^*_{2,\Lambda_*}(-H).
\]
Since \(\psi^*_{2,\Lambda_*}\) is even,
\(\psi^*_{2,\Lambda_*}(-H)\) and
\(\psi^*_{2,\Lambda_*}(H)\) have the same distribution. Likewise, (\ref{equ:c23}), (\ref{equ:c25}), and continuity
of \(\Lambda\mapsto\eta_\alpha(\Lambda)\) imply
\[
\hat\eta_\alpha(\hat\Lambda)
\overset{p}{\rightarrow}
\eta_\alpha(\Lambda_*).
\]
The distribution function of
\(\psi^*_{2,\Lambda_*}(H)\) is continuous at
\(\eta_\alpha(\Lambda_*)\). Consequently,
\[
\begin{split}
\Pr\left\{
\beta_*
\in
\hat C_{1-\alpha}(\hat\Lambda)
\right\}
&=
\Pr\left[
\hat\psi^*_{2,\hat\Lambda}
\left\{
\hat\Sigma
\sqrt n
(\beta_*-\hat\beta_{\mathrm{ERM}})
\right\}
\leq
\hat\eta_\alpha(\hat\Lambda)
\right]
\\
&\rightarrow
\Pr\left\{
\psi^*_{2,\Lambda_*}(H)
\leq
\eta_\alpha(\Lambda_*)
\right\}
\\
&=
1-\alpha.
\end{split}
\]
This completes the proof.
\end{proof}

\section{Auxiliary Results}
\label{app:lemmas}
\subsection{Convex Conjugate of Quadratic Functions}
\begin{lemma}
\label{lem:1}
Assume $\beta\neq 0$. Let
\[
f_\beta(\Delta)= (\beta^\top \Delta)^2 - 2r(\beta)\beta^\top \Delta,
\quad \Delta\in\mathbb R^d,
\]
where $r(\beta)\in\mathbb R$. Then
\[
f_\beta^*(\Delta^*)=
\begin{cases}
\dfrac{(\beta^\top \Delta^* + 2r(\beta)\|\beta\|_2^2)^2}{4\|\beta\|_2^4},
& \text{if } \Delta^* \in \col{\beta},\\[2mm]
\infty, & \text{otherwise.}
\end{cases}
\]
Therefore,
\[
f_\beta^{**}(\Delta)=\sup_{\alpha\in\mathbb R}
\left(
\alpha\beta^\top\Delta-\frac{(\alpha+2r(\beta))^2}{4}
\right).
\]
\end{lemma}
\begin{proof}
The convex conjugate $f_\beta^*(\Delta^*)$ is defined as
\[
f_\beta^*(\Delta^*) \coloneqq \sup_{\Delta\in\mathbb R^d}
\bigl( \Delta^{*\top}\Delta - (\beta^\top\Delta)^2 + 2r(\beta)\beta^\top\Delta \bigr),
\]
where $\Delta^*,\beta\in\mathbb R^d$ and $r(\beta)\in\mathbb R$ are fixed. Decompose
\[
\Delta = a\beta + \omega,
\]
with $a\in\mathbb R$ and $\omega\in\mathbb R^d$ such that $\beta^\top\omega=0$. Then
\[
\Delta^{*\top}\Delta = a\Delta^{*\top}\beta + \Delta^{*\top}\omega,
\]
and hence
\begin{align*}
f_\beta^*(\Delta^*)
&= \sup_{a,\omega}
\bigl(
a\Delta^{*\top}\beta + \Delta^{*\top}\omega
- a^2\|\beta\|_2^4
+2ar(\beta)\|\beta\|_2^2
\bigr) \\
&\text{s.t. } \beta^\top\omega=0.
\end{align*}
For fixed $\omega$, the objective is a concave quadratic function of $a$, hence bounded above in $a$. If $\Delta^*$ is not orthogonal to some $\omega$ satisfying $\beta^\top\omega=0$, then
\[
\sup_{\omega:\beta^\top\omega=0}\Delta^{*\top}\omega = \infty,
\]
so $f_\beta^*(\Delta^*)=\infty$. This happens exactly when $\Delta^*\notin\col{\beta}$. Thus, restricting to $\Delta^*=\alpha\beta$ for some $\alpha\in\mathbb R$, we obtain
\begin{align*}
f_\beta^*(\Delta^*)
&= \sup_a
\bigl(
a\Delta^{*\top}\beta
-a^2\|\beta\|_2^4
+2r(\beta)a\|\beta\|_2^2
\bigr) \\
&=
\frac{\bigl(\Delta^{*\top}\beta+2r(\beta)\|\beta\|_2^2\bigr)^2}{4\|\beta\|_2^4},
\end{align*}
with optimizer
\[
a^*=
\frac{\Delta^{*\top}\beta+2r(\beta)\|\beta\|_2^2}{2\|\beta\|_2^4}.
\]The biconjugate is
\[
f_\beta^{**}(\Delta)=\sup_{\Delta^*}
\bigl(\Delta^\top\Delta^* - f_\beta^*(\Delta^*)\bigr).
\]
Since $f_\beta^*(\Delta^*)=\infty$ whenever $\Delta^*\notin\col{\beta}$, only $\Delta^*\in\col{\beta}$ contribute. Writing $\Delta^*=\alpha\beta$, we get
\begin{align*}
f_\beta^{**}(\Delta)
&= \sup_{\alpha\in\mathbb R}
\left(
\Delta^\top(\alpha\beta)
-\frac{(\beta^\top(\alpha\beta)+2r(\beta)\|\beta\|_2^2)^2}{4\|\beta\|_2^4}
\right) \\
&= \sup_{\alpha\in\mathbb R}
\left(
\alpha\beta^\top\Delta
-\frac{(\alpha+2r(\beta))^2}{4}
\right).
\end{align*}
Since $f_\beta$ is proper, convex, and lower semicontinuous,  Fenchel-Moreau yields
\[
f_\beta^{**}(\Delta)=f_\beta(\Delta).
\]
\end{proof}

\subsection{A Lemma on Schur Complement}
\begin{lemma}
\label{lem:schurcomplement}
Assume $\Sigma\succ 0$ and $\mathbb B\in\mathbb R^{d\times m}$ has full column rank, then \[\Sigma^{-1} - \mathbb{B}(\mathbb{B}^\top\Sigma\mathbb{B})^{-1}\mathbb{B}^\top \succeq 0\]
\end{lemma}
\begin{proof}
Define
\[
L \coloneqq
\begin{bmatrix}
\mathbb B^\top \Sigma^{1/2}\\
\Sigma^{-1/2}
\end{bmatrix}
\in \mathbb R^{(m+d)\times d},
\qquad
K\coloneqq LL^\top
=
\begin{bmatrix}
\mathbb B^\top \Sigma \mathbb B & \mathbb B^\top\\
\mathbb B & \Sigma^{-1}
\end{bmatrix}.
\]
Since $K\succeq 0$ and $\mathbb B^\top\Sigma\mathbb B\succ 0$, the Schur complement gives
\[
\Sigma^{-1}-\mathbb B(\mathbb B^\top\Sigma\mathbb B)^{-1}\mathbb B^\top \succeq 0.
\]
\end{proof}

\subsection{Properties of the Function \texorpdfstring{$\varphi_{p,\Lambda}^2$}{READ seminorm}}
\begin{lemma}
\label{lem:diff_varphi}
Let $1 < p < \infty$, let $q$ be its H\"older conjugate, and let
\[
\varphi_{p,\Lambda}^2(x)
\coloneqq
\inf_{\kappa\in\mathbb{R}^m}
\Bigl\{
\|x-\Theta\kappa\|_p^2 + \|\kappa\|_{\Lambda^{-1}}^2
\Bigr\},
\qquad x\in\mathbb{R}^d,
\]
where $\Theta\in\mathbb{R}^{d\times m}$ has full column rank and $\Lambda\succ 0$ is diagonal. Then the following hold.

\begin{enumerate}[label=(\arabic*)]
    \item The map $x\mapsto \varphi_{p,\Lambda}^2(x)$ is differentiable on $\mathbb{R}^d$.
    \item If one interprets $\Lambda=\infty$ on the diagonal, so that
    \[
    \varphi_{p,\infty}^2(x)
    \coloneqq
    \inf_{\kappa\in\mathbb{R}^m}\|x-\Theta\kappa\|_p^2,
    \]
    then $x\mapsto \varphi_{p,\infty}^2(x)$ is also differentiable on $\mathbb{R}^d$.
    \item In both cases,
    \[
    \|\nabla_x \varphi_{p,\Lambda}^2(x)\|_q \le 2\|x\|_p,
    \qquad x\in\mathbb{R}^d.
    \]
\end{enumerate}
\end{lemma}
\begin{proof}
Define
\[
g(z)\coloneqq \|z\|_p^2
=
\Bigl(\sum_{i=1}^d |z_i|^p\Bigr)^{2/p},
\qquad z\in\mathbb{R}^d.
\]
Since $1 < p < \infty$, the map $g$ is differentiable on $\mathbb{R}^d$, with
\[
\nabla g(z)
=
\begin{cases}
2\|z\|_p^{2-p}\bigl(|z_i|^{p-2}z_i\bigr)_{i=1}^d, & z\neq 0,\\
0, & z=0.
\end{cases}
\]
Moreover,
\[
\|\nabla g(z)\|_q = 2\|z\|_p,
\qquad z\in\mathbb{R}^d.
\]

We first consider the case $\Lambda\succ 0$. Define
\[
\Phi_\Lambda(x,\kappa)
\coloneqq
g(x-\Theta\kappa)+\|\kappa\|_{\Lambda^{-1}}^2.
\]
For each fixed $x\in\mathbb{R}^d$, the map $\kappa\mapsto \Phi_\Lambda(x,\kappa)$ is continuous, coercive, and strictly convex. Indeed, $g$ is convex, and the term $\|\kappa\|_{\Lambda^{-1}}^2$ is strictly convex because $\Lambda^{-1}\succ 0$. Hence there exists a unique minimizer, denoted by $\kappa_\Lambda(x)$, such that
\[
\varphi_{p,\Lambda}^2(x)=\Phi_\Lambda(x,\kappa_\Lambda(x)).
\]
Since $\Phi_\Lambda$ is jointly continuous and coercive in $\kappa$, standard arguments imply that $x\mapsto \kappa_\Lambda(x)$ is continuous. Therefore Danskin's Theorem \citep[Theorem 9.26]{shapiro2021lectures} yields
\[
\nabla_x \varphi_{p,\Lambda}^2(x)
=
\nabla_x \Phi_\Lambda(x,\kappa_\Lambda(x))
=
\nabla g\bigl(x-\Theta\kappa_\Lambda(x)\bigr),
\]
which proves differentiability in part (1).

Now consider the case $\Lambda=\infty$, namely
\[
\varphi_{p,\infty}^2(x)
=
\inf_{\kappa\in\mathbb{R}^m} g(x-\Theta\kappa).
\]
For each fixed $x$, the map $\kappa\mapsto g(x-\Theta\kappa)$ is continuous and coercive, because $\Theta$ has full column rank. It is also strictly convex, since $\|\cdot\|_p^2$ is strictly convex for $1<p<\infty$. Thus there exists a unique minimizer $\kappa_\infty(x)$. By the same continuity argument and Danskin's theorem,
\[
\nabla_x \varphi_{p,\infty}^2(x)
=
\nabla g\bigl(x-\Theta\kappa_\infty(x)\bigr),
\]
so part~(2) follows.

It remains to prove the gradient bound. Let $\kappa_*(x)$ denote either $\kappa_\Lambda(x)$ or $\kappa_\infty(x)$, and set
\[
r(x)\coloneqq x-\Theta\kappa_*(x).
\]
From the previous formulas,
\[
\nabla_x \varphi_{p,\Lambda}^2(x)=\nabla g(r(x)),
\]
hence
\[
\|\nabla_x \varphi_{p,\Lambda}^2(x)\|_q
=
\|\nabla g(r(x))\|_q
=
2\|r(x)\|_p.
\]
On the other hand, by optimality of $\kappa_*(x)$ and the admissible choice $\kappa=0$,
\[
\varphi_{p,\Lambda}^2(x)
\le
\|x\|_p^2.
\]
Also, by definition,
\[
\varphi_{p,\Lambda}^2(x)\ge \|r(x)\|_p^2,
\]
since the objective is $\|r(x)\|_p^2$ plus a nonnegative term in the finite-$\Lambda$ case, and equals $\|r(x)\|_p^2$ in the case $\Lambda=\infty$. Therefore
\[
\|r(x)\|_p \le \|x\|_p,
\]
and so
\[
\|\nabla_x \varphi_{p,\Lambda}^2(x)\|_q
=
2\|r(x)\|_p
\le
2\|x\|_p.
\]
This proves part (3).
\end{proof}

\begin{lemma}
\label{lem:penalty_comparison}
Fix \(p\in[1,\infty]\). Let \(\Lambda \in[0,\infty]^m\), and define
\[
\varphi_{p,\Lambda}^2(\beta)
\coloneqq
\inf_{\kappa\in\mathbb{R}^m}
\left\{ \|\beta-\Theta\kappa\|_p^2 + \|\kappa\|_{\Lambda^{-1}}^2\right\},
\]
with the conventions \(0^{-1}=\infty\), \(\infty^{-1}=0\), and \(\infty\cdot 0=0\). Then there exist constants \(0<a_p\le b_p<\infty\), depending only on \(p\) and \(d\), such that for every \(\Theta\in\mathbb{R}^{d\times m}\) and every \(\beta\in\mathbb{R}^d\),
\[
a_p\varphi_{2,a_p\Lambda}^2(\beta)
\le
\varphi_{p,\Lambda}^2(\beta)
\le
b_p\varphi_{2,b_p\Lambda}^2(\beta),
\]where \(c\Lambda\coloneqq(c\lambda_1,\dots,c\lambda_m)\) with \(c\cdot\infty=\infty\) for \(c>0\).
\end{lemma}
\begin{proof}
Since all norms on \(\mathbb{R}^d\) are equivalent, there exist constants \(0<a_p\le b_p<\infty\) such that
\[
a_p\|x\|_2^2\le \|x\|_p^2\le b_p\|x\|_2^2, \quad x\in\mathbb{R}^d.
\]Hence, for every \(\kappa\in\mathbb{R}^m\),
\[
a_p\|\beta-\Theta\kappa\|_2^2+\|\kappa\|_{\Lambda^{-1}}^2
\le
\|\beta-\Theta\kappa\|_p^2+\|\kappa\|_{\Lambda^{-1}}^2
\le
b_p\|\beta-\Theta\kappa\|_2^2+\|\kappa\|_{\Lambda^{-1}}^2,
\]
where all quantities are understood in \([0,\infty]\). Taking the infimum over \(\kappa\) gives
\[
\inf_\kappa
\Bigl\{a_p\|\beta-\Theta\kappa\|_2^2+\|\kappa\|_{\Lambda^{-1}}^2\Bigr\}
\le
\varphi_{p,\Lambda}^2(\beta)
\le
\inf_\kappa \Bigl\{b_p\|\beta-\Theta\kappa\|_2^2+\|\kappa\|_{\Lambda^{-1}}^2 \Bigr\}.
\]Now fix \(c>0\). Since \((c\Lambda)^{-1}=c^{-1}\Lambda^{-1}\) coordinatewise in the extended sense,
\[
c\|\kappa\|_{(c\Lambda)^{-1}}^2
= \|\kappa\|_{\Lambda^{-1}}^2.
\]Therefore,
\[
\inf_\kappa
\Bigl\{
c\|\beta-\Theta\kappa\|_2^2+\|\kappa\|_{\Lambda^{-1}}^2
\Bigr\}
=c\inf_\kappa
\Bigl\{\|\beta-\Theta\kappa\|_2^2+\|\kappa\|_{(c\Lambda)^{-1}}^2\Bigr\}
= c\varphi_{2,c\Lambda}^2(\beta).
\]Applying this with \(c=a_p\) and \(c=b_p\) proves the claim.
\end{proof}

\subsection{Lemmas Supporting the Proof of \texorpdfstring{\Cref{thm:rwpi}}{Thm: RWPI}}

\begin{lemma}
\label{lem:coercivity_psi}
Fix \(\Lambda\in[0,\infty]^m\), and define
\[
\psi_{p,\Lambda}(\xi)
\coloneqq
\frac14 \mathbb{E}_{\mathbb{P}_*}\bigl[\varphi_{p,\Lambda}^2(\Xi\xi)\bigr],
\qquad
\xi\in\mathbb{R}^d.
\]
Then \(\psi_{p,\Lambda}\) is even, convex, continuous, and positively homogeneous of degree \(2\). In addition,
\[
K\coloneqq \min_{\|\xi\|_p=1}\psi_{p,\Lambda}(\xi)>0,
\]
and therefore
\[
\psi_{p,\Lambda}(\xi)\ge K\|\xi\|_p^2,
\qquad
\xi\in\mathbb{R}^d.
\]
Consequently, \(\psi_{p,\Lambda}^*\) is finite and continuous on \(\mathbb{R}^d\), and for every \(M,R>0\) there exists \(b=b(M,R)<\infty\) such that
\begin{equation}
\label{eq:exact_truncation}
\psi_{p,\Lambda}^*(z-\Sigma u)
=
\max_{\|\xi\|_p\le b}
\Bigl\{
\xi^\top(z-\Sigma u)-\psi_{p,\Lambda}(\xi)
\Bigr\}
\end{equation}
for all \(\|z\|_2\le M\) and \(\|u\|_2\le R\).
\end{lemma}

\begin{proof}
We have \(\psi_{p,\Lambda}\) is finite and continuous because \(\varphi_{p,\Lambda}(x)\le \|x\|_p\) for every \(x\). Evenness, convexity, and \(2\)-homogeneity are immediate from the definition.

Now fix \(\xi\neq 0\). By \cref{asm:pd}, 
\[
\mathbb{P}_*\bigl(\varphi_{p,\Lambda}^2(\Xi\xi)>0\bigr)>0.
\]
Since the integrand is nonnegative, this implies \(\psi_{p,\Lambda}(\xi)>0\) for every \(\xi\neq 0\). Continuity on the compact unit \(p\)-sphere then yields
\[
K=\min_{\|\xi\|_p=1}\psi_{p,\Lambda}(\xi)>0.
\]
By \(2\)-homogeneity,
\[
\psi_{p,\Lambda}(\xi)
= \|\xi\|_p^2\psi_{p,\Lambda}\left(\frac{\xi}{\|\xi\|_p}\right)
\ge
K\|\xi\|_p^2,
\]
which proves the coercive lower bound. For the truncation claim, let
\[
L_{M,R}\coloneqq \sup_{\|z\|_2\le M,\ \|u\|_2\le R}\|z-\Sigma u\|_q<\infty.
\]
Then, for any \(\xi\in\mathbb{R}^d\), H\"older's inequality and the previous lower bound give
\[
\xi^\top(z-\Sigma u)-\psi_{p,\Lambda}(\xi)
\le
L_{M,R}\|\xi\|_p-K\|\xi\|_p^2.
\]
Hence, if \(\|\xi\|_p>L_{M,R}/K\), then the right-hand side is strictly negative, whereas the value at \(\xi=0\) equals \(0\). Therefore every maximizer in the definition of \(\psi_{p,\Lambda}^*(z-\Sigma u)\) lies in the ball \(\{\|\xi\|_p\le L_{M,R}/K\}\), which proves \eqref{eq:exact_truncation}. Finiteness and continuity of \(\psi_{p,\Lambda}^*\) follow immediately from the same coercive bound.
\end{proof}

Since the target $Y$ is fixed under the Wasserstein perturbation, we collapse the pair $(X,Y)$ to just the covariate $X$ in what follows. It follows from the semi-infinite duality \citep[Proposition 3]{blanchet2021statistical}, that we may write 
\[
nR_n(\beta_*+n^{-1/2}u) = \sup_{\xi \in \mathbb{R}^d}\left\{\xi^\top H_n - \frac1n\sum_{i=1}^n m_{n,i}(\xi,u)\right\},
\]
where $H_n = n^{-1/2}\sum_{i=1}^n h(x_i;\beta_*)$,
\begin{equation}
\label{eq:semi_infinite_dual}
m_{n,i}(\xi,u) \coloneqq
\sup_{\Delta\in\mathbb R^d}
\left\{
\sqrt{n}\xi^\top
\Bigl(h(x_i;\beta_*)-h(x_i+n^{-1/2}\Delta;\beta_*+n^{-1/2}u)) 
\Bigr) - c_{q,\Lambda}(\Delta)
\right\}.
\end{equation}

\begin{lemma}
\label{lem:outer_localization}
Fix \(\Lambda\in[0,\infty]^m\) and \(K>0\). Then there exists \(b_0<\infty\) depending on $K$ such that, for every
\(b\ge b_0\),
\[
\sup_{\|u\|_2\le K}
\left|
nR_n(\beta_*+n^{-1/2}u)
- \sup_{\|\xi\|_p\le b}\left\{\xi^\top H_n - M_n(\xi,u) \right\}\right|
\to 0
\qquad\text{in probability,}
\]
where \(M_{n}(\xi,u) =  \frac1n\sum_{i=1}^n m_{n,i}(\xi,u)\) is the dual function defined in
\eqref{eq:semi_infinite_dual}.
\end{lemma}

\begin{proof}
    The truncation of the maximization $\sup_\xi \{\xi^\top H_n - M_n(\xi,u)\}$ follows directly from the proof of \citet[Lemma A6]{blanchet2022confidence}, as we may simply apply the bound $c_{q,\Lambda}(\Delta)\ge \|\Delta\|_q^2$.
\end{proof}

\begin{lemma}
\label{lem:local_dual}
Fix \(\Lambda\in[0,\infty]^m\) and \(K>0\). If \(b\ge b_0(K)\) is as in
\cref{lem:outer_localization}, then
\[
\sup_{\|u\|_2\le K}
\bigl|
nR_n(\beta_*+n^{-1/2}u)-\Phi_{b}(H_n,u)
\bigr|
\to 0 \qquad\text{in probability,}
\]
where
\[
\Phi_{b}(z,u)
\coloneqq
\max_{\|\xi\|_p\le b}
\Bigl\{ \xi^\top(z-\Sigma u)-\psi_{p,\Lambda}(\xi) \Bigr\}.
\]
\end{lemma}

\begin{proof}
Write
\[
\beta_{n}(u)\coloneqq \beta_*+n^{-1/2}u,
\qquad
A_i\coloneqq \nabla_\beta h(x_i;\beta_*),
\qquad
\Xi_i^\top\coloneqq \nabla_x h(x_i;\beta_*).
\]
The dual representation of the robust
Wasserstein profile can be written as
\[
nR_n(\beta_n(u))
= \sup_{\xi\in\mathbb R^d}\left\{ \xi^\top H_n - M_n(\xi,u)\right\},
\]
where
\[
M_n(\xi,u) = \frac{1}{n}\sum_{i=1}^n m_{n,i}(\xi,u)
\]
with
\[
m_{n,i}(\xi,u)
\coloneqq
\sup_{\Delta\in\mathbb R^d}
\left\{
\sqrt{n} \xi^\top \Bigl(
h(x_i;\beta_*)-h(x_i+n^{-1/2}\Delta;\beta_n(u))
 \Bigr) - c_{q,\Lambda}(\Delta) 
\right\}.
\]
Let $D_n(\xi,u) = \xi^\top H_n - M_n(\xi,u)$. By \cref{lem:outer_localization}, it is enough to prove that
\begin{equation}
\label{eq:local_goal}
\sup_{\|\xi\|_p\le b,\ \|u\|_2\le K}
\left|
D_{n}(\xi,u)
-
\bigl\{
\xi^\top(H_n-\Sigma u)-\psi_{p,\Lambda}(\xi)
\bigr\}
\right|
\to 0
\qquad\text{in probability.}
\end{equation}
We split the proof into three steps.
\newline
\noindent\emph{Step 1: linearization of the inner increment.}
By the fundamental theorem of calculus,
\[
\sqrt{n}(h(x_i;\beta_*)-h(x_i+n^{-1/2}\Delta;\beta_n(u)) )
=
-\Xi_i^\top\Delta - A_i u + r_{n,i}(\Delta,u),
\]
where
\begin{align*}
r_{n,i}(\Delta,u)
&=
\int_0^1
\Bigl[
\nabla_x h(x_i+t n^{-1/2}\Delta;\beta_*+t n^{-1/2}u) - \nabla_x h(x_i;\beta_*) \Bigr]^\top \Delta dt
\\
&\quad
+ \int_0^1
\Bigl[
\nabla_\beta h(x_i+t n^{-1/2}\Delta;\beta_*+t n^{-1/2}u) - \nabla_\beta h(x_i;\beta_*) \Bigr]u \,dt.
\end{align*}
Hence, by \cref{asm:regularity_iii}, there exists a measurable envelope \(L_i\coloneqq
K(X_i)+K'(X_i)\) such that whenever \(\|\Delta\|_q+\|u\|_q\le n^{1/2}\),
\begin{equation}
\label{eq:remainder_bound}
\|r_{n,i}(\Delta,u)\|_q
\le n^{-1/2}L_i\bigl(\|\Delta\|_q+\|u\|_q\bigr)^2.
\end{equation}
Now fix \(b\ge b_0(K)\). Since \(c_{q,\Lambda}(\Delta)\ge \|\Delta\|_q^2\), the same coercivity argument as
in the $\Lambda = 0$ case shows that, uniformly over \(\|\xi\|_p\le b\) and \(\|u\|_2\le K\), the
supremum in the definition of \(m_{n,i}(\xi,u)\) may be restricted, at a loss \(o_{p}(1)\), to \(\Delta\) satisfying
\[
\|\Delta\|_q \le C_b(1+\|x_i\|_q)
\]
for a deterministic constant \(C_b<\infty\). On this localized set,
\eqref{eq:remainder_bound} implies
\[
\sup_{\|\xi\|_p\le b,\ \|u\|_2\le K}
\frac1n\sum_{i=1}^n
\sup_{\|\Delta\|_q\le C_b(1+\|x_i\|_q)}
\bigl|
\xi^\top r_{n,i}(\Delta,u)
\bigr|
\to 0
\qquad\text{in probability,}
\]
because \(\mathbb E_{\mathbb P_*}[L_i(1+\|X\|_q)^2]<\infty\) by assumptions.
Therefore
\begin{align}
\label{eq:m_linearized}
\sup_{\|\xi\|_p\le b,\ \|u\|_2\le K}
\Biggl| \frac1n\sum_{i=1}^n m_{n,i}(\xi,u)
- \frac1n\sum_{i=1}^n
\sup_{\Delta\in\mathbb R^d}
\Bigl\{ \xi^\top(\Xi_i^\top\Delta + A_i u) -c_{q,\Lambda}(\Delta) \Bigr\}
\Biggr| \to 0
\end{align}
in probability.
\newline
\noindent\emph{Step 2: evaluation of the linearized inner problem.}
Since the term \(\xi^\top A_i u\) does not depend on \(\Delta\), we obtain
\[
\sup_{\Delta\in\mathbb R^d}
\Bigl\{
\xi^\top(\Xi_i^\top\Delta + A_i u)-c_{q,\Lambda}(\Delta)
\Bigr\}
= \xi^\top A_i u + \sup_{\Delta\in\mathbb R^d}
\Bigl\{ (\Xi_i\xi)^\top\Delta-c_{q,\Lambda}(\Delta) \Bigr\}.
\]
By \Cref{lem:regularizer},
\[
\sup_{\Delta\in\mathbb R^d}
\Bigl\{ (\Xi_i\xi)^\top\Delta-c_{q,\Lambda}(\Delta)
\Bigr\}
= \frac14 \varphi_{p,\Lambda}^2(\Xi_i\xi).
\]
Consequently, if we define
\[
\Sigma_n\coloneqq -\frac1n\sum_{i=1}^n A_i,
\qquad
\psi_{n,p,\Lambda}(\xi)\coloneqq
\frac1{4n}\sum_{i=1}^n \varphi_{p,\Lambda}^2(\Xi_i\xi),
\]
then \eqref{eq:m_linearized} yields
\begin{equation}
\label{eq:boundary_Dn_empirical}
\sup_{\|\xi\|_p\le b,\ \|u\|_2\le K}
\left|
D_{n}(\xi,u) - \bigl\{
\xi^\top(H_n-\Sigma_n u)-\psi_{n,p,\Lambda}(\xi)
\bigr\}\right| \to 0
\qquad\text{in probability.}
\end{equation}
\newline
\noindent\emph{Step 3: law of large numbers.}
Because \(A_i=\nabla_\beta h(x_i;\beta_*)\) is integrable, the law of large numbers implies
\[
\|\Sigma_n-\Sigma\|_{\op}\to 0
\qquad\text{in probability.}
\]
Also, for \(\|\xi\|_p\le b\),
\[
0\le \varphi_{p,\Lambda}^2(\Xi_i\xi)\le \|\Xi_i\xi\|_p^2,
\]
and the growth condition in \cref{asm:regularity_iii} implies that
\[
\sup_{\|\xi\|_p\le b}\varphi_{p,\Lambda}^2(\Xi_i\xi)
\le C_b(1+\|X_i\|_q)^2
\]
for some deterministic \(C_b<\infty\). Hence a uniform law of large numbers on the compact ball
\(\{\|\xi\|_p\le b\}\) gives
\[
\sup_{\|\xi\|_p\le b}
\bigl| \psi_{n,p,\Lambda}(\xi)-\psi_{p,\Lambda}(\xi) \bigr| \to 0
\qquad\text{in probability.}
\]
Therefore
\[
\sup_{\|\xi\|_p\le b,\ \|u\|_2\le K}
\left|
\xi^\top(H_n-\Sigma_n u)-\psi_{n,p,\Lambda}(\xi)
- \bigl\{ \xi^\top(H_n-\Sigma u)-\psi_{p,\Lambda}(\xi) \bigr\} \right| \to 0
\]
in probability, and together with \eqref{eq:boundary_Dn_empirical} this proves
\eqref{eq:local_goal}. Finally, using the elementary bound
\[
\left|
\sup_{\|\xi\|_p\le b} f_n(\xi,u)-\sup_{\|\xi\|_p\le b} f(\xi,u)
\right|
\le
\sup_{\|\xi\|_p\le b}|f_n(\xi,u)-f(\xi,u)|,
\]
we conclude from \eqref{eq:local_goal} that
\[
\sup_{\|u\|_2\le K} \left| \sup_{\|\xi\|_p\le b}D_{n}(\xi,u) - \Phi_{b}(H_n,u) \right|\to 0\qquad\text{in probability.}
\]
Combining this with \cref{lem:outer_localization} completes the proof.
\end{proof}

\subsection{Lemmas Supporting the Proof of \texorpdfstring{\Cref{thm:asymptotic_dro}}{Thm: Asymptotics}}

\begin{lemma}
\label{lem:bias_ulln}
Fix \(\Lambda\in[0,\infty]^m\), and suppose the same assumptions of \cref{thm:asymptotic_dro} hold. Define
\[
\mathcal V(\beta)
\coloneqq
\Bigl(\mathbb E_{\mathbb P_*}\bigl[\varphi_{p,\Lambda}^2(\nabla_x\ell(X,Y;\beta))\bigr] \Bigr)^{1/2},
\qquad 
\mathcal V_n(\beta)
\coloneqq
\Bigl(\mathbb E_{\mathbb P_n}\bigl[\varphi_{p,\Lambda}^2(\nabla_x\ell(X,Y;\beta))\bigr] \Bigr)^{1/2}.
\]
Then there exists \(r>0\) such that
\[
\sup_{\|\beta-\beta_*\|_2\le r}
\bigl|
\mathcal V_n(\beta)-\mathcal V(\beta)
\bigr|
\to 0,
\qquad
\sup_{\|\beta-\beta_*\|_2\le r}
\bigl\|
\nabla_\beta \mathcal V_n(\beta)-\nabla_\beta \mathcal V(\beta)
\bigr\|_2
\to 0
\]
in probability.
\end{lemma}

\begin{proof}
\cref{asm:pd_2} implies that $\mathcal{V}(\beta_*)$ (or $\mathcal{V}^2(\beta_*)$) is differentiable and strictly positive. Write
\[
F_\beta(X,Y)
\coloneqq
\varphi_{p,\Lambda}^2\bigl(\nabla_x\ell(X,Y;\beta)\bigr),
\qquad
\beta\in B_r(\beta_*) \coloneqq \{\beta: \|\beta - \beta_*\|_2 \leq r\}.
\]
Then
\[
\mathcal V_n^2(\beta)=\mathbb E_{\mathbb P_n}[F_\beta(X,Y)],
\qquad
\mathcal V^2(\beta)=\mathbb E_{\mathbb P_*}[F_\beta(X,Y)].
\]
By the bound \(\varphi_{p,\Lambda}^2(x)\le \|x\|_p^2\), we obtain for some constant upper bound $G_0 > 0$
\[
|F_\beta(X,Y)|
\le
\|\nabla_x\ell(X,Y;\beta)\|_p^2
\le
G_0<\infty
\]
uniformly over \(\beta\in B_r(\beta_*)\) by \cref{asm:regularity_ii}. Since \(\beta\mapsto \nabla_x\ell(X,Y;\beta)\) is almost surely continuous, the map \(\beta\mapsto F_\beta(X,Y)\) is almost surely continuous on the compact set \(B_r(\beta_*)\). Hence the class
\[
\mathcal F_r
\coloneqq
\bigl\{
F_\beta:\beta\in B_r(\beta_*)
\bigr\}
\]
is Glivenko--Cantelli, and therefore
\[
\sup_{\|\beta-\beta_*\|_2\le r}
\bigl|
\mathcal V_n^2(\beta)-\mathcal V^2(\beta)
\bigr|
\to 0
\]
in probability.

Next, by the chain rule,
\[
\nabla_\beta F_\beta(X,Y)
=
\nabla_{\beta x}^2\ell(X,Y;\beta)^\top
\nabla\varphi_{p,\Lambda}^2\bigl(\nabla_x\ell(X,Y;\beta)\bigr).
\]
Using the growth bound on \(\|\nabla\varphi_{p,\Lambda}^2(x)\|_q \leq 2\|x\|_p\) from \cref{lem:diff_varphi}, and with \cref{asm:regularity_iii},
\begin{align*}
    \|\nabla_\beta F_\beta(X,Y)\|_2
    &\le
    \|\nabla_{\beta x}^2\ell(X,Y;\beta)\|_{2\to p}\bigl\|\nabla\varphi_{p,\Lambda}^2\bigl(\nabla_x\ell(X,Y;\beta)\bigr)
    \bigr\|_q\\
    &\le
    2G_1 \|\nabla_{\beta x}^2\ell(X,Y;\beta)\|_q \|\nabla_x\ell(X,Y;\beta)\|_p,\\
    &\le
    2G_1\cdot M'(1+\|X\|_q)\cdot G_0
\end{align*}
uniformly over \(\beta\in B_r(\beta_*)\), where $G_1>0$ is the constant that put the $2\to p$ and $q$ norm equivalent. By square-integrability of $\|X\|_q$, the vector-valued class
\[
\mathcal G_r
\coloneqq
\bigl\{
\nabla_\beta F_\beta:\beta\in B_r(\beta_*)
\bigr\}
\]
is Glivenko--Cantelli. Hence
\[
\sup_{\|\beta-\beta_*\|_2\le r}
\bigl\|
\nabla_\beta \mathcal V_n^2(\beta)
-
\nabla_\beta \mathcal V^2(\beta)
\bigr\|_2
\to 0
\]
in probability, where
\[
\nabla_\beta \mathcal V_n^2(\beta)
=
\mathbb E_{\mathbb P_n}\bigl[\nabla_\beta F_\beta(X,Y)\bigr],
\qquad
\nabla_\beta \mathcal V^2(\beta)
=
\mathbb E_{\mathbb P_*}\bigl[\nabla_\beta F_\beta(X,Y)\bigr].
\]

By \cref{asm:pd_2}, $\mathcal V(\beta_*)>0$. Since \(\beta\mapsto \mathcal V^2(\beta)\) is continuous, after shrinking \(r\) if necessary there exists \(c>0\) such that
\[
\inf_{\|\beta-\beta_*\|_2\le r}\mathcal V(\beta)\ge c.
\]
From the already proved uniform convergence of \(\mathcal V_n^2\) to \(\mathcal V^2\), it follows that $\inf_{\|\beta-\beta_*\|_2\le r}\mathcal V_n(\beta)\ge c/2$,with probability tending to one. Now uniformly over \(\beta\in B_r(\beta_*)\)
\[
\bigl|\mathcal V_n(\beta)-\mathcal V(\beta)\bigr|
=
\frac{
\bigl|\mathcal V_n^2(\beta)-\mathcal V^2(\beta) \bigr|
}{\mathcal V_n(\beta)+\mathcal V(\beta)} \le \frac{1}{c} \bigl|\mathcal V_n^2(\beta)-\mathcal V^2(\beta)\bigr|
\]
with probability tending to one. Therefore $\sup_{\|\beta-\beta_*\|_2\le r} \bigl| \mathcal V_n(\beta)-\mathcal V(\beta) \bigr| \to 0$ in probability.

Finally, on the event \(\inf_{\|\beta-\beta_*\|_2\le r}\mathcal V_n(\beta)\ge c/2\), the chain rule gives
\[
\nabla_\beta \mathcal V_n(\beta)
= \frac{\nabla_\beta \mathcal V_n^2(\beta)}{2\mathcal V_n(\beta)},
\qquad
\nabla_\beta \mathcal V(\beta)
= \frac{\nabla_\beta \mathcal V^2(\beta)}{2\mathcal V(\beta)}.
\]
Hence
\begin{align*}
&\sup_{\|\beta-\beta_*\|_2\le r}
\bigl\| \nabla_\beta \mathcal V_n(\beta)-\nabla_\beta \mathcal V(\beta) \bigr\|_2 \\
&\qquad\le
\frac{1}{c} \sup_{\|\beta-\beta_*\|_2\le r}
\bigl\| \nabla_\beta \mathcal V_n^2(\beta)-\nabla_\beta \mathcal V^2(\beta) \bigr\|_2 \\
&\qquad\quad+ \frac{2}{c^2}
\sup_{\|\beta-\beta_*\|_2\le r}
\bigl\| \nabla_\beta \mathcal V^2(\beta) \bigr\|_2
\sup_{\|\beta-\beta_*\|_2\le r}
\bigl| \mathcal V_n(\beta)-\mathcal V(\beta) \bigr|.
\end{align*}
The first term converges to \(0\) in probability by the uniform convergence of \(\nabla_\beta \mathcal V_n^2\), and the second term converges to \(0\) in probability by the uniform convergence of \(\mathcal V_n\). This proves
\[
\sup_{\|\beta-\beta_*\|_2\le r}
\bigl\|
\nabla_\beta \mathcal V_n(\beta)-\nabla_\beta \mathcal V(\beta)
\bigr\|_2
\to 0
\]
in probability, as claimed.
\end{proof}

Define
\[
\Psi_n(\beta;\delta)\coloneqq \sup_{\mathbb{P}\in \mathcal{B}_\delta(\mathbb{P}_n;c_{q,\Lambda})}\mathbb{E}_{\mathbb{P}}[\ell(X,Y;\beta)],
\]
the worst-case objective at \(\beta\) for a fixed radius $\delta > 0$.

\begin{lemma}
\label{lem:worst_case_expansion}
Let \(\delta_n=\eta/n\) with $\eta > 0$ fixed. For every compact set \(B_0\subset B_r(\beta_*)\), we have
\[
\sup_{\beta\in B_0}
\left| n^{1/2} \Bigl( \Psi_n(\beta;\delta_n)-\mathbb E_{\mathbb P_n}[\ell(X,Y;\beta)] \Bigr) - \eta^{1/2}\mathcal V_n(\beta) \right|
\to 0
\]
in probability.
\end{lemma}

\begin{proof}
Fix a compact set \(B_0\subset B_r(\beta_*)\). For \(x\in\mathbb R^d\), \(y\in\mathbb R\), \(\beta\in B_0\), and \(\lambda\ge 0\), define
\[
r_n(x,y,\beta,\lambda)
\coloneqq
\sup_{\Delta\in\mathbb R^d}
\left\{ n^{1/2} \bigl(\ell(x+n^{-1/2}\Delta,y;\beta)-\ell(x,y;\beta)\bigr) - \lambda c_{q,\Lambda}(\Delta) \right\}.
\]
By the same strong duality argument used in the proof of \citet[Proposition 6]{blanchet2022confidence},
\begin{equation}
\label{eq:boundary_dual_expansion}
n^{1/2} \Bigl( \Psi_n(\beta;\delta_n)-\mathbb E_{\mathbb P_n}[\ell(X,Y;\beta)] \Bigr)
=\inf_{\lambda\ge 0} \left\{ \eta\lambda+\mathbb E_{\mathbb P_n}\bigl[r_n(X,Y,\beta,\lambda)\bigr] \right\}.
\end{equation}
Now let \(C_{B_0}:=\sup_{\beta\in B_0}M(\beta)\), where \(M\) is the envelope from \cref{asm:regularity_ii}. By Taylor's theorem, for every \(\beta\in B_0\),
\[
n^{1/2} \bigl( \ell(x+n^{-1/2}\Delta,y;\beta) 
-\ell(x,y;\beta) \bigr)
=\nabla_x\ell(x,y;\beta)^\top \Delta
+\frac12 n^{-1/2}\Delta^\top \nabla_x^2\ell(x+\theta n^{-1/2}\Delta,y;\beta) \Delta
\]
for some \(\theta=\theta(x,y,\beta,\Delta,n)\in(0,1)\). Since all norms are equivalent in finite dimension, there exists a constant, still denoted \(C_{B_0}\) (by taking maximum), such that
\[
\left| \Delta^\top
\nabla_x^2\ell(x+\theta n^{-1/2}\Delta,y;\beta)
\Delta \right|
\le
C_{B_0}\|\Delta\|_q^2
\le
C_{B_0}c_{q,\Lambda}(\Delta),
\]
where the last inequality follows from definition of $c_{q,\Lambda}$. Therefore, whenever \(\lambda>C_{B_0}n^{-1/2}\),
\[
\nabla_x\ell(x,y;\beta)^\top\Delta- (\lambda+C_{B_0}n^{-1/2})c_{q,\Lambda}(\Delta)
\le
n^{1/2} \bigl( \ell(x+n^{-1/2}\Delta,y;\beta)-\ell(x,y;\beta) \bigr) - \lambda c_{q,\Lambda}(\Delta)
\]
and
\[
n^{1/2} \bigl( \ell(x+n^{-1/2}\Delta,y;\beta)-\ell(x,y;\beta) \bigr) - \lambda c_{q,\Lambda}(\Delta)
\le
\nabla_x\ell(x,y;\beta)^\top\Delta-(\lambda-C_{B_0}n^{-1/2})c_{q,\Lambda}(\Delta).
\]
Taking suprema over \(\Delta\in\mathbb R^d\), we obtain
\[
\frac{1}{4(\lambda+C_{B_0}n^{-1/2})}
\varphi_{p,\Lambda}^2(\nabla_x\ell(x,y;\beta))
\le
r_n(x,y,\beta,\lambda)
\le
\frac{1}{4(\lambda-C_{B_0}n^{-1/2})}
\varphi_{p,\Lambda}^2(\nabla_x\ell(x,y;\beta)).
\]
Averaging with respect to \(\mathbb P_n\) and writing \(a_n(\beta)\coloneqq \mathcal V_n^2(\beta)\), this yields
\begin{equation}
\label{eq:boundary_squeeze}
\begin{aligned}
\inf_{\lambda>C_{B_0}n^{-1/2}} \left\{ \eta\lambda+\frac{a_n(\beta)}{4(\lambda+C_{B_0}n^{-1/2})} \right\}
&\le
n^{1/2} \Bigl( \Psi_n(\beta;\delta_n)-\mathbb E_{\mathbb P_n}[\ell(X,Y;\beta)] \Bigr)
\\
&\le
\inf_{\lambda>C_{B_0}n^{-1/2}}
\left\{ \eta\lambda+\frac{a_n(\beta)}{4(\lambda-C_{B_0}n^{-1/2})} \right\}.
\end{aligned}
\end{equation}
By \cref{asm:pd_2}, \(\mathcal V(\beta_*)>0\). Since \(\mathcal V(\cdot)\) is continuous on \(B_r(\beta_*)\), after shrinking \(B_0\) if necessary there exist constants \(0<m<M<\infty\) such that
\[
m \le \inf_{\beta\in B_0}\mathcal V(\beta)
\le \sup_{\beta\in B_0}\mathcal V(\beta) \le M.
\]
Then \cref{lem:diff_varphi} implies that, with probability tending to one,
\[
\frac m2 \le
\inf_{\beta\in B_0}\mathcal V_n(\beta)
\le
\sup_{\beta\in B_0}\mathcal V_n(\beta)
\le 2M.
\]
On this event, for \(n\) sufficiently large, the one-dimensional infima in \eqref{eq:boundary_squeeze} can be computed explicitly. Indeed, with \(a_n(\beta)=\mathcal V_n^2(\beta)\),
\[
\inf_{\lambda>C_{B_0}n^{-1/2}}
\left\{
\eta\lambda+\frac{a_n(\beta)}{4(\lambda+C_{B_0}n^{-1/2})}
\right\}
= -\eta C_{B_0}n^{-1/2}+\eta^{1/2}\mathcal V_n(\beta),
\]
and
\[
\inf_{\lambda>C_{B_0}n^{-1/2}}
\left\{
\eta\lambda+\frac{a_n(\beta)}{4(\lambda-C_{B_0}n^{-1/2})}
\right\}
= \eta C_{B_0}n^{-1/2}+\eta^{1/2}\mathcal V_n(\beta).
\]
Substituting these expressions into \eqref{eq:boundary_squeeze} shows that, uniformly over \(\beta\in B_0\),
\[
\left| n^{1/2} \Bigl( \Psi_n(\beta;\delta_n)-\mathbb E_{\mathbb P_n}[\ell(X,Y;\beta)] \Bigr) - \eta^{1/2}\mathcal V_n(\beta) \right|
\le \eta C_{B_0}n^{-1/2},
\]
with probability tending to one. This proves the claim.
\end{proof}

\begin{lemma}
\label{lem:bias_linearization}
For every \(K>0\),
\[
\sup_{\|u\|_2\le K}
\left| \sqrt n \Bigl( \mathcal V_n(\beta_*+n^{-1/2}u)-\mathcal V_n(\beta_*) \Bigr) - \nabla_\beta \mathcal V(\beta_*)^\top u \right|
\to 0
\]
in probability.
\end{lemma}

\begin{proof}
Fix \(K>0\). For each \(u\) with \(\|u\|_2\le K\), the mean-value theorem gives
\[
\sqrt n
\Bigl(
\mathcal V_n(\beta_*+n^{-1/2}u)-\mathcal V_n(\beta_*)
\Bigr)
= \nabla_\beta \mathcal V_n(\tilde\beta_{n,u})^\top u,
\]
for some \(\tilde\beta_{n,u}\) on the line segment joining \(\beta_*\) and \(\beta_*+n^{-1/2}u\). The differentiation is justified because $\mathcal{V}_n^2(\beta_*) > 0$ with probability tending to 1, and $\varphi_{p,\Lambda}^2$ is differentiable everywhere by \cref{lem:diff_varphi}. For \(n\) sufficiently large, \(\tilde\beta_{n,u}\in B_r(\beta_*)\) uniformly over \(\|u\|_2\le K\). Hence
\begin{align*}
&\sup_{\|u\|_2\le K}
\left|
\nabla_\beta \mathcal V_n(\tilde\beta_{n,u})^\top u - \nabla_\beta \mathcal V(\beta_*)^\top u
\right|\\
&\le
K \sup_{\|\beta-\beta_*\|_2\le r}
\bigl\|
\nabla_\beta \mathcal V_n(\beta)-\nabla_\beta \mathcal V(\beta)
\bigr\|_2
+ K \sup_{\|\beta-\beta_*\|_2\le n^{-1/2}K}
\bigl\|
\nabla_\beta \mathcal V(\beta)-\nabla_\beta \mathcal V(\beta_*)
\bigr\|_2.
\end{align*}
The first term converges to \(0\) in probability by \cref{lem:bias_ulln}, and the second converges to \(0\) by continuity of \(\nabla_\beta \mathcal V(\,\cdot\,)\) at \(\beta_*\). This proves the lemma.
\end{proof}

\subsection{Lemmas Supporting the Proofs of Statements in \texorpdfstring{\Cref{sec:robustness}}{Section: Robustness}}
Let $U$ and $V$ be two random variables, we say $U$ \emph{stochastically dominates} $V$, denoted by \[U\succeq V,\quad (\text{ or } V \preceq U),\]if $F_U(x) \leq F_V(x)$ for all $x\in\mathbb{R}$. If we further require that $F_U(x) < F_V(x)$ for some $x\in \mathbb{R}$, then we say $U$ \emph{strictly} stochastically dominates $V$, written as $U \succ V$.

\begin{lemma}
\label{lem:stoc_dom_additivity}
    If $A \succeq B$ and $W$ is independent of both $A$ and $B$, then \[
    A+W\succeq B+W.
    \]
    \end{lemma}
    \begin{proof}
    First-order stochastic dominance means $
    F_A(t) \leq F_B(t)$, for all $t\in \mathbb{R}$. For every $x \in \mathbb{R}$,
    \[
    \mathbb{P}(A + W \le x) = \mathbb{E}\big[\mathbb{P}(A \le x - W \mid W)\big] = \mathbb{E}\big[F_A(x - W)\big],
    \]where we used the independence of $A$ and $W$. Similarly,
    \[
    \mathbb{P}(B + W \le x) = \mathbb{E}\big[F_B(x - W)\big].
    \]Since $F_A(\cdot) \le F_B(\cdot)$ pointwise, we have
    \[
    \mathbb{E}\big[F_A(x - W)\big] \le \mathbb{E}\big[F_B(x - W)\big] \quad \forall x\in \mathbb{R},
    \]hence,
    \[
    F_{A+W}(x) \le F_{B+W}(x) \quad \forall x \in \mathbb{R},
    \]i.e. $A+W \succeq B+W.$
\end{proof}

\begin{proposition}
\label{prop:stoc_dom_additivity}
Let $U,V,X,Y$ be random variables such that $X$ is independent of $U$ and $Y$ is both independent of $U$ and $V$. If $X \succeq Y$ and $U \succeq V$, then \[
X+U \succeq Y+V.
\]
\end{proposition}
\begin{proof}
    Apply \cref{lem:stoc_dom_additivity} with $(A,B,W)=(X,Y,U)$ we obtain,\[
    X+U \succeq Y+U,
    \]apply \cref{lem:stoc_dom_additivity} with $(A,B,W)=(U,V,Y)$ again we obtain,\[
    Y+U \succeq Y+V.
    \]By transitivity of stochastic dominance we conclude \[
    X+U\succeq Y+U \succeq Y+V.
    \]
\end{proof}

\begin{remark}
    The same arguments also yield strict stochastic dominance for \cref{lem:stoc_dom_additivity,prop:stoc_dom_additivity} whenever at least one premise is strict.
\end{remark}

\begin{lemma}
\label{lem:quantile_interchange}
Let $U$ be a real-valued random variable and let
\[
q_p(U)=F_U^{-1}(p):=\inf\{x\in\mathbb R:F_U(x)\ge p\}, \quad p\in(0,1).
\]
If $g:\mathbb R\to\mathbb R$ is strictly increasing and left-continuous, then $q_p(g(U))=g(q_p(U)).$
\end{lemma}
\begin{proof}
    The statement is well-known and the proof is omitted.
\end{proof}

\begin{lemma}
\label{lem:prob_bound}
Suppose that $X$ and $Y$ are independent, that $Y\ge 0$ almost surely, and that $X$ admits a bounded density $f_X$ with
\[
M_X \coloneqq \sup_{x\in\mathbb{R}} f_X(x) < \infty.
\]
For $t\in\mathbb{R}$ and $b\ge 0$, set $p(b,t) = \mathbb{P}(X +bY \leq t)$, then we have
\[
0\le p(0,t)-p(b,t)
\le \mathbb{E}\big[\min\{bM_XY,1\}\big]
\le 1.
\]
Moreover, for each fixed $y\ge 0$,
\[
F_X(t)-F_X(t-by)
\le
\min\left\{by\sup_{u\in[t-by,t]} f_X(u),1\right\},
\]
so the constant $M_X$ may be replaced by a sharper local bound.
\end{lemma}
\begin{proof}
    Fix $t\in\mathbb{R}$ and define $p(b)\coloneqq \mathbb{P}(X+bY \le t)$ for $b\ge 0$.
    By the tower property and independence of $X$ and $Y$,
    \[
    p(b)=\mathbb{E}\big[\mathbb{P}(X\le t-bY\mid Y)\big]=\mathbb{E}\big[F_X(t-bY)\big],
    \quad\text{and}\quad p(0)=F_X(t).
    \]
    Since $F_X$ is absolutely continuous with density $f_X$, for each $y\ge 0$ we have
    \[
    F_X(t)-F_X(t-by)=\int_{t-by}^{t} f_X(u)du,
    \]
    hence
    \[
    F_X(t) - F_X(t-by) = \int_{t-by}^{t} f_X(u)du
    \leq \min{\left\{by \sup_{u\in[t-by,t]} f_X(u),1\right\}}
    \leq \min{\left\{byM_X, 1\right\}}.
    \]Taking expectations gives
    \[
    p(0)-p(b) =\mathbb{E}[F_X(t)-F_X(t-bY)]
    \leq \mathbb{E}_Y\big(\min\{ b M_XY, 1 \}\big).
    \]
\end{proof}

\begin{proposition}
\label{prop:prob_bound_i}
    Suppose the assumptions of \cref{lem:prob_bound} holds. Let $t_1 > t_0$ and set $\Delta \coloneqq F_X(t_1)-F_X(t_0)>0$, suppose that $q_{1-\Delta/2}(Y)>0$, then whenever \[
    b < \frac{\Delta}{2(1+s)M_X q_{1-\Delta/2}(Y)},
    \]we have \[
    p(b(1+s),t_1) > p(b,t_0),
    \]for any $s > 0$. If, in addition, $\mathbb E(Y) < \infty$, then we may replace the sufficient condition by \[
    b < \frac{\Delta}{(1+s)M_X\mathbb{E}(Y)}.
    \]
\end{proposition}
\begin{proof}
    Set $p_1(b) = p(b,t_1)$ and $p_0(b) = p(b,t_0)$. Apply \cref{lem:prob_bound} to $p_1(b(1+s))$, we have \[
    p_1(0) - p_1(b(1+s)) \leq \mathbb{E} \min{\{b(1+s)M_XY,1\}}.
    \]
    Observe that \begin{align*}
        p_1(b(1+s)) - p_0(b) &=(p_1(b(1+s)) - p_1(0)) - (p_0(b) - p_0(0)) + p_1(0) - p_0(0)\\
    &\geq -\mathbb{E}\min{\{b(1+s)M_X Y,1\}} + 0 + \Delta.\\
    &=\Delta - \mathbb{E}\min{\{b(1+s)M_X Y,1\}}
    \end{align*}Let $c \coloneqq b(1+s)$, then for any $0<\delta<1$, we have\[
    \mathbb{E}\min{\{cM_X Y,1\}} \leq \min{\{c M_X q_{1-\delta}(Y),1\}} + \delta,
    \]choose $\delta = \Delta/2$, then the assumption $cM_X q_{1-\Delta/2}(Y) < \Delta/2$ implies \[
    \mathbb{E}\min\{cM_X Y,1\} \leq c M_X q_{1-\Delta/2}(Y) +\Delta/2 < \Delta,
    \] hence \[
    p_1(b(1+s)) - p_0(b) > 0,
    \]completing the proof. The replacement of the sufficient condition when $\mathbb E(Y) < \infty$ is straightforward.
\end{proof}

\begin{lemma}
\label{lem:noncentral_chi}
    Let $W \sim \chi_\nu^2(\lambda)$ be a non-central $\chi$-squared distribution with $\nu > 2$ degrees of freedom and non-centrality parameter $\lambda > 0$. Let $f_W$ denote the density function of $W$ supported on $[0,\infty)$, then \[
    \sup_{w\geq 0} f_W(w) \leq \frac{1}{2\sqrt{\pi}}\left[ \frac{e^{-\lambda/2}}{\sqrt{\nu-2}} + \frac{1-e^{-\lambda/2}}{\sqrt{\nu}}\right].
    \]
\end{lemma}
\begin{proof}
    Let $X \sim \chi_\nu^2$ be the standard $\chi$-squared distribution with $\nu > 2$ degrees of freedom, and set $k = \nu/2$. The mode of the density of $X$ occurs at \[x_* = \argmax_x f_X(x) = \nu - 2,\]and therefore \[
    \sup_x f_X(x) = f_X(x_*) = \frac{(k-1)^{k-1}e^{-(k-1)}}{2\Gamma(k)}.
    \]Stirling's lower bound entails for all $k > 1$:\[
    \Gamma(k) \geq \sqrt{2\pi (k-1)}(k-1)^{k-1}e^{-(k-1)},
    \]from this we obtain\[
    \sup_x f_X(x) \leq \frac{1}{2\sqrt{\pi(\nu-2)}}.
    \]Now let $W \sim \chi_\nu^2(\lambda)$, where $\lambda>0$ is a non-centrality parameter. It is well known that \[
    W\mid J=j \sim \chi_{\nu+2j}^2, \quad J\sim \rm{Pois}(\mu),\quad\mu=\lambda/2 .
    \]Meaning that, \[
    f_W(w) = \sum_{j=0}^\infty \pi_jf_{\chi^2_{\nu+2j}}(w), \quad \pi_j = e^{-\mu}\frac{\mu^j}{j!}.
    \]Hence,\[
    \sup_w f_W(w) \leq \sum_{j=0}^\infty \pi_j \sup_{x}f_{\chi^2_{\nu+2j}}(x)\leq \sum_{j=0}^\infty \pi_j \frac{1}{2\sqrt{\pi (\nu+2j-2)}},
    \]since the function $j \mapsto (\nu+2j-2)^{-1/2}$ is decreasing in $j$, we obtain for any $J_0 \in \{0,1,2,\ldots\}$ \[
    \sup_w f_W(w) \leq \frac{1}{2\sqrt{\pi}}\left[\sum_{j=0}^{J_0} \pi_j \frac{1}{\sqrt{\nu+2j-2}} + \mathbb{P}(J>J_0) \frac{1}{\sqrt{\nu + 2J_0}}\right]. 
    \]Take $J_0 = 0$, we obtain \[
    \sup_w f_W(w) \leq \frac{1}{2\sqrt{\pi}}\left[\frac{e^{-\lambda/2}}{\sqrt{\nu-2}} + \frac{1-e^{-\lambda/2}}{\sqrt{\nu}}\right].
    \]
\end{proof}

Here \(\kappa_m := \pi^{m/2}/\Gamma(m/2+1)\) denotes the volume
of the unit ball in \(\mathbb{R}^m\).

\begin{lemma}
\label{lem:prob_bound_2}
Assume \(d \ge 2\). Let \(X\) and \(Y\) be independent random vectors in \(\mathbb{R}^d\). Assume that \(X\) has a density \(f_X\) such that
\[
M_X \coloneqq \sup_x f_X(x) < \infty,
\]
and that \(\mathbb{E}\|Y\|^2 < \infty\). For \(b \ge 0\) and \(t>0\), define
\[
p(b,t) \coloneqq \mathbb{P}(\|X+bY\|^2+\|bY\|^2 \le t).
\]
Then
\[
|p(b,t)-p(0,t)|
\le M_X \kappa_d \left(2dt^{(d-1)/2} b\mathbb{E}\|Y\|
+\left(\frac d2+1\right)t^{d/2-1} b^2 \mathbb{E}\|Y\|^2\right),
\]
where
\[
\kappa_d \coloneqq \vol(B(0,1)) = \frac{\pi^{d/2}}{\Gamma(d/2+1)}.
\]
\end{lemma}

\begin{proof}
Fix \(t>0\), and write \(p(b)=p(b,t)\). Conditioning on \(Y=y\), we get
\[
p(b|y)
= \mathbb{P}(\|X+by\|^2 \le t-b^2\|y\|^2)
= \mathbb{P}(X \in B(-by,r(y))),
\]
where
\[
r(y)\coloneqq \sqrt{(t-b^2\|y\|^2)_+}.
\]
Also,
\[
p(0,t)=\mathbb{P}(X\in B(0,\sqrt t)).
\]Therefore,
\begin{align*}
|p(b,t)-p(0,t)|
&=\left|\mathbb{E}\bigl[p(b\mid Y)-p(0,t)\bigr]\right| \\
&\le\mathbb{E}\bigl|p(b\mid Y)-p(0,t)\bigr| \\
&\le M_X\mathbb{E}\left[\vol\bigl(B(-bY,r(Y)) \oplus B(0,\sqrt t)\bigr)\right],
\end{align*}
where \(\oplus\) denotes symmetric difference. Next,
\[
B(-bY,r(Y)) \oplus B(0,\sqrt t)
\subset \bigl(B(-bY,r(Y)) \oplus B(0,r(Y))\bigr)
\cup \bigl(B(0,r(Y)) \oplus B(0,\sqrt t)\bigr),
\]
hence
\begin{align*}
&\vol\bigl(B(-bY,r(Y)) \oplus B(0,\sqrt t)\bigr)
\le\\
&\vol\bigl(B(-bY,r(Y)) \oplus B(0,r(Y))\bigr) +
\vol\bigl(B(0,r(Y)) \oplus B(0,\sqrt t)\bigr).
\end{align*}

Using
\[
\vol\bigl(B(h,r)\oplus B(0,r)\bigr) \le 2d\kappa_dr^{d-1}\|h\|,
\quad
\vol\bigl(B(0,r)\oplus B(0,R)\bigr) = \kappa_d |r^d-R^d|,
\]
we obtain
\[
\vol\bigl(B(-bY,r(Y)) \oplus B(0,\sqrt t)\bigr)
\le 2d \kappa_d  r(Y)^{d-1} b\|Y\| + \kappa_d|r(Y)^d-t^{d/2}|.
\]
Since \(r(Y)\le \sqrt t\),
\[
\mathbb{E}\bigl[r(Y)^{d-1} b\|Y\|\bigr] \le t^{(d-1)/2} b\mathbb{E}\|Y\|.
\]
For the second term, let \(s=b^2\|Y\|^2\). Then
\[
|r(Y)^d-t^{d/2}| = t^{d/2}-(t-s)_+^{d/2}.
\]
Since \(d\ge 2\), the map \(u\mapsto u^{d/2}\) is \(C^1\) on \([0,t]\), so by the mean-value theorem,
\[
t^{d/2}-(t-s)^{d/2} \le \frac d2 t^{d/2-1}s \quad\text{on }\{s\le t\}.
\]
Therefore,
\[
t^{d/2}-(t-s)_+^{d/2} \le
\frac d2 t^{d/2-1}s + t^{d/2}\mathbf{1}_{\{s>t\}}.
\]
Taking expectations and using Markov's inequality,
\[
\mathbb{E}\bigl[t^{d/2}\mathbf{1}_{\{s>t\}}\bigr]
\le t^{d/2}\mathbb{P}(s>t)
\le t^{d/2-1}\mathbb{E}[s]
=t^{d/2-1}b^2\mathbb{E}\|Y\|^2.
\]
Hence,
\[
\mathbb{E}|r(Y)^d-t^{d/2}|
\le \left(\frac d2+1\right)t^{d/2-1}b^2\mathbb{E}\|Y\|^2.
\]
Combining the above estimates yields
\[
|p(b,t)-p(0,t)|
\le M_X \kappa_d
\left( 2d t^{(d-1)/2} b \mathbb{E}\|Y\|
+ \left(\frac d2+1\right)t^{d/2-1} b^2 \mathbb{E}\|Y\|^2 \right),
\]
which proves the claim.
\end{proof}

\begin{remark}
The same proof shows that independence is not needed. It suffices that \(X\) admits a conditional density given \(Y\), say \(f_{X\mid Y=y}\), such that
\[
M_{X\mid Y}
\coloneqq
\esssup_{y} \sup_x f_{X\mid Y=y}(x)
<\infty.
\]
Then the same bound holds with \(M_X\) replaced by \(M_{X\mid Y}\).
\end{remark}

\begin{corollary}
\label{corollary:prob_bound_2}
Under the assumptions of the \cref{lem:prob_bound_2}, with $X \in \mathbb{R}^d$, $Y\in \mathbb{R}^m$ and \(A\in \mathbb{R}^{d\times m}\) and \(\delta\ge 0\). Define
\[
p(b,\delta,t) \coloneqq
\mathbb{P}\bigl(\|X+bAY\|^2+\delta\|bY\|^2\le t\bigr),
\quad
r_\delta(y)\coloneqq \sqrt{(t-\delta b^2\|y\|^2)_+}.
\]
Then
\begin{align*}
|p(b,\delta,t)-p(0,0,t)|
&\le \mathbb{E}\left[
\min\left\{M_X\vol\bigl(B(-bAY,r_\delta(Y))\oplus B(0,\sqrt t)\bigr),1
\right\}\right] \\
&\le
M_X \kappa_d \left(
2d t^{(d-1)/2} b\mathbb{E}\|AY\|
+\left(\frac d2+1\right)t^{d/2-1}\delta b^2 \mathbb{E}\|Y\|^2\right).
\end{align*}
\end{corollary}

\section{Simulation Details}
\label{sec:details_sim}

Recall the notation introduced in \Cref{sec:simulation}: \(d\) denotes the dimension of \(X\) and \(\beta_*\), \(m\) the number of sources, \(n\) the sample size, \(\rho\) the correlation parameter, and \(C\) the parameter controlling the expected magnitude of each \(\theta_j\), \(j\in[m]\). The experiments in \Cref{sec:simulation} are conducted over different combinations of these parameters, and each configuration is repeated 60 times to obtain Monte Carlo averages of the performance metric (MSE).

The data-generating mechanism for \(\Theta\) and \(\beta_*\) is as follows. Let \(\theta_j \overset{\mathrm{iid}}{\sim} N(0,\sigma^2\mathbb{I}_d)\) for \(j\in[m]\), where \(\sigma^2=C/d\) for a fixed constant \(C>0\), and define \(\Theta=[\theta_1,\ldots,\theta_m]\). Let \(\kappa\in\mathbb{R}^m\) be a random unit vector. We then generate \(\beta_*\) as
\begin{equation}
\label{eq:beta_Theta}
\beta_* = \rho \Theta\kappa + \sqrt{1 - \rho^2}\,\varepsilon,
\qquad
\varepsilon\sim N(0, Cd^{-1}\mathbb{I}_d).
\end{equation}

We now specify the conditional generation of \((X,Y)\) given \(\beta_*\). In particular, conditional on \(\beta_*\), we generate
\begin{equation}
\label{eq:data_generating}
Y = X^\top \beta_* + \epsilon,
\qquad X \sim N(0,\mathbb{I}_d),
\qquad \epsilon \sim N(0,1).
\end{equation}
All simulation experiments are based on this data-generating model.

\subsection{Methods \& Hyperparameter Tunings}
\label{sec:benchmark_methods}
In each experiment, we generate a validation dataset with the same sample size as the training dataset and select the hyperparameter that minimizes the mean squared error on this validation set. We also account for computational complexity, so that the total number of grid-search evaluations is roughly comparable across estimators.

\paragraph{TransLasso}
The distance-based transfer learning benchmark, denoted \texttt{Dist-Based}, is based on \citet{li2021translasso}. Since we observe only the coefficients, we replace Step 1 of their algorithm with a naive domain-adaptation step that learns \(\hat \omega^\mathcal{A}\) on the target sample under the linear constraint \(\hat \omega^\mathcal{A} \in \col(\Theta)\). In this step, \(\hat \omega^\mathcal{A}\) estimates \(\Theta\kappa\). This leaves a single hyperparameter, denoted \(\lambda_\delta\) in their notation. It controls an \(\ell_1\)-type penalty and is selected over a grid of 30 log-spaced points on \([0.001,10]\).

\paragraph{Angle-Based.}
The \texttt{Angle-Based} transfer learning method of \citet{gu2025angle-based} aggregates the \(\theta_j\)'s using their Algorithm 2. This leaves two hyperparameters in the objective function, denoted by \(\lambda\) and \(\eta\) in their notation. These hyperparameters are tuned jointly: \(\lambda\) is selected over 8 log-spaced points on \([0.001,10]\), and \(\eta\) is selected over 8 equally spaced points on \([-5,5]\), yielding 64 hyperparameter combinations.

\paragraph{Representation-Based.}
The \texttt{Rep-Based} method implements the learning-to-learn framework of \citet{tian2025similar-rep}, where the aggregation of the \(\theta_j\)'s is obtained through a low-rank representation estimated by the spectral method in their Algorithm 2. In Algorithm 2, the intrinsic dimension \(r\) is treated as a hyperparameter and selected from three equally spaced integer values in \([1,0.75\,m]\). The transfer step in Algorithm 4 introduces an additional hyperparameter \(\gamma\), which is selected over 20 log-spaced points on \([0.001,10]\).

\paragraph{READ.}
We implement \eqref{obj:read} with \(p=2\). The selection of \(\Lambda\) is done using \Cref{alg:lambda-cv} as follows. Compute the initialization
\[
\kappa_{\rm init} = \argmin_{\kappa}\|\hat{\beta} -\Theta\kappa\|_p,
\]
then reparameterize $\Lambda$ in $a$ by $\Lambda(a)
= a\cdot \diag(|\kappa_{\rm init}|^2).$ We then tune \(a\) over 30 equally spaced points on $(\max(\kappa_{\rm init}^2))^{-1}\big[0.001,300\big]$, so that the largest diagonal entry of \(\Lambda\) can reach \(300\). We then select $\delta$ by \Cref{alg:delta} given $\Lambda(\hat a)$.

\paragraph{DRO.}
The vanilla DRO baseline corresponds to \(\Lambda=0\), the Wasserstein radius is selected according to \Cref{alg:delta}.

\subsection{Results of Task 1 Point Estimation}

\begin{table}[H]
\centering
\renewcommand{\arraystretch}{0.65}
\begin{tabular}{lccccc}
\hline
Setting & \texttt{DRO} & \texttt{TransLasso} & \texttt{Angle} & \texttt{Rep} & \texttt{READ} \\
\hline
\multicolumn{6}{l}{Varying \(\rho\) \((n=70,\, m=25)\)} \\
0.65 & 1.882 & 1.978 & 2.127 & 1.768 & 1.756 \\
0.75 & 1.882 & 1.985 & 2.127 & 1.735 & 1.684 \\
0.85 & 1.884 & 1.984 & 2.110 & 1.692 & 1.582 \\
0.95 & 1.887 & 1.992 & 2.096 & 1.622 & 1.388 \\
\hline
\multicolumn{6}{l}{Varying \(n\) \((\rho=0.85,\, m=35)\)} \\
60  & 2.154 & 2.071 & 2.139 & 1.781 & 1.633 \\
75  & 1.847 & 1.956 & 2.182 & 1.756 & 1.562 \\
90  & 1.626 & 1.746 & 1.989 & 1.553 & 1.429 \\
105 & 1.550 & 1.612 & 1.884 & 1.449 & 1.402 \\
\hline
\multicolumn{6}{l}{Varying \(m\) \((\rho=0.85,\, n=90)\)} \\
15 & 1.626 & 1.764 & 1.962 & 1.470 & 1.431 \\
25 & 1.626 & 1.739 & 1.968 & 1.522 & 1.428 \\
35 & 1.626 & 1.746 & 1.989 & 1.553 & 1.429 \\
45 & 1.626 & 1.735 & 1.978 & 1.587 & 1.429 \\
\hline
\end{tabular}
\caption{Task 1 high-\(m\) results. Reports out-of-sample MSE for Panels 1.1--1.3 of \Cref{fig:sim_1}.}
\label{table:task1_highm}
\end{table}

\begin{table}[H]
\centering
\renewcommand{\arraystretch}{0.65}
\begin{tabular}{lccccc}
\hline
Setting & \texttt{DRO} & \texttt{TransLasso} & \texttt{Angle} & \texttt{Rep} & \texttt{READ} \\
\hline
\multicolumn{6}{l}{Varying \(\rho\) \((n=70,\, m=7)\)} \\
0.65 & 1.860 & 1.993 & 2.057 & 1.659 & 1.702 \\
0.75 & 1.865 & 2.003 & 2.034 & 1.605 & 1.609 \\
0.85 & 1.870 & 2.015 & 2.003 & 1.514 & 1.476 \\
0.95 & 1.877 & 2.021 & 1.990 & 1.404 & 1.268 \\
\hline
\multicolumn{6}{l}{Varying \(n\) \((\rho=0.85,\, m=7)\)} \\
60  & 2.191 & 2.181 & 2.093 & 1.617 & 1.541 \\
75  & 1.845 & 1.981 & 2.027 & 1.553 & 1.478 \\
90  & 1.654 & 1.802 & 1.967 & 1.448 & 1.415 \\
105 & 1.563 & 1.639 & 1.852 & 1.403 & 1.394 \\
\hline
\multicolumn{6}{l}{Varying \(m\) \((\rho=0.85,\, n=90)\)} \\
4  & 1.625 & 1.778 & 1.825 & 1.424 & 1.358 \\
8  & 1.643 & 1.778 & 1.956 & 1.394 & 1.391 \\
12 & 1.620 & 1.773 & 1.953 & 1.429 & 1.415 \\
16 & 1.629 & 1.729 & 1.993 & 1.480 & 1.458 \\
\hline
\end{tabular}
\caption{Task 1 low-\(m\) results. Reports out-of-sample MSE for Panels 1.4--1.6 of \Cref{fig:sim_1}.}
\label{table:task1_lowm}
\end{table}

\begin{figure}[!htb]
    \centering
    \includegraphics[width=1.0\linewidth]{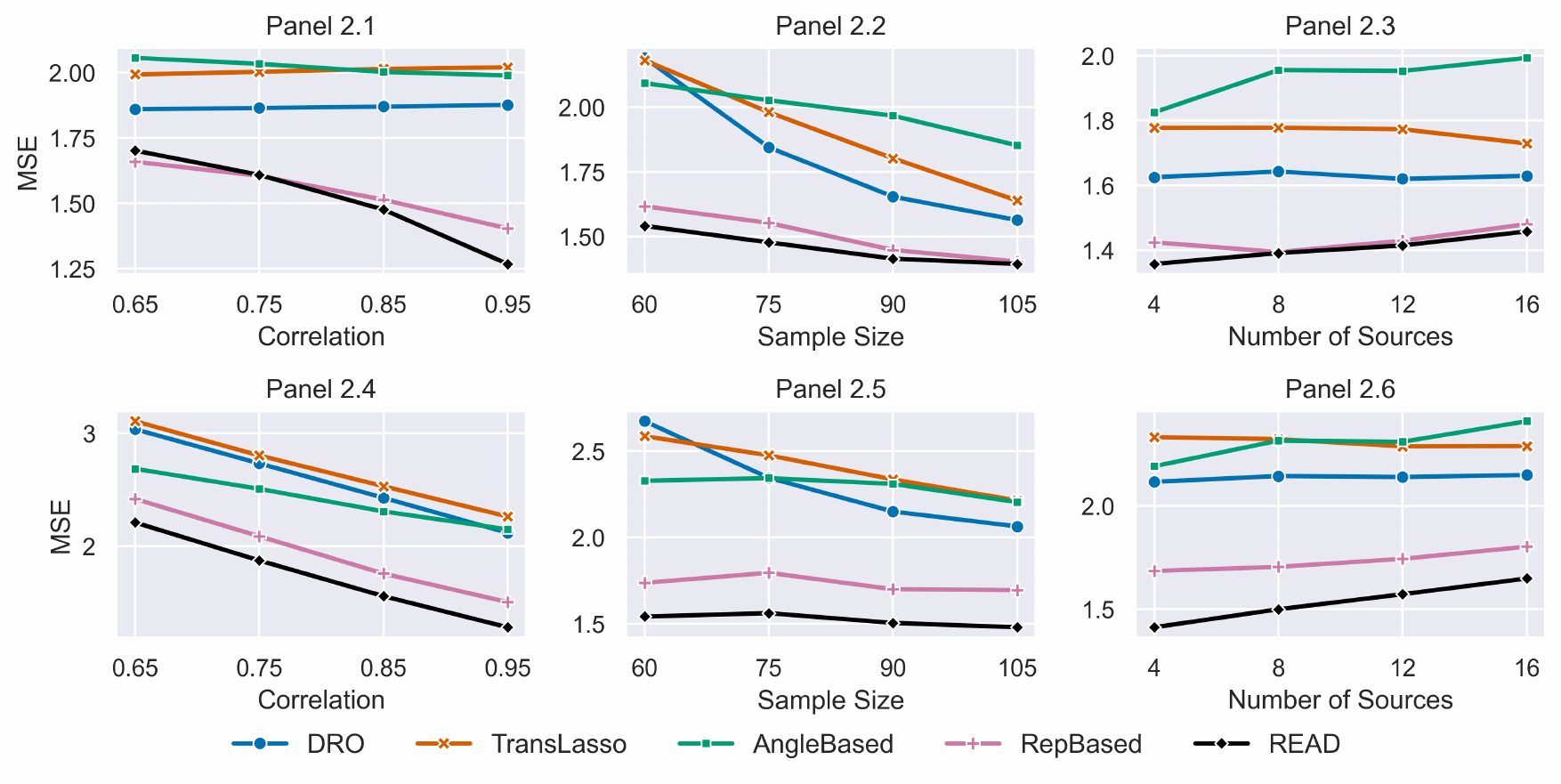}
    \caption{Results for the high-\(m\) regime. Panels 1.1--1.3 report Task 1 point-estimation performance under the target distribution, while Panels 1.4--1.6 report Task 2 OOD performance under future distributions generated according to \eqref{asm:drift_model}.} 
    \label{fig:sim_3}
\end{figure}

\subsection{Results of Task 2 Out-of-Distribution Generalization}

\begin{table}[H]
\centering
\renewcommand{\arraystretch}{0.65}
\begin{tabular}{lccccc}
\hline
Setting & \texttt{DRO} & \texttt{TransLasso} & \texttt{Angle} & \texttt{Rep} & \texttt{READ} \\
\hline
\multicolumn{6}{l}{Varying \(\rho\) \((n=70,\, m=25)\)} \\
0.65 & 2.988 & 3.037 & 2.788 & 2.640 & 2.414 \\
0.75 & 2.686 & 2.738 & 2.573 & 2.322 & 2.065 \\
0.85 & 2.384 & 2.477 & 2.403 & 2.040 & 1.693 \\
0.95 & 2.081 & 2.174 & 2.238 & 1.711 & 1.391 \\
\hline
\multicolumn{6}{l}{Varying \(n\) \((\rho=0.85,\, m=35)\)} \\
60  & 2.700 & 2.491 & 2.455 & 2.150 & 1.848 \\
75  & 2.329 & 2.367 & 2.429 & 2.067 & 1.708 \\
90  & 2.124 & 2.256 & 2.350 & 1.979 & 1.597 \\
105 & 2.030 & 2.143 & 2.274 & 1.847 & 1.546 \\
\hline
\multicolumn{6}{l}{Varying \(m\) \((\rho=0.85,\, n=90)\)} \\
15 & 2.154 & 2.314 & 2.355 & 1.786 & 1.539 \\
25 & 2.140 & 2.275 & 2.362 & 1.922 & 1.578 \\
35 & 2.124 & 2.256 & 2.350 & 1.979 & 1.597 \\
45 & 2.108 & 2.241 & 2.350 & 2.052 & 1.611 \\
\hline
\end{tabular}
\caption{Task 2 high-\(m\) results. Reports out-of-dist MSE for Panels 2.1--2.3 of \Cref{fig:sim_2}.}
\label{table:task2_highm}
\end{table}

\begin{table}[H]
\centering
\renewcommand{\arraystretch}{0.65}
\begin{tabular}{lccccc}
\hline
Setting & \texttt{DRO} & \texttt{TransLasso} & \texttt{Angle} & \texttt{Rep} & \texttt{READ} \\
\hline
\multicolumn{6}{l}{Varying \(\rho\) \((n=70,\, m=7)\)} \\
0.65 & 3.036 & 3.107 & 2.686 & 2.419 & 2.211 \\
0.75 & 2.734 & 2.804 & 2.508 & 2.088 & 1.874 \\
0.85 & 2.429 & 2.531 & 2.309 & 1.758 & 1.559 \\
0.95 & 2.119 & 2.263 & 2.149 & 1.506 & 1.283 \\
\hline
\multicolumn{6}{l}{Varying \(n\) \((\rho=0.85,\, m=7)\)} \\
60  & 2.672 & 2.585 & 2.328 & 1.737 & 1.543 \\
75  & 2.345 & 2.475 & 2.343 & 1.795 & 1.562 \\
90  & 2.150 & 2.335 & 2.309 & 1.700 & 1.506 \\
105 & 2.064 & 2.213 & 2.204 & 1.696 & 1.480 \\
\hline
\multicolumn{6}{l}{Varying \(m\) \((\rho=0.85,\, n=90)\)} \\
4  & 2.118 & 2.335 & 2.194 & 1.685 & 1.411 \\
8  & 2.146 & 2.326 & 2.318 & 1.705 & 1.499 \\
12 & 2.141 & 2.290 & 2.313 & 1.744 & 1.572 \\
16 & 2.151 & 2.292 & 2.413 & 1.803 & 1.648 \\
\hline
\end{tabular}
\caption{Task 2 low-\(m\) results. Reports out-of-dist MSE for Panel 2.4--2.6 of \Cref{fig:sim_2}.}
\label{table:task2_lowm}
\end{table}

\subsection{Results of Task 3: Coverage Post \texorpdfstring{$\Lambda$}{Lambda}-Selection}
\label{sec:cr_coverage}
We present a simulation study to assess the coverage probability of the READ confidence region \eqref{eq:confidence_region} in \cref{prop:linearmodel} when \(\Lambda\) is selected by the data-driven procedure in \Cref{alg:lambda-cv}. Given the population parameter \(\beta_*\), we generate samples from the linear model
\[
Y = X^\top \beta_* + \epsilon,
\qquad X \sim N(0,\mathbb{I}_d),
\qquad \epsilon \sim N(0,1).
\]
The population parameters \((\beta_*,\Theta)\) are generated according to \eqref{eq:beta_Theta} with \(C=1\). We consider three different dimension settings: \((d,m)=(5,2)\), \((10,4)\), and \((15,7)\). For each setting, we vary the sample-to-dimension ratio \(n/d \in \{15,30,45\}\) to assess convergence as the sample size increases. For each pair of dimension and \(n/d\) ratio, we repeat the coverage experiment 1000 times and report the empirical coverage probability at each nominal level \(1-\alpha\). Ideally, the empirical coverage should be close to the nominal coverage.

For each pair of dimension and \(n/d\) ratio, we estimate coverage over
\[
\rho \in \{0.6,0.7,0.8,0.9,0.95\},
\qquad
1-\alpha \in \{0.95,0.9,0.85,0.8\}.
\]
In total, we estimate \(3\times 3\times 5\times 4\) coverage probabilities, each based on 1000 trials.

\paragraph{\(\Lambda\)-Selection.}
For each simulated dataset, we first compute the full-sample ERM estimator
\(\hat\beta_{\erm}\). We then compute
\[
\kappa_{\rm init} = \argmin_\kappa
\left\{ \|\hat\beta_{\erm}-\Theta\kappa\|_2^2 \right\},
\]
and use this initialization to define the one-dimensional family
\[
\Lambda(a) = a\diag(\kappa_{\init}^2).
\]
The scale parameter \(a\) is selected by \(K\)-fold cross-validation over a grid of 9 equally spaced points in
\[
(\min_j \kappa_{{\rm init},j}^2)^{-1}[0.01,400],
\]
augmented with the limiting candidate \(a=\infty\), which corresponds to
\texttt{KGDRO}. After selecting \(\hat a\), we set \(\hat\Lambda=\Lambda(\hat a)\) and construct the READ confidence region on the full sample using \(\hat\Lambda\).

The results are reported in \Cref{table:coverage_1,table:coverage_2_3}. Overall, the empirical coverage supports the validity of the proposed confidence region after data-driven selection of \(\Lambda\). When the sample-to-dimension ratio is small, \(n/d=15\), the procedure slightly under-covers. This is expected from a large-sample approximation. As \(n/d\) increases from \(15\) to \(30\) and then to \(45\), the empirical coverage moves closer to the nominal level across all three dimension settings. At \(n/d=45\), the \(\rho\)-averaged empirical coverage is within one percentage point of the target level for all values of \(1-\alpha\) and all choices of \((d,m)\). The coverage is sometimes slightly above and sometimes slightly below the nominal level, with no systematic dependence on \(\rho\). These results indicate that the RWPI-based confidence region remains well calibrated after the data-driven selection of \(\Lambda\), and that the calibration improves as the sample size grows.

\begin{table}[H]
\centering
\renewcommand{\arraystretch}{0.65}
\begin{tabular}{lcccccc}
\hline
\(1-\alpha\) & \(\rho=0.60\) & \(\rho=0.70\) & \(\rho=0.80\) & \(\rho=0.90\) & \(\rho=0.95\) & Average \\
\hline
\multicolumn{7}{l}{Experiment 1: \((d,m,n/d)=(5,2,15)\)} \\
0.95 & 0.941 & 0.938 & 0.919 & 0.939 & 0.936 & 0.935 \\
0.90 & 0.891 & 0.886 & 0.863 & 0.887 & 0.880 & 0.881 \\
0.85 & 0.843 & 0.833 & 0.819 & 0.833 & 0.820 & 0.830 \\
0.80 & 0.784 & 0.777 & 0.767 & 0.790 & 0.767 & 0.777 \\
\hline
\multicolumn{7}{l}{Experiment 2: \((d,m,n/d)=(5,2,30)\)} \\
0.95 & 0.949 & 0.942 & 0.952 & 0.950 & 0.941 & 0.947 \\
0.90 & 0.884 & 0.897 & 0.903 & 0.900 & 0.886 & 0.894 \\
0.85 & 0.828 & 0.851 & 0.853 & 0.848 & 0.830 & 0.842 \\
0.80 & 0.783 & 0.805 & 0.801 & 0.798 & 0.781 & 0.794 \\
\hline
\multicolumn{7}{l}{Experiment 3: \((d,m,n/d)=(5,2,45)\)} \\
0.95 & 0.932 & 0.945 & 0.951 & 0.935 & 0.942 & 0.941 \\
0.90 & 0.882 & 0.891 & 0.907 & 0.886 & 0.885 & 0.890 \\
0.85 & 0.837 & 0.846 & 0.873 & 0.832 & 0.842 & 0.846 \\
0.80 & 0.785 & 0.796 & 0.817 & 0.779 & 0.797 & 0.795 \\
\hline
\end{tabular}
\caption{Empirical coverage probabilities of the READ confidence region after data-driven selection of \(\Lambda\) for \((d,m)=(5,2)\). Each entry is based on 1000 simulation trials.}
\label{table:coverage_1}
\end{table}

\begin{table}[H]
\centering
\renewcommand{\arraystretch}{0.65}
\begin{tabular}{lcccccc}
\hline
\(1-\alpha\) & \(\rho=0.60\) & \(\rho=0.70\) & \(\rho=0.80\) & \(\rho=0.90\) & \(\rho=0.95\) & Average \\
\hline
\multicolumn{7}{l}{Panel A: \((d,m)=(10,4)\)} \\
\multicolumn{7}{l}{\(n/d=15\)} \\
0.95 & 0.934 & 0.940 & 0.925 & 0.934 & 0.951 & 0.937 \\
0.90 & 0.882 & 0.898 & 0.868 & 0.879 & 0.888 & 0.883 \\
0.85 & 0.827 & 0.845 & 0.814 & 0.823 & 0.826 & 0.827 \\
0.80 & 0.773 & 0.780 & 0.755 & 0.781 & 0.767 & 0.771 \\
\hline
\multicolumn{7}{l}{\(n/d=30\)} \\
0.95 & 0.941 & 0.955 & 0.945 & 0.947 & 0.934 & 0.944 \\
0.90 & 0.892 & 0.896 & 0.893 & 0.899 & 0.870 & 0.890 \\
0.85 & 0.842 & 0.843 & 0.836 & 0.840 & 0.807 & 0.834 \\
0.80 & 0.803 & 0.784 & 0.789 & 0.791 & 0.755 & 0.784 \\
\hline
\multicolumn{7}{l}{\(n/d=45\)} \\
0.95 & 0.945 & 0.943 & 0.948 & 0.950 & 0.957 & 0.949 \\
0.90 & 0.894 & 0.881 & 0.898 & 0.905 & 0.904 & 0.896 \\
0.85 & 0.847 & 0.838 & 0.840 & 0.858 & 0.844 & 0.845 \\
0.80 & 0.802 & 0.797 & 0.786 & 0.798 & 0.781 & 0.793 \\
\hline
\multicolumn{7}{l}{Panel B: \((d,m)=(15,7)\)} \\
\multicolumn{7}{l}{\(n/d=15\)} \\
0.95 & 0.926 & 0.938 & 0.935 & 0.935 & 0.930 & 0.933 \\
0.90 & 0.862 & 0.885 & 0.880 & 0.889 & 0.875 & 0.878 \\
0.85 & 0.809 & 0.839 & 0.841 & 0.832 & 0.818 & 0.828 \\
0.80 & 0.776 & 0.797 & 0.787 & 0.777 & 0.751 & 0.778 \\
\hline
\multicolumn{7}{l}{\(n/d=30\)} \\
0.95 & 0.942 & 0.938 & 0.938 & 0.953 & 0.944 & 0.943 \\
0.90 & 0.884 & 0.902 & 0.894 & 0.904 & 0.890 & 0.895 \\
0.85 & 0.830 & 0.863 & 0.842 & 0.838 & 0.831 & 0.841 \\
0.80 & 0.774 & 0.819 & 0.797 & 0.781 & 0.787 & 0.792 \\
\hline
\multicolumn{7}{l}{\(n/d=45\)} \\
0.95 & 0.939 & 0.958 & 0.948 & 0.957 & 0.953 & 0.951 \\
0.90 & 0.885 & 0.895 & 0.899 & 0.901 & 0.891 & 0.894 \\
0.85 & 0.838 & 0.852 & 0.854 & 0.849 & 0.842 & 0.847 \\
0.80 & 0.786 & 0.798 & 0.801 & 0.798 & 0.796 & 0.796 \\
\hline
\end{tabular}
\caption{Empirical coverage probabilities of the READ confidence region after data-driven selection of \(\Lambda\) for \((d,m)=(10,4)\) and \((d,m)=(15,7)\). Each entry is based on 1000 simulation trials.}
\label{table:coverage_2_3}
\end{table}

\subsection{Results of Task 4: Coverage of Future \texorpdfstring{$\Theta$}{Theta}-Invariant Shifts}

\begin{table}[H]
\centering
\renewcommand{\arraystretch}{0.65}
\begin{tabular}{llccccccc}
\hline
Setting & Method & 0.00 & 0.05 & 0.10 & 0.15 & 0.20 & 0.25 & 0.30 \\
\hline
\multicolumn{9}{l}{\((d,m)=(5,2)\)} \\
& \texttt{READ} & 0.9520 & 0.9440 & 0.9054 & 0.8659 & 0.8305 & 0.7896 & 0.7482 \\
& \texttt{DRO}  & 0.9520 & 0.6960 & 0.6506 & 0.5942 & 0.5461 & 0.5231 & 0.4902 \\
& \texttt{ERM}  & 0.9560 & 0.7375 & 0.6934 & 0.6406 & 0.5932 & 0.5671 & 0.5330 \\
\hline
\multicolumn{9}{l}{\((d,m)=(10,4)\)} \\
& \texttt{READ} & 0.9600 & 0.9148 & 0.8887 & 0.8264 & 0.7664 & 0.6986 & 0.6426 \\
& \texttt{DRO}  & 0.9480 & 0.6419 & 0.5902 & 0.5486 & 0.4885 & 0.4307 & 0.3929 \\
& \texttt{ERM}  & 0.9680 & 0.7104 & 0.6612 & 0.6198 & 0.5610 & 0.4973 & 0.4581 \\
\hline
\multicolumn{9}{l}{\((d,m)=(15,7)\)} \\
& \texttt{READ} & 0.9440 & 0.9162 & 0.8351 & 0.7674 & 0.6846 & 0.6267 & 0.5272 \\
& \texttt{DRO}  & 0.9480 & 0.6340 & 0.5608 & 0.4978 & 0.4191 & 0.3860 & 0.3051 \\
& \texttt{ERM}  & 0.9720 & 0.7404 & 0.6635 & 0.6080 & 0.5216 & 0.4861 & 0.4006 \\
\hline
\end{tabular}
\caption{Task 4 OOD-coverage results. Columns correspond to the signal ratio \(\tau/\sigma\), with the \(0.00\) column denoting the target-inference case \(\tau=\sigma=0\). Rows report empirical coverage for \texttt{READ}, \texttt{DRO}, and \texttt{ERM} under the three dimension settings. The nominal coverage level is \(0.95\).}
\label{table:task4_ood_coverage}
\end{table}

\section{Additional Real Data Details}
\subsection{Selections of ADTs via Screening Data}
\label{sec:screen}

We first preprocess the screening data by retaining the top 4000 highly variable genes (HVG) from the gene expression (GEX) features using Scanpy's implementation of the Seurat v3 method for HVG selection \citep{wolf2018scanpy,stuart2019comprehensive}, and denote the resulting features by \(X_\hvg\). We then apply the \texttt{log1p} transformation to both \(Y\) (ADT) and \(X_\hvg\) (GEX). The HVG selection is performed \emph{solely} on the \texttt{CD14+ Mono} cells. For each ADT target, for example \(Y_{\texttt{CD41}}\), we estimate out-of-sample predictiveness from \(X_\hvg\) by cross-validation. Specifically, we split \((X_\hvg, Y_{\texttt{CD41}})\) into training and hold-out sets. On the training set, we fit an inner cross-validation pipeline consisting of standardization and elastic net regression \citep{zou2005elasticnet}, and then evaluate the resulting predictor on the hold-out set using out-of-sample \(R^2\). Repeating this procedure over different hold-out splits provides an estimate of predictive performance for each ADT target. We use 5 folds in both the outer and inner cross-validation loops.

The screening results led us to identify a top-ranked contiguous block of 9 cluster-of-differentiation (CD) ADTs, followed next by \texttt{HLA-DR}. We restrict attention to the 9 CD markers in order to study a more homogeneous family of surface protein targets in the multi-task learning analysis. Unlike the CD markers, \texttt{HLA-DR} is an MHC class II antigen rather than a CD-numbered differentiation marker. The selected CD markers are \texttt{CD32}, \texttt{CD35}, \texttt{CD41}, \texttt{CD9}, \texttt{CD172a}, \texttt{CD54}, \texttt{CD14}, \texttt{CD13}, and \texttt{CD45}. Although these targets are not all monocyte markers in exactly the same sense, they form a biologically informative surface-protein panel for \texttt{CD14+ Mono} cells, capturing monocyte identity, innate immune function, adhesion and activation state, and platelet-associated inflammatory interactions. The out-of-sample values of \(R^2\), as well as the Pearson and Spearman correlations, are reported in \Cref{table:screen_values}.

\begin{table}[ht]
\centering
\renewcommand{\arraystretch}{0.65}
\begin{tabular}{lccc}
\hline
Protein & $R^2_{\rm{oof}}$ & Pearson $r$ & Spearman $\rho$ \\
\hline
CD32    & 0.516 & 0.719 & 0.679 \\
CD35    & 0.485 & 0.697 & 0.610 \\
CD41    & 0.427 & 0.654 & 0.511 \\
CD172a  & 0.399 & 0.633 & 0.594 \\
CD54    & 0.391 & 0.626 & 0.624 \\
CD9     & 0.390 & 0.626 & 0.513 \\
CD14    & 0.372 & 0.610 & 0.546 \\
CD13    & 0.346 & 0.589 & 0.560 \\
CD45    & 0.341 & 0.585 & 0.550 \\
HLA-DR  & 0.309 & 0.557 & 0.497 \\
\hline
\end{tabular}
\caption{Top RNA-to-protein predictability results within the screening batch \texttt{s3d7}, evaluated using out-of-fold $R^2$ and correlation metrics.}
\label{table:screen_values}
\end{table}

\subsection{Details of Scenario 1: Multi-Source Batch Transfer}
\label{sec:details_task_i}

In this task, we apply HVG selection to \(X\), reducing its dimension to \(400\). We denote the resulting GEX matrix by \(X_\hvg\). Both \(Y\) and \(X_\hvg\) are then transformed using \texttt{log1p}. Recall that \(Y\) contains 9 targets. To remove the intercept effect, we standardize \(X_\hvg\) separately for each of the four source datasets and demean \(Y\) accordingly. We then fit an elastic net model without an intercept to each scaled source dataset \(\{(\mathbf{X}_\hvg^{(j)},\mathbf{Y}^{(j)})\}_{j=1}^4\). The experiment results and the source-domain \(R^2\) values for each ADT target are reported in \Cref{table:task1_r2,table:task1_source_r2}.

The transfer learning procedure is described in \Cref{sec:data}. The aggregation steps for \texttt{TransLasso} and \texttt{AngleBased} follow the simulation setup; see \Cref{sec:simulation}. For each subsample of 360 training observations from the 1{,}460 observations in the target dataset, we split the data evenly into 180 training observations and 180 validation observations. We fit a standardization transform for \(X_\hvg\) and a demeaning transform for \(Y\) using only the training split. These fitted transformations are then applied to the validation and test data. Hyperparameters are selected by minimizing the mean squared error on the validation set. We use a validation split of the same size as the training split, mimicking the leave-one-out cross-validation procedure used in the simulations. We summarize the hyperparameter grid below, using the notation of \Cref{sec:details_sim}:
\begin{itemize}
    \item \texttt{TransLasso}: \(\lambda_\delta\) is selected over a grid of 15 log-spaced points on \([0.001,10]\).
    
    \item \texttt{AngleBased}: \(\lambda\) is selected over a grid of 7 log-spaced points on \([0.001,10]\), and \(\eta\) is selected over a grid of 7 equally spaced points on \([-5,5]\).
    
    \item \texttt{READ}: We implements \Cref{alg:lambda-cv} by first computing
    \[
    \kappa_{\rm init} = \argmin_\kappa \|\hat{\beta} - \Theta\kappa\|_p,
    \]
    where \(\hat{\beta}\) is the baseline \texttt{DRO} estimator. We then validate over $\Lambda(a) = a\cdot \diag(\kappa_{\rm init}^2)$, where \(a\) is selected over 20 equally spaced points on
    \[
    {\max(\kappa_{\rm init}^2)}^{-1}\left[0.001, {300}\right],
    \]
    together with the additional choice \(a=\infty\). The Wasserstein radius \(\delta\) is selected over a grid of 7 log-spaced points on \([0.001,10]\).

    \item \texttt{RepBased}: The intrinsic dimension $r$ is selected in $\{1,3\}$, with $\gamma$ selected over a grid of 10 equally spaced points in $[0.001,10]$.
\end{itemize}
The test-set $R^2$ are presented in \Cref{table:task1_r2}.

\begin{table}[ht]
\centering
\renewcommand{\arraystretch}{0.65}
\begin{tabular}{lccccc}
\hline
Protein & \texttt{DRO} & \texttt{READ} & \texttt{TransLasso} & \texttt{AngleBased} & \texttt{RepBased} \\
\hline
CD32   & 0.506 & \textbf{0.616} & 0.573 & 0.612 & 0.588 \\
CD35   & 0.325 & \textbf{0.527} & 0.470 & 0.517 & 0.449 \\
CD41   & 0.186 & 0.319 & 0.296 & \textbf{0.324} & 0.132 \\
CD9    & 0.167 & 0.288 & 0.257 & \textbf{0.304} & 0.157 \\
CD172a & 0.400 & \textbf{0.488} & 0.467 & 0.462 & 0.463 \\
CD54   & 0.373 & 0.443 & 0.416 & \textbf{0.454} & 0.412 \\
CD14   & 0.490 & \textbf{0.597} & 0.530 & 0.594 & 0.578 \\
CD13   & 0.551 & 0.570 & \textbf{0.573} & 0.537 & 0.547 \\
CD45   & 0.283 & 0.381 & 0.364 & 0.296 & \textbf{0.389} \\
\hline
\end{tabular}
\caption{Test-set $R^2$ values for each method across the 9 ADT targets in Task 1.}
\label{table:task1_r2}
\end{table}

\begin{table}[ht]
\centering
\renewcommand{\arraystretch}{0.65}
\begin{tabular}{lcccc}
\hline
Protein & Source 1 & Source 2 & Source 3 & Source 4 \\
\hline
CD32   & 0.560 & 0.390 & 0.530 & 0.568 \\
CD35   & 0.393 & 0.449 & 0.350 & 0.651 \\
CD41   & 0.401 & 0.202 & 0.290 & 0.488 \\
CD9    & 0.412 & 0.222 & 0.286 & 0.449 \\
CD172a & 0.452 & 0.425 & 0.422 & 0.510 \\
CD54   & 0.379 & 0.429 & 0.340 & 0.443 \\
CD14   & 0.476 & 0.271 & 0.383 & 0.456 \\
CD45   & 0.453 & 0.414 & 0.415 & 0.456 \\
CD13   & 0.187 & 0.342 & 0.108 & 0.403 \\
\hline
\end{tabular}
\caption{Source-domain \(R^2\) values for each ADT target across the four source datasets in Task 1.}
\label{table:task1_source_r2}
\end{table}

\subsection{Details of Scenario 2: Multi-Task Learning of ADT}
\label{sec:details_task_ii}

In Task 2 of \Cref{sec:data}, we consider a multi-task learning setting in which we first learn coefficients for 8 of the 9 ADT targets and then transfer this information to the remaining held-out target. We apply HVG selection to reduce \(X\) to a 400-dimensional feature matrix \(X_\hvg\), and the experiment is conducted using only the target dataset. The full target dataset is used to fit the 8 source coefficients that form \(\Theta\). For the held-out target, we repeatedly subsample 180 observations as the training split and 180 observations as validations, then use the remaining observations as the test set. The fitting procedure and hyperparameter tuning are the same as in Task I; see \Cref{sec:details_task_i}. The in-sample \(R^2\) values of the auxiliary ADT prediction tasks used to construct \(\Theta\), together with the test-set \(R^2\) values for the held-out-target prediction task, are reported in \Cref{table:task2_results}.

\begin{table}[ht]
\centering
\renewcommand{\arraystretch}{0.65}
\begin{tabular}{lcccccc}
\hline
Protein & Auxiliary Task & \texttt{DRO} & \texttt{READ} & \texttt{TransLasso} & \texttt{AngleBased} & \texttt{RepBased} \\
\hline
CD32   & 0.679 & 0.506 & \textbf{0.604} & 0.575 & 0.577 & 0.587 \\
CD35   & 0.595 & 0.325 & \textbf{0.509} & 0.447 & 0.469 & 0.428 \\
CD41   & 0.421 & 0.186 & \textbf{0.378} & 0.314 & 0.341 & 0.185 \\
CD9    & 0.418 & 0.167 & \textbf{0.363} & 0.307 & 0.327 & 0.221 \\
CD172a & 0.559 & 0.400 & \textbf{0.503} & 0.474 & 0.412 & 0.474 \\
CD54   & 0.554 & 0.373 & \textbf{0.501} & 0.452 & 0.439 & 0.455 \\
CD14   & 0.684 & 0.490 & 0.649 & 0.607 & \textbf{0.656} & 0.626 \\
CD13   & 0.672 & 0.551 & \textbf{0.608} & 0.582 & 0.523 & 0.570 \\
CD45   & 0.498 & 0.283 & \textbf{0.435} & 0.321 & 0.309 & 0.435 \\
\hline
\end{tabular}
\caption{Task 2 results across the 9 ADT targets. The second column reports the in-sample \(R^2\) values of the auxiliary ADT prediction tasks used to construct \(\Theta\), and the remaining columns report test-set \(R^2\) values for the downstream held-out-target prediction problem.}
\label{table:task2_results}
\end{table}

\subsection{Scenario 3: Inference for Invariant Coordinates}
\label{sec:additional_task}

In this additional real-data experiment, we compare the interval lengths of the READ confidence region in \eqref{eq:confidence_region} with those of the vanilla DRO confidence region, corresponding to the special case \(\Lambda=0\). The goal is to examine whether the representation-aware geometry induced by READ leads to shorter projected intervals in directions informed by the external representation.

We first use the screening dataset described in \Cref{sec:screen} to select a common set of GEX features that are jointly predictive of the ADT targets. Specifically, we fit a multi-task elastic-net model using \texttt{MultiTaskElasticNetCV} from the \texttt{sklearn} package, with the nine selected ADT targets as multivariate responses. We then rank genes by the \(\ell_2\)-norm of their coefficient vectors across the nine tasks, and retain the top 20 GEX features. This gives a low-dimensional feature set on which the empirical covariance matrix is well-conditioned.

Using these 20 selected GEX features, we next fit separate linear regressions on each of the four source datasets described in \Cref{sec:details_task_i}. For each ADT target, the four resulting source coefficient vectors are collected to form the estimated representation matrix \(\hat\Theta\in\mathbb R^{20\times 4}\). We then apply the same \(\Lambda\)-selection procedure as in Task 1, using \Cref{alg:lambda-cv}, to obtain the data-adaptive alignment matrix \(\hat\Lambda\). Based on this choice, we construct the READ confidence region \(\hat{\mathcal C}_{0.95}(\hat\Lambda)\), as well as the vanilla DRO confidence region \(\hat{\mathcal C}_{0.95}(0)\).

We compare the two regions through two types of projected coordinate intervals. First, we project each confidence region onto the representation subspace \(\col{(\hat\Theta)}\subset\mathbb R^{20}\), and then compute the marginal interval length for each of the 20 original GEX coefficient coordinates after this projection. This measures how much READ reduces uncertainty in the feature-coordinate representation of the source-informed subspace. Second, we consider the invariant representation coordinates \(\kappa\in\mathbb R^4\), defined through the least-squares projection map
\[
\beta \mapsto (\hat\Theta^\top \hat\Theta)^{-1}\hat\Theta^\top \beta .
\]
From here we compute the four marginal interval length for each coordinate of \(\kappa\). For each of the nine ADT targets, we report the average percentage reduction in marginal interval length achieved by READ relative to vanilla DRO. Specifically, for a collection of projected coordinate intervals, we compute
\[
100\times
\left(
1-\frac{\text{average interval length under READ}}
{\text{average interval length under vanilla DRO}}
\right).
\]
Positive values therefore indicate that the READ confidence region yields shorter projected intervals than the vanilla DRO region.

The results in \Cref{fig:additional_data,table:projected_interval_reduction} show that READ generally produces shorter projected intervals than vanilla DRO. The reduction is especially consistent for the invariant coordinates \(\kappa\), where READ improves upon vanilla DRO for all nine ADT targets, with average reductions ranging from \(2.67\%\) to \(23.88\%\). For the projected 20-dimensional coefficient intervals, the reductions are also positive for eight of the nine targets, with only a small increase for CD9. This pattern is consistent with the theoretical message of \Cref{sec:projected_inference}: representation-aware alignment is most directly beneficial after projecting onto the stable invariant coordinates, where increasing alignment strength is expected to sharpen the projected confidence region while preserving coverage.

\begin{table}[ht]
\centering
\renewcommand{\arraystretch}{0.70}
\begin{tabular}{lcc}
\hline
Protein & \(\col{(\hat\Theta)}\) projection & \(\kappa\) projection \\
\hline
CD32   & 5.04  & 12.08 \\
CD35   & 5.06  & 5.99  \\
CD41   & 1.53  & 2.67  \\
CD9    & -2.83 & 3.36  \\
CD172a & 6.55  & 11.86 \\
CD54   & 10.46 & 16.63 \\
CD14   & 18.21 & 23.88 \\
CD13   & 18.54 & 22.38 \\
CD45   & 7.76  & 16.13 \\
\hline
\end{tabular}
\caption{Average percentage reduction in marginal interval length of READ relative to vanilla DRO for the two projected confidence regions in the additional real-data experiment.}
\label{table:projected_interval_reduction}
\end{table}


\end{document}